\documentclass[11pt]{article}
\pdfoutput=1
\usepackage[utf8]{inputenc}
\usepackage{cite}
\usepackage{amsmath,amssymb,amsbsy,amstext,amsthm,simplewick,amsfonts}
\usepackage{graphicx}
\usepackage{hyperref}
\hypersetup{colorlinks=true,linkcolor=teal,citecolor=orange3,urlcolor=blue,pdfencoding=auto}
\usepackage{cleveref}
\usepackage{wrapfig}
\usepackage{upgreek}
\usepackage{bm} 
\usepackage{framed}
\usepackage{bbm}
\usepackage{textcomp}
\usepackage{tikz}
\usepackage{pifont}
\usetikzlibrary{matrix,shapes,fit,tikzmark,calc}
\usepackage{adjustbox}
\usepackage{makecell}
\usepackage{tcolorbox}
\usepackage{physics}
\usepackage{empheq}
\usepackage[normalem]{ulem}
\usepackage{enumitem}
\usepackage{array}
\usepackage{dsfont}
\usepackage{ulem}
\usepackage{geometry}
\usepackage{floatrow}
\usepackage{layout}
\usepackage{multirow}
\usepackage{authblk}
\usepackage{indentfirst}
\usepackage{mdframed}
\usepackage{xcolor}
\usepackage{xspace}
\usepackage{cancel}
\usepackage[bottom]{footmisc}

\numberwithin{equation}{section}

\usepackage{caption}
\usepackage{subcaption}

\definecolor{blue3}{RGB}{31,119,180}
\definecolor{red3}{RGB}{214,39,40}
\definecolor{orange3}{RGB}{255,127,14}
\definecolor{green3}{RGB}{44,160,44}

\newcommand{\at}[1]{\textcolor{red3}{AT: #1}}
\newcommand{\gar}[1]{\textcolor{blue3}{GAR: #1}}

\usepackage{colortbl}
\definecolor{lightgreen}{cmyk}{0.2, 0, 0.2, 0.2}
\definecolor{lightgray}{cmyk}{0.1,0.2,0,0.1}
\definecolor{lightgray2}{cmyk}{0.1,0.1,0,0.1}

\newcommand*\diff{\mathop{}\!\mathrm{d}}

\makeatletter
\newlength{\apb@width}
\newcommand{\autoparbox}[2][c]{\settowidth{\apb@width}{#2}\parbox[#1]{\apb@width}{#2}}

\makeatother


\newcommand{\beq}{\begin{equation}}
\newcommand{\eeq}{\end{equation}}
\newcommand{\bea}{\begin{eqnarray}}
\newcommand{\eea}{\end{eqnarray}}

\def\bfk{\textbf{k}}
\def\bfx{\textbf{x}}
\def\bfy{\textbf{y}}
\def\bfq{\textbf{q}}

\def\h{h}
\def\ha{\mathfrak{h}}

\usepackage{pgfplots}
\pgfplotsset{width=10cm,compat=1.9}
\usepgfplotslibrary{fillbetween}

\allowdisplaybreaks[1]
\begin{document}


\begin{titlepage}
\setcounter{page}{1} \baselineskip=15.5pt 
\thispagestyle{empty}

\begin{center}
{\fontsize{22.5}{22.5}\centering \bf Holographic Cosmology at Finite Time }
\end{center}

\vskip 18pt
\begin{center}
\noindent
{\fontsize{12}{18}\selectfont Gonçalo Araújo-Regado\footnote{\tt \, goncalo-araujo@oist.jp}$^{,a,b}$, Ayngaran Thavanesan\footnote{\tt \, at735@cantab.ac.uk}$^{,a,c,d}$, and Aron C. Wall\footnote{\tt \, aroncwall@gmail.com}$^{,a,e}$}
\end{center}

\begin{center}
\vskip 8pt
$a$\textit{ Department of Applied Mathematics and Theoretical Physics, University of Cambridge, Wilberforce Road, Cambridge, CB3 0WA, UK.} \\
$b$\textit{ Okinawa Institute of Science and Technology,
1919-1 Tancha, Onna, Okinawa 904-0495, Japan.} \\
$c$\textit{ Kavli Institute for Theoretical Physics, Santa Barbara, CA 93106, USA.} \\
$d$\textit{ Laboratory for Theoretical Fundamental Physics, Institute of Physics, École Polytechnique Fédérale de Lausanne (EPFL), CH-1015 Lausanne, Switzerland.}\\
$e$\textit{ School of Natural Sciences, Institute for Advanced Study, Princeton, NJ 08540 USA.}
\end{center}


\vspace{1.4cm}

\noindent We investigate Cauchy Slice Holography in de Sitter spacetime. By performing a $T^2$ deformation of a (bottom-up) dS/CFT model, we obtain a holographic theory living on flat Cauchy slices of de Sitter, for which time is an emergent dimension, associated with an RG flow. In this $T^2$-deformed field theory, the dS/CFT is an IR fixed point rather than a UV fixed point, potentially affecting discussions of naturalness. As in the case of AdS/CFT, the terms in the $T^2$ deformation depend on the dimension and the bulk matter sector; in this article we consider gravity, plus optionally a scalar field of arbitrary mass.  We compute scalar and graviton two-point correlation functions in the deformed boundary theory, and demonstrate precise agreement with finite-time wavefunction coefficients, which we calculate independently on the bulk side.  The results are analytic in the scalar field dimension $\Delta$, and may therefore be continued to arbitrary generic values, including the principal series. Although many aspects of the calculations are similar to the AdS/CFT case, some new features arise due to the complex phases which appear in cosmology.  Our calculations confirm previous expectations that the holographic counterterms are purely imaginary, when expressed in terms of wavefunction coefficients.  But cosmological correlators, calculated by the Born rule, are shifted in a more complicated and nonlinear way.


\end{titlepage} 


\newpage
{
\hypersetup{linkcolor=black}
\tableofcontents
}

\newpage


\section{Introduction}

\subsection{Emergent Space}

In the AdS/CFT holographic correspondence, a $D = d+1$ dimensional quantum gravity theory in AdS emerges from a $d$-dimensional QFT~\cite{Maldacena:1997re,Gubser:1998bc,Witten:1998qj}. The extra emergent dimension is a spatial dimension, typically represented by the coordinate $z$, in e.g. the AdS-Poincar\'{e} metric:
\begin{equation}
    ds^2 = \frac{L^2_{\text{AdS}}}{z^2}(dz^2 + \eta_{ab}\,dx^a \,dx^b),
\end{equation}
where $L_{\text{AdS}}$ is the AdS length scale $\eta^{ab}$ is the Minkowski metric in $d$ dimensions. The CFT provides boundary conditions at $z = 0$, while the dual bulk is in the region $z > 0$.  Thus, there is an emergent space dimension.\footnote{In addition, there are typically one or more large compact dimensions that emerge from the $R$-symmetries of the boundary CFT, e.g.~the $S_5$ direction in $\text{AdS}_5 \times \text{S}_5$.  But as unbroken supersymmetry is incompatible with unitarity in de Sitter spacetime, such additional emergent dimensions are less likely to play a role in the cosmological dualities considered in this paper.}  This metric has a scale invariance symmetry of the form: $z \to az, x^a \to ax^a$, corresponding to the dilatation symmetry of the dual CFT.  

If we now turn on a single-trace source $J$ in the dual CFT, the boundary field theory can have non-trivial beta functions, leading to a new metric which may not be scale-invariant.  These beta functions may be calculated (up to scheme-dependence) using the bulk equations of motion.  This is called \emph {holographic RG flow}~\cite{Susskind:1998dq,Alvarez:1998wr,Balasubramanian:1999jd,Skenderis:2002wp,Myers:2010xs,Shyam:2016zuk,Shyam:2017znq,Shyam:2020ocx}. At leading order, such sources may be classified by their dimension $\Delta_J$.  In Wilsonian language:
\begin{itemize}
    \item if $\Delta_J > 0$, the source is \textbf{relevant} (decaying at $z \to 0$), so that the theory remains a CFT in the UV limit;
    \item if $\Delta_J < 0$, it is \textbf{irrelevant} (blowing up at $z \to 0$, corresponding to possibly bad UV behaviour in the dual field theory);
    \item if $\Delta_J = 0$, the source is \textbf{marginal}, and one must look at nonlinear beta functions to determine the direction of flow (if any).
\end{itemize}
In all of these cases, the boundary conditions are still imposed at $z = 0$.  There is, however, an alternative way of studying the $z > 0$ region, and this is to impose Dirichlet-like boundary conditions at some surface with $z > 0$.  This can be done by doing a $T^2$-\emph{deformation} of the original ``seed'' CFT, so-called because it includes an irrelevant (double-trace) product of the stress-tensor with itself in the special combination:\footnote{Not coincidentally, this is the same combination as the kinetic term $\Pi_{ab}\Pi^{ab} - \Pi^2/(d-1)$ of the ADM Hamiltonian constraint.  As discussed in \cite{Araujo-Regado:2022gvw}, if one attempts to deform by a different linear combination than the one in \eqref{eqn:TTterm}, the resulting bulk Hamiltonian constraint becomes nonlocal even at the AdS scale.  This happens in any case where you try to aim for a Hamiltonian constraint in the bulk that doesn't close on the constraint algebra.}
\begin{equation}\label{eqn:TTterm}
    \lambda \left(T_{ab}T^{ab} - \frac{1}{d-1}T^2\right),
\end{equation}
along with other irrelevant and marginal terms, whose form depends on the dimension $d$ and the matter content~\cite{Taylor:2018xcy,Hartman:2018tkw,Shyam:2018sro,Caputa:2019pam,Shyam:2020ocx,Araujo-Regado:2022gvw}.  (This is a generalisation of the $T\overline{T}$ deformation in $d=2$ originally proposed by Zamolodchikov~\cite{Zamolodchikov:2004ce,Smirnov:2016lqw,Aharony:2018vux}, which is the correct deformation to use if the bulk AdS$_3$ theory looks like pure gravity at low energies~\cite{McGough:2016lol,Shyam:2017znq,Kraus:2018xrn,Donnelly:2018bef,Guica:2019nzm,Donnelly:2019pie,Mazenc:2019cfg,Coleman:2020jte,Shyam:2021ppn,AliAhmad:2025kki,Callebaut:2025thw,Callebaut:2025uye,Guica:2025jkq}.)  

In order to go to $z > 0$, it is important to use the sign of $\lambda$ (for us $\lambda > 0$) for which the deformation is not unitary, for which the theory has a complex spectrum at high energies.  The reason for this choice is clarified by thinking about cases in which the $T^2$ theory instead has an emergent time dimension.

\subsection{Emergent Time}  What if instead we have an emergent time dimension?  In~\cite{Araujo-Regado:2022gvw} (and follow-up work~\cite{Araujo-Regado:2022jpj,Khan:2023ljg,Soni:2024aop,Khan:2025gld,Shyam:2025ttb}), it was shown that the exact same $T^2$ deformation can be used to study a holographic theory living on a time slice $\Sigma$ of an asymptotic AdS theory.  This was dubbed ``Cauchy Slice Holography'' (CSH)\footnote{Using the appropriate analogue of global hyperbolicity in an asymptotically AdS spacetime.}  It requires starting with the Euclidean CFT on some hyperbolic geometry $g_{ab}$ and deforming it to a critical value of $\lambda$ (depending on $g_{ab}$) at which the theory crosses a phase transition from real to imaginary values of the stress-tensor.  After this phase transition, the slice $\Sigma$ is embedded in a Lorentzian signature spacetime.

An alternative approach to CSH is to start with an asymptotically de Sitter spacetime.  The study of the $T^2$ deformation is simpler, because there is no need to cross through a phase transition to obtain a theory living on a Cauchy slice, as $\mathcal{I}^+$ is already (the conformal limit of) a Cauchy slice.  This also more closely matches our observed universe, which appears to be asymptotically de Sitter at late times~\cite{SupernovaSearchTeam:1998fmf,SupernovaCosmologyProject:1998vns,SDSS:2006lmn,BOSS:2016wmc,Planck:2018vyg}, as well as approximately so during inflation~\cite{Starobinsky:1980te,Guth:1980zm,Linde:1981mu}.    

But the price we pay for this, is that we would have to start with a ``dS/CFT'' seed CFT, whose dual bulk spacetime (when evaluated on a flat Euclidean metric) corresponds to the dS-Poincar\'{e} metric:
\begin{equation}
    ds^2 = \frac{L^2_{\text{dS}}}{\eta^2}(
    -d\eta^2 + \delta_{ab}\,dx^a \,dx^b
    ),
\end{equation}
where $L_{\text{dS}}$ is the de Sitter length scale (related to the Hubble parameter in cosmology by $H^2=1/L^2_{\text{dS}}$) and $\eta < 0$ is a bulk conformal time coordinate, with $\eta = 0$ at $\mathcal{I}^+$.  More generally, if you turn on single-trace boundary sources, it should be possible to get a QFT at $\mathcal{I}^+$ which is dual to some non-trivial FLRW cosmology in which deviations from dS are supported by scalar fields.\footnote{More generally, to get a universe with a thermal matter sector, so as to add a $\rho_\text{matter}$ term to the Friedmann equation,  it would be necessary to turn on ``mixed'' sources involving classical probability distributions, which are only homogeneous and isotropic \emph{on average}.}

In part because of the absence of supersymmetry and/or simple string theory solutions as a guide, such dual CFTs are much more difficult to find. Nevertheless, various proposals have been made, including but not limited to (see e.g.~\cite{Anninos:2012qw,Flauger:2022hie,Harlow:2022qsq,Galante:2023uyf} for a more comprehensive list of references on de Sitter holography):
\begin{itemize}
    \item the Higher-Spin dS/CFT correspondence~\cite{Anninos:2011ui,Ng:2012xp,Anninos:2012ft,Anninos:2017eib,DeLuca:2021pej}.
    \item ``domain wall/cosmology correspondence''~\cite{Cvetic:1996vr,Skenderis:2006fb}
    \item boundary duals for Einstein gravity in $dS_3$ arising from taking a limit of a complex level $k$ in WZW models~\cite{Castro:2011xb,Castro:2011ke,Ouyang:2011fs,Castro:2012gc,Castro:2020smu,Hikida:2022ltr,Castro:2023dxp,Fliss:2023muk}.
    \item the complex Liouville model \cite{Collier:2024kmo,Collier:2024kwt,Collier:2024lys,Collier:2025pbm,Collier:2025lux} with central charge $c = 13 + i\nu$.
    \item holographic duals to dS through double-scaled SYK~\cite{Susskind:2021omt,Rahman:2022jsf,Susskind:2022bia,Narovlansky:2023lfz,Verlinde:2024znh,Yuan:2024utc,Verlinde:2024zrh}.
    \item field theories capturing the microscopic degrees of freedom of dS obtained by deforming appropriate seed holographic CFTs with the $T\overline{T}+\Lambda$ deformation~\cite{Gorbenko:2018oov,Lewkowycz:2019xse,Shyam:2021ciy,Coleman:2021nor,Silverstein:2022dfj,Batra:2024kjl,Silverstein:2024xnr,Philcox:2025faf,Shyam:2025ttb,Chang:2025ays}.
    \item potential string constructions of dS/CFT~\cite{Hull:1998vg,Hull:1998ym,Hull:1999mt,Balasubramanian:2001rb,Dong:2010pm,Dijkgraaf:2016lym,DeLuca:2021pej,Cotler:2024xzz,Thavanesan:2025ibm}.
\end{itemize}
Also, it is known that dS/CFT cannot be the Wick rotation of a normal unitary field theory, as it does not in general satisfy the ``reflection positivity'' property of such field theories.\footnote{This should not be a surprise as the theory is not defined by such a Wick rotation, but is rather a Euclidean boundary of a Lorentzian bulk. This means that there is no reason for the $\dagger$ operation to reverse a dimension of space, as it does in Wick rotations of Lorentzian QFTs.}  (In a recent paper~\cite{Thavanesan:2025ibm}, two of us proposed a general class of large-$N$ gauge theories called ``Kosmic Field Theories'' in which the complex phases of $N$ and $\lambda$ are chosen to satisfy all reality conditions for unitary cosmology, but more work is needed to construct concrete, physically realistic examples of dS/CFT from this paradigm.)

In this article, we set ourselves a different task.  Assuming \emph{as a hypothesis} the existence of some dS/CFT whose bulk dual contains specified bulk fields (namely massive scalar fields and gravitons)\footnote{This is sometimes called the \emph{bottom-up} approach, as opposed to a \emph{top-down} approach where a holographic duality is justified from some UV complete construction in e.g.~string theory.}, we show how to deform it by a $T^2$ deformation in such a way as to impose Dirichlet-like boundary conditions on a slice $\Sigma$ at some finite time $\eta$.  (In other words, the deformed theory matches onto the usual Hamiltonian ADM formulation of General Relativity~\cite{Arnowitt:1962hi,York:1972sj}.)  Despite the theoretical uncertainties about the nature of the initial ``seed'' CFT, in some ways the story is actually much simpler than Cauchy Slice Holography in AdS, because an arbitrarily tiny deformation (any small $\lambda > 0$) suffices to define a theory living at finite time $\eta < 0$ (i.e.~$t = \text{finite}$).  Put another way, the dS/CFT must \emph{already} have an imaginary stress-tensor, and thus it \emph{starts off} in the correct phase to have an emergent time dimension.\footnote{If you start with dS/CFT on a sphere S$_d$, a sufficiently large deformation would eventually take you across a phase transition to the Euclidean part of the geometry. But there is no need to cross this threshold to identify the dual bulk state. In \cite{Araujo-Regado:2022jpj} it was argued that one can recover both the Hartle-Hawking and Vilenkin no-boundary proposals by taking different superpositions of seed CFTs with opposite anomalies.}  

\subsection{New Aspects of the de Sitter $T^2$ deformation} \label{sec:new aspects}

This procedure was already briefly sketched in \cite{Araujo-Regado:2022gvw,Araujo-Regado:2022jpj}, but here we provide more details, and resolve some difficulties.  Specifically, we start with a CFT that is holographically dual to pure gravity, plus (optionally) an additional scalar field $\phi$ with mass $m^2 \ge 0$, which in the holographic dictionary corresponds to conformal dimensions by the relation:
\begin{equation}
\Delta(d-\Delta) = m^2L_{\text{dS}}^2 = \frac{m^2}{H^2},
\end{equation}
where the two solutions correspond to the dimension $\Delta_\mathcal{O} \equiv \Delta = d/2 + \nu$ of a scalar primary CFT operator $\cal O$ and the dimension $\Delta_J = d - \Delta_\mathcal{O} = \Delta = d/2 - \nu$ of its corresponding conjugate source $J$, with $\nu=\sqrt{d^2/4-m^2L_{\text{dS}}^2}$.  Following Hartman et al. \cite{Hartman:2018tkw}, we do a $T^2$ deformation, and compare the resulting two-point function to a bulk version of the same calculation.

\paragraph{Imaginary Counterterms.}  Although broadly speaking our results are similar to \cite{Hartman:2018tkw}, there are some new features not present in the AdS version of finite cutoff holography.  The most straightforward one is that the sign of $\lambda$ (and the other terms in the $T^2$ deformation) is now required to be \emph{imaginary}.  This includes various \emph{holographic counterterms}, which are needed to remove bulk divergences in the CFT limit $\eta \to 0^-$.\footnote{These are IR divergences from a bulk point of view, and UV divergences from a boundary point of view.}  This phenomenon was previously discussed in~\cite{Maldacena:2002vr,McFadden:2009fg,McFadden:2010na,McFadden:2010jw,McFadden:2010vh,McFadden:2011kk,Easther:2011wh,Bzowski:2011ab,Bzowski:2012ih,Bzowski:2013sza,McFadden:2013ria,Afshordi:2016dvb,Afshordi:2017ihr,Nastase:2019rsn,Bzowski:2019kwd,Penin:2021sry,Bzowski:2023nef}.  

\paragraph{Nonlinearity of Cosmological Counterterms.}  This in turn leads to some subtle unexpected consequences when comparing the holographic calculation to the bulk correlator.  Specifically, the holographic duality relates the partition function to the Wheeler-DeWitt (WDW) wavefunction of the universe:
\begin{equation}
Z_{T^2}[g_{ab}, \phi] = \Psi_\text{WDW}[g_{ab}, \phi],
\end{equation}
where on the LHS the variables are interpreted as sources in the $T^2$-deformed field theory, while on the RHS they are interpreted as configuration space variables in the Hamiltonian formalism of $d+1$ dimensional quantum gravity (hence the emergent time dimension).  By differentiating both sides w.r.t.~the same variables, one obtains:
\begin{equation}\label{wavefn_dict}
\langle \mathcal{O}(x_1) \ldots \mathcal{O}(x_1) \rangle_\text{connected} = -c_n(x_1, \ldots, x_n),
\end{equation}
where the LHS is the $n$-pt function of the $T^2$ field theory, and the RHS is the \emph{wavefunction coefficient} which is the Taylor expansion of $-\log \Psi_\text{WDW}$ w.r.t.~variables (and which can therefore be complex).  It is these wavefunction coefficients which are most closely related to the dual field theory, as also observed by \cite{Bzowski:2011ab,Bzowski:2012ih,Bzowski:2013sza,Bzowski:2019kwd,Bzowski:2023nef}.

Probability distributions in cosmology are instead given by \emph{cosmological correlators}, which are obtained from $\Psi_\text{WDW}$ via the Born rule, which is schematically an integral over all sources $J$ of the form:
\begin{equation}
\langle \hat{X} \rangle_\text{cosmo} =\int 
[dJ]\,\Psi^*[J]\,\Psi[J]\,X, 
\end{equation}
where $\hat{X}$ is expressed as a bulk QM operator in the $J$ basis, in terms of $J$ and ${\cal O} = \delta/ \delta J$.  (Notice that the wavefunction comes in \emph{twice} and there is an integral that is simply not present in \eqref{wavefn_dict}.)  The relationship is already nontrivial at the level of the scalar two-point correlators \cite{Sengor:2021zlc,Sengor:2022hfx,Thavanesan:2025kyc}, which  are related to the real and imaginary parts of $c_2$ as follows (suppressing the dependence on $x_1$ and $x_2$):
\begin{align}
    \expval{\Phi \Phi}_{\Psi}&=\frac{1}{2\textbf{Re}[c_2]}, \\
    \expval{\Phi\Pi_\Phi+\Pi_\Phi\Phi}_{\Psi}&=-\frac{\textbf{Im}[c_2]}{\textbf{Re}[c_2]},\label{mixed}
    \\
    \expval{\Pi_\Phi\Pi_\Phi}_{\Psi}&=\frac{|c_2|^2}{2\textbf{Re}[c_2]} 
    = \frac{\textbf{Re}[c_2]}{2} + \frac{\textbf{Im}[c_2]^2}{2\textbf{Re}[c_2]}.\label{pipi}
\end{align}
As we have already mentioned, there exist various holographic counterterms, which are proportional to boundary integrals like e.g.~$\Phi^2$, or $R$ (in the analogous graviton sector). These counterterms act as a linear shift of $\textbf{Im}[c_2]$; however it can be seen above that this acts as a nonlinear correction to \eqref{pipi}. In order for one to obtain convergent $\expval{\Phi\Pi_\Phi+\Pi_\Phi\Phi}_{\Psi}$ and $\expval{\Pi_\Phi\Pi_\Phi}_{\Psi}$ correlators, one must work with the holographically renormalised wavefunction (see Figure \ref{fig:flow_picture}). There is no analogue of this nonlinearity in the AdS case, where the analogous coefficients (which may as well be calculated in Euclidean signature) are strictly real and hence \eqref{pipi} becomes linear in $c_2$ and its counterterm.

\paragraph{Principal Series.}  Another new feature of dS is that for heavy scalar fields with $m^2L_{\text{dS}}^2 > d^2/4$, the primary operators are in the \emph{principal series} which has complex operator dimensions $\Delta = d/2 + i\nu$, with $\nu \in \mathbb{R}$.  If we consider product operators such as $\phi^2$, this implies the possibility of operators that are neither relevant, irrelevant, nor marginal.  In particular, we modify our definitions to cover complex dimensions as follow:
\begin{itemize}
    \item A \textbf{relevant} operator has $\textbf{Re}[\Delta_J] > 0$;
    \item An \textbf{irrelevant} operator has $\textbf{Re}[\Delta_J] < 0$;
    \item A \textbf{marginal} operator has 
    $\textbf{Re}[\Delta_J] = 0$ \emph{and} $\textbf{Im}[\Delta_J] = 0$;    
    \item A \textbf{lateral} operator has 
    $\textbf{Re}[\Delta_J] = 0$ \emph{but} $\textbf{Im}[\Delta_J] \ne 0$.    
\end{itemize}
These cases differ in their convergence properties when included in a $T^2$ deformation.

\subsection{Constructing The Deformation}

As in the analysis of \cite{Araujo-Regado:2022gvw}, we start by writing the bulk Hamiltonian constraint $\mathcal{H}$ as:
\begin{equation}
\mathcal{H} =
\mathcal{H}_\text{rel} + 
\mathcal{H}_\text{marg} + 
\mathcal{H}_\text{irrel} +
\mathcal{H}_\text{lat},
\end{equation}
according to the dimension $\Delta_J$ of the sources when interpreted in CFT language.  The relevant terms in the bulk Hamiltonian constraint would lead to divergences as $\eta \to 0$, and thus need to be eliminated by counterterms, so as to obtain a new Hamiltonian constraint $e^{-\text{CT}}{\cal H} e^{\text{CT}}$ which does not have any relevant terms in it.  Having done this, the marginal terms correspond to \emph{anomaly matching conditions} where some central charge or other coefficient must agree between the bulk and boundary theories, for the duality to be consistent.  And irrelevant terms (e.g~ \eqref{eqn:TTterm}) correspond to the terms included in the $T^2$ deformation itself.\footnote{As in all cases of irrelevant deformations, the resulting theory might require UV completion.  However, this does not prevent the theory from being predictive as long as we work in a holographic large $N$ regime, and we stay away from the Planck scale.}  

The new case, lateral terms, are not permitted in a unitary AdS dual, but are permitted in dS.  These produce an oscillatory integral as $\eta \to 0$, which is on the boundary of convergence.  The resulting integral is well-defined, so long as one treats it with an appropriate $i\epsilon$ damping factor, leading to an unambiguous result.  Hence, unlike the marginal case, there is \emph{no} anomaly-matching coefficient.  

Furthermore, as we shall see, it is actually possible to analytically continue through this boundary in order to allow relevant terms in the $T^2$ deformation.\footnote{This is actually equivalent to the counterterm story, because going around poles in complex analysis is equivalent to subtracting off power law divergences.} This is also quite convenient for studying the complementary series $0 \le \Delta \le d$ because, when following the original prescription of \cite{Araujo-Regado:2022gvw}, the required counterterms differ in different regions of the interval $[0,d]$, breaking up into an infinite number of different subcases (separated by poles on the real axis of $\Delta$, as various operators become marginal).  The use of analytic continuation through complex $\Delta$ allows us to calculate formulae only in a single interval, say $\Delta$ slightly above $d/2$, and then extend the results to generic values of $\Delta$.\footnote{One must, however, be especially careful if $\Delta$ is exactly on one of the ``poles'' when an operator becomes exactly marginal.  Such cases involve trace anomalies that break the exact conformal invariance of the dS/CFT, and require anomaly matching as stated above.}

\begin{figure}[ht]
    \centering
    \includegraphics[width=\textwidth]{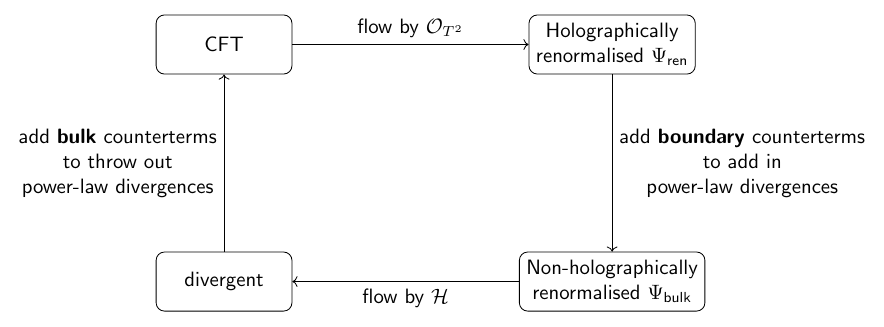}
    \caption{A schematic illustration of how the hypothetical boundary ``dS/CFT’’ theory can be deformed by an appropriately defined \textbf{irrelevant} $T^2$ deformation operator (denoted schematically by $\hat{O}_{T^2}$, whose precise form depends on the bulk field content) to produce the holographically renormalised wavefunction $\Psi_{\text{ren}}$ at finite time. To reconstruct the unrenormalised bulk wavefunction $\Psi_{\text{bulk}}$, one must then reintroduce the power-law divergent terms through the addition of counterterms that are \textbf{relevant} from a boundary RG perspective.}
    \label{fig:flow_picture}
\end{figure}

Figure \ref{fig:flow_picture} summarises the situation with regards to the counterterms $\text{CT}$. One can either think of them from the bulk perspective, as removing power-law divergences from bulk states, or from the dual perspective as adding in the appropriate power-law divergences to recover bulk states.\footnote{Here, by bulk states we mean solutions to the Hamiltonian constraint in its usual form, as obtained from the Einstein-Hilbert action.} The divergences arise when evaluating bulk states on large-volume configurations, here corresponding to $\eta \to 0^-$.

\subsection{Plan of Paper}
The plan of this paper is as follows. In Section~\ref{sec:Wavefunction}, we review the Bunch--Davies wavefunction for free scalar fields and linearised gravitons in de Sitter space. We perform a bulk calculation of the corresponding wavefunctions and their two-point wavefunction coefficients, identifying the (purely \emph{imaginary}) counterterms required to remove divergences in the conformal time variable $\eta$ during holographic renormalisation. Applying the Born rule to the wavefunction then yields the two-point correlation functions of the fields and their conjugate momenta, i.e.~cosmological correlators. This section is largely standard, but establishes notation and a framework that will be used throughout the paper.

In Section~\ref{sec:BulkPropagator}, we provide a complementary derivation of cosmological correlators via the bulk propagator, computed using canonical quantisation methods. This serves both as a consistency check of the wavefunctional approach and as a bridge to later sections, where bulk and boundary perspectives are compared more directly.

Section~\ref{sec:Boundary} introduces the Cauchy Slice Holography dictionary and highlights the features most relevant for a finite-volume formulation of dS/CFT. It reviews the original construction pedagogically, while adapting it for a dual description of the Poincaré slicing of dS, setting the conceptual ground for direct comparison with the bulk computations. 

In Section~\ref{sec:Concrete}, we move to a more concrete implementation of the framework by explicitly constructing the relevant deformation operators, both for pure gravity and for matter fields with various values of $\Delta$. We also study the analytic properties of the deformation in the complex-$\Delta$ plane.  This section develops the structural input required to define and analyse the boundary flow, and adapts known perturbative ingredients to the finite-volume setting considered here.

Section~\ref{sec:BoundaryFlow} contains several of the main technical results of the paper. Using the deformation operators constructed earlier, we compute the flow equations for correlation functions and show how they encode bulk physics at finite time. In particular, specific results such as \eqref{eqn:dSCSH2ptCorrelatorFlowEquationSolutionCT} admit a direct interpretation as finite-time bulk two-point functions, illustrating how time-dependent bulk observables are captured within the flowed boundary description.

Finally, in Section~\ref{sec:Discussion} we discuss our results and outline possible future directions. We comment on the limitations of the present analysis, potential extensions to more concrete dS/CFT models, and broader implications --- including naturalness considerations --- of viewing cosmological time evolution as an RG flow.

\section{Bulk Wavefunction}\label{sec:Wavefunction}
We will outline the formalism in general $D=d+1$ spacetime dimensions, but quote certain results for the case most relevant to our Universe of $\, 3+1 \,$ spacetime dimensions. We will assume a perfect de Sitter (dS) background spacetime on which we allow for matter and graviton fluctuations. This case is supposed to model either the early inflationary history of the Universe (neglecting the slow-roll parameter), or its late-time behaviour after the end of the matter-dominated era~\cite{Starobinsky:1980te,Guth:1980zm,Linde:1981mu,Mukhanov:2005sc,Weinberg:2008zzc,Baumann:2009ds,Baumann:2022mni}.

\begin{figure}[ht]
    \centering
    \includegraphics[width=0.8\textwidth]{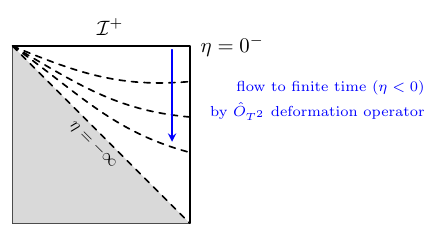}
    \caption{\small The patch covered by our planar coordinates. The bulk Hilbert space will live on these slices. Notice that as $\eta \to-\infty$ the slice approaches the null horizon, which agrees with the d-volume $a \to 0$. In Section \ref{sec:Boundary} we will show how to interpret this time flow as a $T^2$ deformation on the dual boundary side, as indicated by the blue arrow here.}
    \label{fig:dS}
\end{figure}

Because dS is a maximally symmetric spacetime, it admits homogeneous and isotropic foliations. Thus, we can write for the background metric:
\begin{equation}\label{eqn:dSmetricCosmic}
    \textbf{g}_{\mu\nu}\diff\textbf{x}^\mu \diff\textbf{x}^\nu=-\diff t^2 + a^2(t) \delta_{ab} \diff x^a \diff x^b \, ,
\end{equation}
which in conformal time, where $\diff t= a(\eta)\diff \eta$
\begin{equation}\label{eqn:dSmetricConformal}
    \textbf{g}_{\mu\nu} \diff \textbf{x}^\mu \diff \textbf{x}^\nu=a^2(\eta)(-\diff \eta^2+\delta_{ab}\diff x^adx^b) \, ,
\end{equation}
where the scale factor takes the form:
\begin{equation}
    a(\eta)=-\frac{1}{H \eta} \, ,
\end{equation}
where the Hubble constant $H$ is related to the cosmological constant via $\Lambda=d(d-1)H^2/2=d(d-1)/(2L_{\text{dS}}^2)$. Notice that these \emph{conformally flat} coordinates only cover half of a global dS space, with $\eta \to 0^-$ corresponding to future infinity and $\eta \to-\infty$ corresponding to the past horizon (see Figure \ref{fig:dS}). This is referred to as the Poincaré (or inflationary) patch in the literature.  Also note that the future-oriented unit normal to this foliation is given by 
\begin{equation}
    n=\frac{1}{a}\frac{\partial}{\partial \eta} \, .
\end{equation}

We would now like to quantise the dynamical fields perturbatively on this background. The quantisation of fields in de Sitter space has a long and rich history~\cite{Chernikov:1968zm,Bunch:1978yq,Birrell:1982ix,Allen:1985ux,Allen:1986ta,Allen:1986dd,Allen:1986tt,Allen:1987tz,Fulling:1972md,Schomblond:1976xc}\footnote{For modern pedagogical reviews see e.g.~\cite{Spradlin:2001pw,Anninos:2012qw}.}. The universe we are considering is expanding, and one of the main novelties of quantum field theory in curved spacetime is the \emph{non-uniqueness} of the vacuum state, even for free fields.\footnote{This feature is intimately related to the absence of a preferred foliation of spacetime. Each foliation generally defines a distinct Hilbert space, and these are, in general, unitarily inequivalent. However, for ``good'' foliations --- that is, those admitting a nowhere-vanishing, everywhere-timelike vector field $n^{\mu}$ (as depicted schematically in Figure~\ref{fig:dS}) --- there exists a canonical construction of a Hilbert space. For weakly coupled theories, one may then build a Fock space from the special state annihilated by the corresponding lowering operators. The quantum state of the dynamical fields will, in general, be an \emph{excited} state in this Fock space, defined on a particular slice within the foliation.} The Bunch-Davies state is the vacuum state defined with respect to the conformal time foliation in Figure \ref{fig:dS}.

We consider small fluctuations of weakly-coupled fields on the dS background. In this regime, we can ignore any back-reaction on the geometry, to leading order. To be more precise, we will be considering the regime in which both:
\begin{align}
    |\partial\Phi|&\ll \sqrt{\frac{\Lambda}{G_N}}\\
    |\Phi|&\ll \frac{1}{m}\sqrt{\frac{\Lambda}{G_N}}
\end{align}
where $m$ is the mass of the field $\Phi$. When both of these hold, the stress-energy tensor of $\Phi$ is subdominant compared to $\Lambda$. 

Of course, a general quantum gravitational state $\Psi[g,\Phi]$ has support outside of this range. We should not trust the prescription described below to evaluate the state on those configurations, since there will be strong back-reaction effects, resulting in the assumption that the background spacetime is dS with metric \eqref{eqn:dSmetricCosmic}-\eqref{eqn:dSmetricConformal} becoming invalid. If we are to stay in the semiclassical regime of the theory, there is only so many questions we can ask. 

\subsection{The Wavefunction}
Cosmological correlation functions (referred to as ``cosmological correlators'' in the literature) can be computed by means of the so-called ``wavefunction of the universe". The wavefunction is a richer object than correlators themselves, as it contains all the information of the cosmological correlators — the relation between them is roughly the same as that between the $S$-matrix and scattering cross-sections. However, the wavefunction is somewhat simpler than correlators in certain ways that we will see. In this section, we will describe the wavefunction approach and show, in specific examples, how it reproduces the in-in results in Section \ref{sec:BulkPropagator}.

Consider a set of bulk fields, labelled generically as $\Phi(\eta,x)$, which can include both matter fields (such as the inflaton field) and metric fluctuations (such as the graviton). We are interested in their spatial correlations. The information about these correlations is contained in the quantum-mechanical state of the system at a given time. It is convenient to describe states in the basis of field eigenstates, which (in the Heisenberg picture) satisfy
\begin{equation}
    \hat{\Phi}(\eta,\bfx)\ket{\Phi(\bfx)}=\Phi(\bfx)\ket{\Phi(\bfx)} \, ,
\end{equation}
where $\hat{\Phi}$ is a field operator and $\Phi(\bfx) \equiv \Phi(\eta,\bfx)$ is the spatial profile of the field at a given time $\eta$. Any state of the system can be projected onto this basis, but we will primarily be interested in the (interacting) vacuum state of the system. Writing this state in the basis of field eigenstates, the overlap coeﬃcients are
\begin{equation}\label{eqn:WFUdefintion}
    \Psi_\eta[{\Phi}] \equiv \langle \Phi;\eta | \Omega \rangle = \int_{\phi(-\infty)=0}^{\phi(\eta)=\Phi} \mathrm{D}\phi \, e^{iS[\phi]} \, .
\end{equation}

\subsubsection{Wavefunction Coefficients}

It is generally assumed in the literature that the wavefunction is an analytic function of its arguments. This means that it allows for an expansion of the form
\begin{equation}\label{eqn:ParameterisedWFUposition}
    \Psi_\eta[\Phi]=\exp{\;-\sum_{n=2}^{\infty} \frac{1}{n!}\int \; \left[\prod_{i=1}^n \diff^dx_i \Phi(x_i) \right] c_n(\eta;x_1,...,x_n) } \, ,
\end{equation}
where the complex coefficients $c_n(\eta;x_1,...,x_n)$ are called \emph{wavefunction coefficients}. They are distributions in real space, but we can equivalently write them in momentum space
\begin{equation}\label{eqn:ParameterisedWFUmomentum}
    \Psi_\eta[{\Phi}] = \exp\left\{-\sum_{n=2}^\infty \frac{1}{n!} \int \left[\prod_{a=1}^n  \frac{\diff^dk_a}{(2\pi)^d}{\Phi}_{\bfk_a} \right] c_n(\eta;\bfk_1,\dots,\bfk_n)\, \delta^d\left(\sum_{a=1}^n \textbf{k}_a\right)\right\} \, .
\end{equation}

There are relationships between the expansion in \eqref{eqn:ParameterisedWFUposition} and other objects in quantum field theory. For a review on the relation between the path integral and the in-in formalism see e.g. Appendix A of~\cite{Goodhew:2020hob} and~\cite{Abolhasani:2022twf,Goodhew:2024eup}.

\subsubsection{Relation to Cosmological Correlators}
After having made a choice of quantisation procedure, as outlined above, we can compute expectation values using the Born rule, with standard rules encoded in the examples below
\begin{align}
    \expval{\Phi(x_1)...\Phi(x_n)}_{\Psi_\eta} &= \frac{1}{||\Psi_\eta||^2}\int \mathrm{D} \Phi \;\Psi^*_\eta[\Phi]\;\Phi(x_1)...\Phi(x_n)\;\Psi_\eta[\Phi] \, , \\
    \expval{\Pi_\Phi(x_1)...\Pi_\Phi(x_n)}_{\Psi_\eta} &= \frac{1}{||\Psi_\eta||^2}\int \mathrm{D} \Phi\; \Psi^*_\eta[\Phi]\; (-i)^n \frac{\delta}{\delta\Phi(x_1)}...\frac{\delta}{\delta\Phi(x_n)}\;\Psi_\eta[\Phi] \, ,
\end{align}
where $*$ denotes complex conjugation. The norm is the standard $L^2$ norm. These are called \emph{cosmological correlators}. 

In particular the two-point correlators, which we will use in the body of the paper, are related to the two-point wavefunction coefficients as follows \cite{Sengor:2021zlc,Sengor:2022hfx,Thavanesan:2025kyc}
\begin{align}
    \expval{\Phi(x_1)\Phi(x_2)}_{\Psi_\eta}&=\frac{1}{2\textbf{Re}[c_2(x_1,x_2)]}; \label{eqn:BornRule1}\\
    \expval{\Pi_\Phi(x_1)\Pi_\Phi(x_2)}_{\Psi_\eta}&=\frac{|c_2(x_1,x_2)|^2}{2\textbf{Re}[c_2(x_1,x_2)]};  \label{eqn:BornRule2}\\
    \expval{\Phi(x_1)\Pi_\Phi(x_2)+\Pi_\Phi(x_2)\Phi(x_1)}_{\Psi_\eta}&=-\frac{\textbf{Im}[c_2(x_1,x_2)]}{\textbf{Re}[c_2(x_1,x_2)]}.  \label{eqn:BornRule3}
\end{align}

Thus, to obtain the cosmological correlators, all we need to do is compute the wavefunction coefficient $c_2$. We will start by deriving it from bulk computations in Sections \ref{sec:FreeScalardS} and \ref{sec:FreeGravitondS} (as well as a complementary method in Section \ref{sec:BulkPropagator}), after which Sections \ref{sec:Boundary}, \ref{sec:Concrete} and \ref{sec:BoundaryFlow} are entirely dedicated to deriving the same result from the $T^2$-deformed field theory. But first, we take a small detour as a warm-up.

\subsection{Harmonic Oscillator Warmup}\label{sec:HarmonicOscillator}
Before tackling quantum fields in de Sitter space, we follow closely the approach adopted in~\cite{BaumannJoyce:2023Lecs} and recall the analogous calculation for a single harmonic oscillator, since the structure of the de Sitter wavefunction and its perturbative corrections carry over directly. This serves as an intuitive starting point for the wavefunction approach, allowing us to first explore it in the simpler context of quantum mechanics.

Consider a simple harmonic oscillator, whose action is
\begin{equation}\label{eqn:HOaction}
    S[\Phi] = \int dt \, \left(\frac{1}{2} \dot{\Phi}^2 - \frac{1}{2} \omega^2 \Phi^2 \right) \,,
\end{equation}
where $\Phi$ is the deviation from equilibrium and $\omega$ is the constant frequency of the oscillator. Using integration by parts, the action can equivalently be written as a pure boundary term on-shell:
\begin{equation}
    S[\Phi_{\mathrm{cl}}] = \int_{t_i}^{t_f} dt \, \Bigg[\frac{1}{2} \partial_t \left( \Phi_{\mathrm{cl}} \, \dot{\Phi}_{\mathrm{cl}} \right) - \frac{1}{2} \Phi_{\mathrm{cl}} \underbrace{\left( \ddot{\Phi}_{\mathrm{cl}} + \omega^2 \Phi_{\mathrm{cl}} \right)}_{=0} \Bigg] \, ,
\end{equation}
where we have integrated by parts and used the fact that the classical solution satisfies the equation of motion
\begin{equation}
    \ddot{\Phi}_{\mathrm{cl}} + \omega^2 \Phi_{\mathrm{cl}} = 0 \, .
\end{equation}
Thus, the on-shell action reduces to
\begin{equation}\label{eqn:HOonshell}
    S[\Phi_{\mathrm{cl}}] = \frac{1}{2} \Phi_{\mathrm{cl}} \, \dot{\Phi}_{\mathrm{cl}} \Big|_{t=t_f} \, .
\end{equation}

To determine the classical solution $\Phi_{\mathrm{cl}}(t)$, we must specify two boundary conditions:  
\begin{align}
    &\text{1) The late-time value of the oscillator position is:} \quad &\Phi_{\mathrm{cl}}(t_f \equiv 0) = \phi \, , \\
    &\text{2) The early-time behaviour is:} \quad &\Phi_{\mathrm{cl}}(t \to -\infty) \sim e^{i(\omega+i\epsilon) t} \, ,
\end{align}
where the $i\epsilon$ prescription ensures that we are in the ground state, and permits us to neglect the boundary term contribution from early times ($t \to -\infty)$.  This also ensures that the wavefunction is normalisable.

The unique solution satisfying these conditions is (after taking $\epsilon \to 0$ at finite $t$):
\begin{equation}
    \Phi_{\mathrm{cl}}(t) = \phi \, e^{i\omega t} \,.
\end{equation}
Substituting into \eqref{eqn:HOonshell}, we find
\begin{equation}
    S[\phi] = \frac{i\omega}{2} \phi^2 \quad \implies \quad \Psi[\phi]=e^{iS[\phi]} = \exp\left( -\frac{\omega}{2} \phi^2 \right) \quad \implies \quad |\Psi[\phi]|^2 = e^{-\rm Re(\omega) \, \phi^2} \,.
\end{equation}
We have recovered the familiar fact that the ground state wavefunction\footnote{The overall normalisation of the wavefunction has been omitted, since it plays no role in computing expectation values or correlation functions.} of the simple harmonic oscillator is a Gaussian.\footnote{The width of this Gaussian determines the size of the zero-point fluctuations of the oscillator.}
The variance is
\begin{equation}
    \langle \phi^2 \rangle = \frac{1}{2 \, \mathrm{Re}(\omega)}.
\end{equation}

\paragraph{Free scalar field in Fourier space.}
Free quantum field theory in Fourier space is essentially equivalent to a set of harmonic oscillators indexed by the continuous wavenumber $\mathbf{k}$. To see this, we write the action of a free scalar in Fourier space:
\begin{align}
    S[\Phi] &= \frac{1}{2} \int dt \, d^dx \left[ \dot{\Phi}^2 - (\nabla \Phi)^2 - m^2 \Phi^2 \right] \nonumber \\
    &= \frac{1}{2} \int dt \, \frac{d^dk}{(2\pi)^d} \left[ \dot{\Phi}_{\mathbf{k}} \, \dot{\Phi}_{-\mathbf{k}} - (k^2 + m^2) \Phi_{\mathbf{k}} \Phi_{-\mathbf{k}} \right] \,,
\end{align}
where $\Phi_{\mathbf{k}}(t)$ are the Fourier modes of the field.

We see that each Fourier mode satisfies the equation of a simple harmonic oscillator with frequency
\begin{equation}
    \omega_k \equiv \sqrt{k^2 + m^2} \,.
\end{equation}
The wavefunction, and the associated zero-point fluctuations, can then be written as a sum over the corresponding results for each harmonic oscillator:
\begin{equation}
    \Psi[\phi] = \exp\left[ -\frac{1}{2} \int \frac{d^dk}{(2\pi)^d} \, \omega_k \, \phi_{\mathbf{k}} \phi_{-\mathbf{k}} \right] 
    \quad \implies \quad \langle \phi_{\mathbf{k}} \phi_{-\mathbf{k}} \rangle' = \frac{1}{2 \, \rm Re(\omega_k)} \, .
\end{equation}
The treatment of free fields in quantum field theory is therefore as simple as that of an infinite collection of uncoupled harmonic oscillators in quantum mechanics. Thus the Gaussian width $\Omega$ is determined entirely by the classical mode functions. This observation will be central when we compute the free-field wavefunction coefficient in de Sitter space. In Sections \ref{sec:FreeScalardS} and \ref{sec:FreeGravitondS}, we will apply this harmonic oscillator intuition directly to fields in de Sitter, where each Fourier mode is described by a time-dependent oscillator as studied in Section \ref{sec:TimeDependentOscillator}.

\subsection{Time-dependent Oscillator}\label{sec:TimeDependentOscillator}
A free field in de Sitter space (and also the fluctuations during inflation) satisfies the equation of motion of a harmonic oscillator with a time-dependent frequency (for each Fourier mode). It is therefore useful to consider the following time-dependent oscillator:
\begin{equation}
    S[\Phi] = \int dt \, \left[\frac{1}{2} A(t) \dot{\Phi}^2 - \frac{1}{2} B(t)\Phi^2 \right] \, ,
\label{eqn:tdHOaction}
\end{equation}
where $A(t)$ and $B(t)$ are arbitrary functions of time. On-shell, this action can again be written as a boundary term:
\begin{align}\label{eqn:tdHOonshell}
    S[\Phi_{\rm cl}] &= \frac{1}{2} \left[\int_{t_i}^{t_f} dt \, \partial_t\!\big(A(t) \Phi_{\rm cl}\dot{\Phi}_{\rm cl}\big) 
     - \frac{1}{2} \Phi_{\rm cl}\underbrace{\big( \partial_t(A \dot{\Phi}_{\rm cl}) + B \Phi_{\rm cl}\big) }_{0}\right] \\
     &= \frac{1}{2} A(t)\Phi_{\rm cl}(t)\dot{\Phi}_{\rm cl}(t)\big|_{t=t_f} \, ,
\end{align}
where we have used that the classical solution satisfies
\begin{equation}
    \partial_t(A(t) \dot{\Phi}_{\rm cl}) + B(t) \Phi_{\rm cl} = 0 \, .
\end{equation}

We again need to find the classical solution with the correct boundary conditions to describe the ground state. We can write the desired solution abstractly as
\begin{equation}
    \Phi_{\rm cl}(t) = \phi\, K(t)\,.
\end{equation}
Substituting this into the equation of motion, we obtain a differential equation for $K(t)$:
\begin{equation}\label{eqn:Kdiff}
    \partial_t\!\left(A(t) \dot{K}(t)\right) + B(t)K(t) = 0 \,.
\end{equation}
The boundary conditions that we impose on the solution are
\begin{equation}\label{eqn:TDOscillatorBC}
    K(t_f) = 1 \, , \qquad K(t\to -\infty) \sim e^{i\omega t} \, .
\end{equation}
Of course, the precise form of $K(t)$ depends on $A(t)$ and $B(t)$, but given those functions we can then solve the equation of motion subject to the prescribed boundary conditions to obtain $K(t)$.

Substituting this solution into \eqref{eqn:tdHOonshell}, we have the following expression for the ground state wavefunction:
\begin{equation}\label{eqn:tdHOwfu}
    \Psi[\phi] = e^{iS[\phi]} = \exp\!\left[\frac{i}{2} \, \left(A(t) \,\partial_t \log K(t) \right)\Big|_{t=t_f}\, \phi^2 \right]=\exp\!\left[-\frac{\psi_2}{2} \, \phi^2 \right] \, .
\end{equation}
Hence we can read off the two-point wavefunction coefficient as
\begin{equation}\label{eqn:tdHOwfucoefficient}
    \psi_2 = - i\!\left(A(t)\partial_t\log K(t)\right)\Big|_{t=t_f} \,.
\end{equation}
and compute the two-point correlator using the usual for
\begin{equation}\label{eqn:tdHOcorr}
    |\Psi[\phi]|^2 = \exp\!\left[-\,{\rm Re}(\psi_2) \, \phi^2 \right] \quad \Longrightarrow \quad \langle \phi^2 \rangle = \frac{1}{2\,{\rm Re}(\psi_2)} \,.
\end{equation}
As we will see in Sections \ref{sec:FreeScalardS} and \ref{sec:FreeGravitondS}, this result will be directly applicable to fields in de Sitter space.

\subsection{Free Scalars in de Sitter}\label{sec:FreeScalardS}
Having gained some intuition for the wavefunction in the quantum-mechanical setting in Sections \ref{sec:HarmonicOscillator} and \ref{sec:TimeDependentOscillator}, we will now use it to compute the de Sitter scalar wavefunction and later the graviton wavefunction in Section \ref{sec:FreeGravitondS}.

Let us start with the action of a free massive scalar field in a curved background:
\begin{equation}
    S = \int d^{d+1}\bfx \sqrt{-g} \left[ -\frac{1}{2} g^{\mu\nu}\partial_\mu \Phi \partial_\nu \Phi - \frac{1}{2} m^2 \Phi^2 \right].
\end{equation}
Substituting the de Sitter metric (in conformal time) from \eqref{eqn:dSmetricConformal}, this becomes
\begin{align}
    S[\Phi] &= \frac{1}{2} \int_{-\infty}^{\eta_0} d\eta\, d^{d}x \, a^{d+1}(\eta)\, \frac{1}{a^2(\eta)}\left[ (\Phi')^2 - (\nabla \Phi)^2 - m^2 a^2(\eta)\Phi^2 \right] \nonumber \\
    &= \frac{1}{2} \int_{-\infty}^{\eta_0} d\eta \,\frac{d^d k}{(2\pi)^d} \, \frac{1}{(-H\eta)^{d-1}} \left[ \Phi'_{\mathbf{k}}\Phi'_{-\mathbf{k}} - \left(k^2 + \frac{m^2}{(-H\eta)^2}\right) \Phi_{\mathbf{k}}\Phi_{-\mathbf{k}} \right],
\label{eqn:dSactionmodes}
\end{align}
where in the second line we have written the field in terms of its Fourier components and used $a(\eta) = 1/(-H\eta)$.  

We see that the action for each Fourier mode is the same as that of the time-dependent oscillator in \eqref{eqn:tdHOaction}, with the identifications
\begin{equation}
    A(\eta) = \frac{1}{(-H\eta)^{d-1}} \; , 
    \qquad B(\eta) = \frac{1}{(-H\eta)^{d-1}}\left(k^2 + \frac{m^2}{(-H\eta)^2}\right) \, .
\end{equation}
We can therefore apply the result \eqref{eqn:tdHOcorr} to compute the wavefunction and the variance of the field.

The classical equation of motion is
\begin{equation}\label{eqn:phiEOMdS}
    \left(A(\eta)\Phi'_{\rm cl}\right)' + B(\eta)\Phi_{\rm cl} = 0
    \quad \Longrightarrow \quad 
    \Phi_{\rm cl}'' - \frac{d-1}{\eta}\Phi_{\rm cl}' + \left(k^2 + \frac{m^2}{H^2\eta^2}\right)\Phi_{\rm cl} = 0 \, .
\end{equation}
On small scales, $|k\eta|\gg 1$, the equation reduces to that of a simple harmonic oscillator with frequency $k$. On large scales, $|k\eta|\to 0$, the harmonic oscillator becomes over-damped and the fluctuations freeze out (for $m=0$) or decay slowly (for $m>0$).

We write the solution to \eqref{eqn:phiEOMdS} as
\begin{equation}
    \Phi_{\rm cl}(\eta) \equiv \phi\, K(\eta) \, ,
\end{equation}
where the function $K(\eta)$ approaches unity at late times and oscillates like $e^{ik\eta}$, since at early times the equation of motion for each Fourier mode becomes the same as that of a harmonic oscillator
\begin{equation}\label{eqn:phiEOMdSLateTime}
    \lim_{\eta \to -\infty} \left[\Phi_{\rm cl}'' - \frac{d-1}{\eta}\Phi_{\rm cl}' + \left(k^2 + \frac{m^2}{H^2\eta^2}\right)\Phi_{\rm cl} \right] = \Phi_{\rm cl}''  + k^2 \Phi_{\rm cl} = 0 \, .
\end{equation}
One can write the quantum free field operator as 
\begin{equation} \label{eqn:FreeScalarField}
    \hat{\phi}({\bf k}, \eta) = \phi^{-}(k, \eta) a_\bfk + \phi^{+}(k,\eta) a_{-\bfk}^\dagger \, ,
\end{equation}
where the mode functions $\phi^{\pm} (k, \eta)$ correspond to solutions of the free classical equation of motion and are given by
\begin{equation}\label{eqn:ScalarModeFunctionsMassive}
    \phi^+(k,\eta)=i\,\frac{\sqrt{\pi} H}{2} e^{-i\frac{\pi}{2}(\nu+\frac{1}{2})}\, \left( -\eta  \right)^{\frac{d}{2}} H^{(2)}_{\nu}(-k\eta) \, , \qquad \phi^-(k,\eta)=(\phi^+(k,\eta))^* \, ,
\end{equation}
where $\nu=\sqrt{d^2/4-m^2/H^2}$ denotes the order of the Hankel function, and by simply setting $m=0$, one can find the mode functions for massless fields $\phi$ to be
\begin{equation}\label{eqn:ModeFunctionsMassless}
    \phi^+(k,\eta)=i\,\frac{\sqrt{\pi} H}{2} e^{-i\frac{\pi}{2}(\frac{d+1}{2})}\, \left( -\eta  \right)^{\frac{d}{2}} H^{(2)}_{d/2}(-k\eta) \, , \qquad \phi^-(k,\eta)=(\phi^+(k,\eta))^* \, .
\end{equation}
For the purposes of this work, we will impose the boundary condition that the fields started at the beginning of inflation in a non-excited initial state, specifically the Bunch-Davies vacuum. This corresponds to imposing a boundary condition on the fields, that they vanish in the infinite past ($\eta \to -\infty$):
\begin{equation} \label{eqn:BDFreeScalarField}
    \lim_{\eta \to -\infty} \hat{\phi}({\bf k}, \eta) = \phi^{+}(k,\eta) a_{-\bfk}^\dagger \, .
\end{equation}
In reality this leads to attributing an $i\epsilon$-prescription to the magnitude of the spatial momenta, i.e. $k=\sqrt{k \cdot k}=\tilde{k}(1-i\epsilon) \in \mathbb{C}$, where $\tilde{k} \in \mathbb{R} \geq 0$ (see Section 7 of~\cite{Goodhew:2024eup} for more details).

The relevant solution to \eqref{eqn:phiEOMdS}, which satisfies the Bunch-Davies vacuum initial condition, i.e. the curved spacetime version of the boundary condition in \eqref{eqn:TDOscillatorBC}, is thus
\begin{equation}
    K(k,\eta) = \frac{\phi^+(k,\eta)}{\phi^+(k,\eta_0)} = \left(\frac{-\eta}{-\eta_0}\right)^{d/2} \frac{H_\nu^{(2)}(-k\eta)}{H_\nu^{(2)}(-k\eta_0)} \, , 
    \qquad \nu = \sqrt{\frac{d^2}{4} - \frac{m^2}{H^2}} \, ,
\end{equation}
so that
\begin{equation}
    \log K(\eta) = \frac{d}{2}\log(-\eta)-\frac{d}{2}\log(-\eta_0) + \log H_\nu^{(2)}(-k\eta)-\log H_\nu^{(2)}(-k\eta_0) \,.
\end{equation}
Substituting into \eqref{eqn:tdHOwfucoefficient} yields
\begin{equation}\label{eqn:dSscalar2pt}
    \psi_2 = - i\!\left(A(\eta)\partial_{\eta}\log K(\eta)\right)\Big|_{\eta=\eta_0} = \frac{i H}{2} \left(-H \eta_0\right)^{-d} \frac{ 2 (-k\eta_0) H^{(2)}_{\nu-1}(-k\eta_0) + (d - 2\nu) H^{(2)}_{\nu}(-k\eta_0) }{ H^{(2)}_{\nu}(-k\eta_0) } \, ,
\end{equation}
with $H^{(2)}_\nu$ denoting Hankel functions of the second kind. This is the exact tree-level two-point wavefunction coefficient for a scalar of arbitrary mass in the Poincaré patch of de Sitter space.

In order to ensure the wavefunction coefficients are finite in the $\eta \to 0^-$ limit, we must include a multiplicative $(-\eta_0)^{2(d-\Delta)}=(-\eta_0)^{d-2\nu}$ prefactor\footnote{This differs from the power-law divergences which are removed by counterterms during holographic renormalisation. The step carried out here is just a field-renormalisation via a multiplicative factor.} which comes from the late-time limit of the bulk fields, i.e.~ the sources of the dual dS/CFT operators.
\begin{align}
    \psi_2 &= \left(-\eta_0\right)^{d-2\nu} \frac{i H}{2} \left(-H \eta_0\right)^{-d} \frac{ 2 (-k\eta_0) H^{(2)}_{\nu-1}(-k\eta_0) + (d - 2\nu) H^{(2)}_{\nu}(-k\eta_0) }{ H^{(2)}_{\nu}(-k\eta_0) } \, , \\
    \psi_2 &= \frac{i (d - 2\nu) H \left(-\eta_0\right)^{-2\nu} \left(-H \right)^{-d}}{2} + ik H \left(-\eta_0\right)^{1-2\nu} \left(H \right)^{-d}\frac{H^{(2)}_{\nu-1}(-k\eta_0)}{ H^{(2)}_{\nu}(-k\eta_0) } \, , \label{eqn:dSscalar2ptEtaFactors}
\end{align}
which is consistent with the result obtained in \eqref{eqn:dSCSH2ptCorrelatorFlowEquationSolutionCT} of Section \ref{sec:BoundaryFlow}.

The width of the wavefunction is given by
\begin{align}
    {\rm Re}\!\left(-iA(\eta)\partial_\eta \log K(\eta)\right)\Big|_{\eta=\eta_0}
    &= \frac{1}{(-H\eta_0)^{d-1}} \, {\rm Re}\!\left(-i\frac{\partial_\eta H_\nu^{(2)}(-k\eta)}{H_\nu^{(2)}(-k\eta)}\right)\Bigg|_{\eta=\eta_0} \\ 
    &= \frac{1}{(-H\eta_0)^{d-1}}\frac{1}{|H_\nu^{(2)}(-k\eta_0)|^2}{\rm Re}\!\left(-iH_\nu^{(2)*}(-k\eta)\partial_\eta H_\nu^{(2)}(-k\eta)\right)\Bigg|_{\eta=\eta_0} \, .
\end{align}
When $m < \frac{d}{2}H$, $\nu$ is real and $H_\nu^{(2)}$ has the property that $H_\nu^{(2)\,*} = H_\nu^{(1)}$. We then find
\begin{align}
    {\rm Re}\!\left(-iH_\nu^{(2)*}(-k\eta)\partial_\eta H_\nu^{(2)}(-k\eta)\right)\Big|_{\eta=\eta_0}
    &= \frac{1}{2}\left(-iH_\nu^{(1)}(-k\eta)\partial_\eta H_\nu^{(2)}(-k\eta) + iH_\nu^{(2)}(-k\eta) \partial_\eta H_\nu^{(1)}(-k\eta) \right)\Big|_{\eta=\eta_0} \, \label{eqn:variance1} \\
    &= -\frac{i}{2}\left(H_\nu^{(1)}(-k\eta) \partial_\eta H_\nu^{(2)}(-k\eta) - H_\nu^{(2)}(-k\eta) \partial_\eta H_\nu^{(1)}(-k\eta) \right)\Big|_{\eta=\eta_0} \, ,
\end{align}
which is just the Wronskian of Hankel functions,
\begin{equation}
    H_\nu^{(1)}(-k\eta) \partial_\eta H_\nu^{(2)}(-k\eta) - H_\nu^{(2)}(-k\eta) \partial_\eta H_\nu^{(1)}(-k\eta)
    = -\frac{4i}{\pi \eta} \, .
\end{equation}
Hence we find that
\begin{equation}
    {\rm Re}\!\left(-iH_\nu^{(2)*}(-k\eta)\partial_\eta H_\nu^{(2)}(-k\eta)\right)\Big|_{\eta=\eta_0} = -\frac{2}{\pi \eta_0} \, ,
\end{equation}
and thus the width of the wavefunction is given by
\begin{equation}
    {\rm Re}\!\left(-iA(\eta)\partial_\eta \log K(\eta)\right)\Big|_{\eta=\eta_0}
    = -\frac{2}{\pi \eta_0}\frac{1}{(-H\eta_0)^{d-1}}\frac{1}{|H_\nu^{(2)}(-k\eta)|^2} \, .
\end{equation}
Substituting this into the variance \eqref{eqn:tdHOcorr}, we obtain for the two-point cosmological correlator, i.e. the power spectrum
\begin{equation}\label{eqn:2pointcosmocorrelator}
    \langle \phi_{\mathbf{k}}\phi_{-\mathbf{k}}\rangle' = \frac{\pi}{4}H^{d-1} (-\eta_0)^d\,H_\nu^{(1)}(-k\eta_0) H_\nu^{(2)}(-k\eta_0) \, ,
\end{equation}
A similar expression for the two-point cosmological correlator holds for the principal series case $m>\frac{d}{2}H$, where $\nu=i\tilde{\nu} \in i\mathbb{R}$. We can see this from the fact that \eqref{eqn:dSscalar2pt} and \eqref{eqn:dSscalar2ptEtaFactors} were derived for general mass and dimension. Hence, using the expression for the two-point cosmological correlator 
\begin{equation}
    \langle \phi_{\mathbf{k}}\phi_{-\mathbf{k}}\rangle' = \frac{1}{2 \, {\rm Re} (\psi_2)} \, ,
\end{equation}
and the fact that for the principal series when $m > \frac{d}{2}H$, $\nu=i\tilde{\nu}$ is imaginary and $H_\nu^{(2)}$ has the property that $H_{i\tilde{\nu}}^{(2)\,*} = H_{-i\tilde{\nu}}^{(1)}$, we find by following the same steps from \eqref{eqn:variance1} to \eqref{eqn:2pointcosmocorrelator} as above that
\begin{equation}
    \langle \phi_{\mathbf{k}}\phi_{-\mathbf{k}}\rangle' = \frac{\pi}{4}H^{d-1} (-\eta_0)^d \, H_{i\tilde{\nu}}^{(1)}(-k\eta_0) H_{i\tilde{\nu}}^{(2)}(-k\eta_0) = e^{\pi \tilde{\nu}}\frac{\pi}{4}H^{d-1} (-\eta_0)^d \, |H_{i\tilde{\nu}}^{(2)}(-k\eta_0)|^2 \, .
\end{equation}

\subsection*{Examples of $\psi_2$ for certain masses in $d=3$}
For those intimidated by Hankel functions, we note that the formula simplifies for conformally coupled scalars in all $d$, and also for massless scalars when $d = \text{odd}$ in $D=d+1$ spacetime dimensions.  The $D = 3+1$ results are provided below:

\paragraph{Conformally coupled, $m=\sqrt{2}H, \, \nu=1/2$ in $d=3$:}
\begin{equation}\label{conformal-no-CT}
    \psi_2=-i\frac{1+ik\eta_0}{H^2 \eta_0}.
\end{equation}

\paragraph{Massless $m=0, \, \nu=3/2$ in $d=3$:}
\begin{equation}\label{massless-no-CT}
    \psi_2=\frac{-ik^2}{H^2 \eta_0(1-ik\eta_0)}.
\end{equation}
For the two-point cosmological correlator, we obtain
\begin{equation}
    \langle \phi_{\mathbf{k}}\phi_{-\mathbf{k}}\rangle'=\frac{H^2(1+k^2\eta_0^2)}{k^3}
\end{equation}
which is consistent with \eqref{eqn:ScalarCosmoCorr1} and \eqref{eqn:ScalarCosmoCorr1CT} obtained in Section \ref{sec:ScalarCorrelator}. Taking the late-time limit $\eta_0 \to 0$ yields
\begin{equation}
    \lim_{\eta_0 \to 0} \;\langle \phi_{\mathbf{k}}\phi_{-\mathbf{k}}\rangle = \frac{H^2}{2k^3} + \frac{H^2 \eta_0^2}{2k} + \mathcal{O}(\eta_0^3) \, .
\end{equation}

\paragraph{Relation to the EAdS result.}  
The corresponding scalar two-point function at finite cutoff $z_c$ in Euclidean AdS$_{d+1}$ is (see e.g.~\cite{Hartman:2018tkw}):
\begin{equation}
\langle \mathcal{O}(\mathbf{k})\mathcal{O}(\mathbf{k}') \rangle_{\text{EAdS}} 
= -\frac{L^{d-1}_{\text{EAdS}}}{2} (z_c)^{-2\nu} \frac{ 2 (k z_c) K_{\nu-1}(k z_c) + (d - 2\nu) K_\nu(k z_c) }
     { K_\nu(k z_c) } (2\pi)^d \delta(\mathbf{k} + \mathbf{k}') \,,
\end{equation}
where $K_\nu$ are modified Bessel functions, and $z_c$ is the cutoff on the holographic coordinate $z$ in Poincar\'{e} coordinates.  The de Sitter result \eqref{eqn:dSscalar2ptEtaFactors} follows from this expression by the analytic continuation
\begin{equation}
    L_{\text{EAdS}} \to iL_{\text{dS}} \quad \text{and} \quad z_c \to -i \eta_0 \, ,
\end{equation}
and using the identity
\begin{equation}
    K_\nu(z_c)=\frac{\pi}{2} e^{-i\pi(\nu+1)/2} H_\nu^{(2)}(-i z_c) \, ,
\end{equation}
which implements the analytic continuation from the inverse radius coordinate in EAdS to conformal time in Lorentzian dS. This correspondence makes transparent the structural similarity between bulk-to-boundary propagators in the two spaces.

\subsection{Free Gravitons in de Sitter}\label{sec:FreeGravitondS}
We will now quantise gravitons in the Poincaré patch of de Sitter space. The action of gravitons $h_{ab}$ in $D=d+1$-dimensional spacetime, for traceless\footnote{To consider fields with a non-zero trace, one can subtract the trace and treat it as an additional scalar field.} gravitons, is given by:
\begin{equation}\label{eqn:GravitonAction}
    S=\frac{1}{16\pi G_N}\int \diff^D x \, \left[a(\eta)\right]^{d-1}\frac{1}{2}
    \left[\left(h_{ab}' \right)^2-\left(\partial_jh_{ab}\right)^2 + \mathcal{L}_{\text{int}}\right] \, .
\end{equation}
We have once again grouped the interaction terms in the Lagrangian density into $\mathcal{L}_{\text{int}}$ and separated them from the free part, and taken the background metric to be a flat FLRW metric, which we will later specify to be the Poincaré patch of de Sitter with the metric of \eqref{eqn:dSmetricConformal}.

We can Fourier-expand a single polarisation mode (suppressing the polarisation index $\lambda$ and polarisation tensor $\varepsilon_{ij}^{(\lambda)}$),
\begin{equation}\label{eqn:GravitonFourier}
    h_{ij}(\eta,\mathbf{x}) 
    = \sum_{\lambda}\int\frac{\diff^{\,d}k}{(2\pi)^{d}}\,
    \varepsilon_{ij}^{(\lambda)}(\hat{\mathbf{k}})\,
    h_{\mathbf{k}}^{(\lambda)}(\eta)\,
    e^{i\mathbf{k}\cdot\mathbf{x}} \, ,
\end{equation}
where the polarisation tensors satisfy the following conditions:
\begin{align}\label{pol1}
    k^{i}\varepsilon^{(\lambda)}_{ij}({\bf k}) &=0 & \text{(transverse)} \, , \\
    \varepsilon_{ii}^{(\lambda)}({\bf k}) &=0 & \text{(traceless)} \, , \\
    \varepsilon_{ij}^{(\lambda)}({\bf k}) - \varepsilon_{ji}^{(\lambda)}({\bf k}) &=0 & \text{(symmetric)} \, , \\
    \varepsilon_{ij}^{(\lambda)}({\bf k})\varepsilon_{jk}^{(\lambda)}({\bf k}) &=0 & \text{(lightlike)} \, , \\
    e^{(\lambda)}_{ij}({\bf k})e^{h'}_{ij}({\bf k})^{\ast} - 4\delta_{hh'} &=0 & \text{(normalisation)} \, , \label{eqn:normeps}  \\
    \varepsilon_{ij}^{(\lambda)}({\bf k})^{\ast} - \varepsilon_{ij}^{(\lambda)}(-{\bf k}) &=0 &\text{($  h_{ab}(x) $ is real)}  \, .\label{eqn:poln}
\end{align} 
The equation of motion for each mode amplitude $h_k(\eta)$ then follows from the quadratic action:
\begin{equation}\label{eqn:GravitonModeEOM}
    h_k'' + (d-1)\frac{a'}{a}h_k' + k^2 h_k = 0 \, .
\end{equation}
For a massless, minimally-coupled scalar field $\phi$ with action
\begin{equation}\label{eqn:ScalarAction}
    S_\phi = \frac{1}{2}\int \diff^D x \, \left[a(\eta)\right]^{d-1}
    \left[(\phi')^2 - (\partial_j \phi)^2\right] \, ,
\end{equation}
the equation of motion is identical:
\begin{equation}\label{eqn:ScalarModeEOM}
    \phi_k'' + (d-1)\frac{a'}{a}\phi_k' + k^2 \phi_k = 0 \, .
\end{equation}
Hence, the mode functions for each transverse-traceless graviton polarisation are identical to those of a massless scalar field.

Working in momentum space, we can write the quantum free field operator in terms of creation and annihilation operators:
\begin{equation} \label{eqn:FreeGravitonField}
    \hat{h}_{(\lambda)}({\bf k}, \eta) = h_{(\lambda)}^{-}(k, \eta) a_\bfk + h_{(\lambda)}^{+}(k,\eta) a_{-\bfk}^\dagger \, ,
\end{equation}
where the mode functions $h_{(\lambda)}^{\pm} (k, \eta)$ correspond to solutions of the free classical equation of motion and are given by
\begin{equation}\label{eqn:GravitonModeFunctions}
    h_{(\lambda)}^+(k,\eta)=i\,\frac{\sqrt{\pi} H}{2} e^{-i\frac{\pi}{2}(\frac{d+1}{2})}\, \left( -\eta  \right)^{\frac{d}{2}} H^{(2)}_{\frac{d}{2}}(-k\eta) \, , \qquad h_{(\lambda)}^-(k,\eta)=(h_{(\lambda)}^+(k,\eta))^* \, ,
\end{equation}
and we once again impose the boundary condition that the fields started at the beginning of inflation in a non-excited initial state, specifically the Bunch-Davies vacuum. This corresponds to imposing a boundary condition on the fields, that they vanish in the infinite past ($\eta \to -\infty$):
\begin{equation} \label{eqn:BDFreeGravitonField}
    \lim_{\eta \to -\infty} \hat{h}_{(\lambda)}({\bf k}, \eta) = h_{(\lambda)}^{+}(k,\eta) a_{-\bfk}^\dagger \, ,
\end{equation}
which as previously discussed attributes an $i\epsilon$-prescription to the magnitude of the spatial momenta, i.e. $k=\sqrt{k \cdot k}=\tilde{k}(1-i\epsilon) \in \mathbb{C}$, where $\tilde{k} \in \mathbb{R} \geq 0$.

Defining the graviton wavefunction coefficient $b(\bfk)$ via
\begin{equation}\label{eqn:GravitonGaussain}
    \Psi[h]=e^{iS[h]}=\exp\left(-\frac{1}{2}\int d^dk\;b(\bfk)\mathbf{P}^{abcd}\h_{ab}(\bfk)\h_{cd}(-\bfk)\right) \, ,
\end{equation}
we can thus use the results from the analysis in Section \ref{sec:FreeScalardS} to determine that
\begin{equation}
    b(\bfk) = \frac{1}{16\pi G_N}\frac{i H}{2} \left(-H \eta_0\right)^{-d} \frac{ 2 (-k\eta_0) H^{(2)}_{\nu-1}(-k\eta_0) + (d - 2\nu) H^{(2)}_{\nu}(-k\eta_0) }{ H^{(2)}_{\nu}(-k\eta_0) } \, .
\end{equation}

To easily compare with the results obtained in Section \ref{sec:GravitonCorrelator}, we will now quote the results explicitly in $d = 3$
\begin{equation}\label{eqn:b_full}
    b(\bfk)=-\frac{1}{(16\pi G_N)H^2}\left(\frac{ik^2/\eta}{1-ik\eta}\right) \, .
\end{equation}
\begin{equation}\label{eqn:b_real}
    \text{Re}[b(\bfk)]=\frac{1}{(16\pi G_N)H^2}\frac{1}{\left(\frac{1}{k^3}+\frac{\eta^2}{k}\right)}=\frac{1}{H^2(16\pi G_N)}\left(\frac{k^3}{1+k^2\eta^2}\right) \, .
\end{equation}
\begin{equation}\label{eqn:b_imag}
    \text{Im}[b(\bfk)]=-\frac{1}{(16\pi G_N)H^2}\frac{1}{\left(\frac{\eta}{k^2}+\eta^3\right)}=-\frac{1}{H^2(16\pi G_N)}\left(\frac{k^2/\eta}{1+k^2\eta^2}\right) \, .
\end{equation}


\subsection{Imaginary Counterterms in de Sitter}\label{sec:ImaginaryCT}
If we take the bulk results in the previous Section \ref{sec:FreeScalardS} and take the limit $\eta \to 0^-$, they diverge. This is analogous to a similar problem in AdS/CFT, where it is necessary to subtract off divergences as $z_c \to 0$.  This can be done by adding \emph{holographic counterterms} that depend on the boundary sources (and their derivatives, up to some finite order).

A subtle but important point is that in de Sitter space the counterterms required to renormalise the on-shell action are \emph{imaginary} in all dimensions.  This arises from the fact that the de Sitter path integral up to a Cauchy slice $\Sigma$ is Lorentzian, so that the amplitude associated with it scales like $e^{i S_{\text{dS}}}$ with $S_{\text{dS}} \in \mathbb{R}$.  
By contrast, if we take an AdS path integral \emph{also up to a spacelike boundary} $\Sigma$, the bulk path integral will be Euclidean, and scale like $e^{-S_{\text{EAdS}}}$ with $S_{\text{EAdS}} \in \mathbb{R}$.

It is possible to discuss these counterterms independently of holography, where their function is simply to ensure that the wavefunction coefficients remain finite while taking the $\eta \to 0^-$ (or $z \to 0$) limits.  However, if we have a holographic dual theory, these are the same as the \emph{holographic counterterms} which may be interpreted as UV renormalisation of the dual boundary theory.  This is because in AdS/CFT, the correspondence relates a Euclidean bulk path integral
\begin{equation}
    Z_{\text{AdS}} \sim e^{-S_{\text{EAdS}}}
\end{equation}
to the generating functional of a Euclidean CFT (or equivalently in AdS/CFT the Lorentzian bulk path integral is dual to the generating functional of a Lorentzian CFT). By contrast, in dS/CFT the bulk path integral is Lorentzian,
\begin{equation}
    Z_{\text{dS}} \sim e^{i S_{\text{dS}}} \, ,
\end{equation}
but it is still dual to a Euclidean CFT.  However, the counterterms in dS/CFT are imaginary, unlike counterterms in a Wick rotated AdS/CFT which are real (for parity-even terms).


\subsubsection{Wavefunction Coefficients with Counterterms}

We now give the explicit results for conformally coupled scalars, massless scalars and gravitons in $d=3$ including the appropriate counterterms.

\paragraph{Conformally coupled, $m=\sqrt{2}H, \, \nu=1/2$ in $d=3$:}
\begin{equation}
    \psi_{2,\text{CT}}=-i\frac{1+ik\eta_0}{H^2 \eta_0} + \frac{i}{H^2 \eta_0} = \frac{k}{H^2} \, .
\end{equation}
which corresponds to adding a local counterterm to the bulk free action proportional to:
\begin{equation}
    \frac{i}{H^2\eta_0^3}\int \sqrt{h} \, d^3x \, \phi^2 \, .
\end{equation}


\paragraph{Massless $m=0, \, \nu=3/2$ in $d=3$:}
\begin{equation}
    \psi_{2,\text{CT}}=\frac{-ik^2}{H^2 \eta_0(1-ik\eta_0)} + \frac{ik^2}{H^2\eta_0}=\frac{k^3}{H^2(1-ik\eta_0)} \, , \label{eqn:Delta=d=3 bulk}
\end{equation}
which corresponds to adding a local counterterm to the bulk free action proportional to 
\begin{equation}
    \frac{i}{H^2\eta_0}\int \sqrt{h} \, d^3x \, (\partial_{\mu}\phi)^2 \, .
\end{equation}
The reason the massless scalar has this additional counterterm is because \eqref{massless-no-CT} is more divergent than \eqref{conformal-no-CT} as $\eta \to 0^-$.  (For generic values of $m$ near zero, the counterterm with fewer derivatives would also appear, but in the strictly massless case ($m^2 = 0$), the shift symmetry of $\phi$ means that no $\phi^2$ counterterm is required.)

\paragraph{Graviton in $d=3$:}
\begin{equation}\label{eqn:b_real_CT}
    \text{Re}[b_\text{CT}(\bfk)]=\frac{1}{(16\pi G_N)H^2}\frac{1}{\left(\frac{1}{k^3}+\frac{\eta^2}{k}\right)}=\frac{1}{H^2(16\pi G_N)}\left(\frac{k^3}{1+k^2\eta^2}\right)
\end{equation}
\begin{equation}\label{eqn:b_imag_CT}
    \text{Im}[b_\text{CT}(\bfk)]=-\frac{1}{(16\pi G_N)H^2}\frac{1}{\frac{1}{k^4\eta}+\frac{\eta}{k^2}}=-\frac{1}{H^2(16\pi G_N)}\left(\frac{k^4\eta}{1+k^2\eta^2}\right)
\end{equation}
\begin{equation}\label{eqn:b_full_CT}
    b_\text{CT}(\bfk)=\frac{1}{(16\pi G_N)H^2}\left(\frac{k^3}{1-ik\eta}\right)
\end{equation}
where in this case there is a $k^2$ counterterm proportional to the Ricci scalar spatial curvature:
\begin{equation}
    \frac{1}{16\pi G_N}\frac{i}{H^2\eta_0}\int \sqrt{h} \, d^3x \, R^{(3)} \, .
\end{equation}
(Gravity also requires a $\mathrm{Vol}(\Sigma)$ counterterm which will be discussed in Section \ref{sec:Boundary}; however this counterterm does not affect the transverse-traceless modes considered above.)  These results will be calculated in Section \ref{sec:BulkPropagator} by another means.

\section{Cosmological Correlators from the Bulk Propagator}\label{sec:BulkPropagator}
In this section we will calculate the cosmological correlators directly from the bulk propagator.  We will also show that once the counterterms have been added in appropriately through holographic renormalisation, they agree with the correlators obtained by applying the Born rule to the wavefunction coefficients which we obtained in Section \ref{sec:Wavefunction}.

We will do the scalar case first, then the graviton case.

\subsection{Scalar Correlator}\label{sec:ScalarCorrelator}
Recall that we can write a mode expansion for the quantum field on the Cauchy slice $\Sigma$, of constant $\eta$:
\begin{equation}
    \phi(\eta; \bfx)=\int \frac{d^3k}{(2\pi)^{3/2}} [\phi^-(\eta;\bfk)a_{\bfk} + \phi^+(\eta;\bfk)a^\dagger_{\bfk}] \, ,
\end{equation}
where $\bfk$ represents a wavevector on the \emph{flat} slice and we have the usual creation/annihilation algebra. The mode functions $\phi^{\pm} (k, \eta)$ correspond to solutions of the free classical equation of motion and are given by
\begin{equation}\label{eqn:MassiveScalarModeFunctions2}
    \phi^+(k,\eta)=i\,\frac{\sqrt{\pi} H}{2} e^{-i\frac{\pi}{2}(\nu+\frac{1}{2})}\, \left( -\eta  \right)^{\frac{d}{2}} H^{(2)}_{\nu}(-k\eta) \, , \qquad \phi^-(k,\eta)=(\phi^+(k,\eta))^* \, ,
\end{equation}
where $\nu=\sqrt{d^2/4-m^2/H^2}$ denotes the order of the Hankel function, and by simply setting $m=0$, one can find the mode functions for massless fields.

We first want to obtain the scalar propagator in the vacuum state of the Hilbert space corresponding to this flat slicing of the spacetime. For the choice of the Bunch-Davies vacuum a calculation was carried out for the \emph{symmetrised} unequal-time graviton correlator~\cite{Allen:1986dd}, which will be reviewed in Section \ref{sec:GravitonCorrelator}. This will turn out to not be enough to get what we want, but it is instructive to understand why. Following a similar analysis to that of~\cite{Allen:1986dd} one finds the \emph{symmetrised} unequal-time scalar correlator to be, 
\begin{equation}\label{eqn:ScalarSymmetrisedProp}
    \langle \phi(\eta;\bfx)\phi(\eta';\bfx')+\phi(\eta;\bfx)\phi(\eta';\bfx') \rangle=
    \int \frac{d^3k}{(2\pi)^3} \tilde{w}_{\phi}(k,\eta,\eta')e^{i\bfk\cdot(\bfx-\bfx')} \, ,
\end{equation}
and
\begin{equation}
    \tilde{w}_{\phi}(k,\eta,\eta')=H^2 \Biggl(\frac{1}{k^3}\cos[k(\eta-\eta')]+\frac{\eta-\eta'}{k^2}\sin[k(\eta-\eta')]+\frac{\eta \eta'}{k}\cos[k(\eta-\eta')] \Biggr) \, . \label{eqn:symmetric}
\end{equation}
Thus the symmetrised unequal-time correlator is a purely real quantity.\footnote{The imaginary parts of the unequal-time correlator are odd under $\eta \to \eta'$.} This will not provide the correct answer when computing correlators of the conjugate momentum $\Pi_{\phi}$ which involves derivatives of the propagator, since this will be sensitive to the imaginary part of the correlator. 

Equation \eqref{eqn:ScalarSymmetrisedProp} will give the correct answer for the full (unsymmetrised) scalar propagator \emph{only} when evaluated on $\eta=\eta'$, but \emph{not} for $\eta\neq\eta'$. This is because, in general, for free fields we have that on Gaussian states (like the vacuum)
\begin{align}
    \expval{\phi(x)\phi(y)} = \mu(x,y) + i \{\phi(x),\phi(y)\}\,,
\end{align}
where $\mu(x,y)$ is a real symmetric function (to be identified with \eqref{eqn:symmetric}), while $\{\phi(x),\phi(y)\}$ is the Peierls bracket, which is an antisymmetric real function \cite{Wald:1995yp}. If we take $x$ and $y$ to be spacelike separated, the imaginary piece vanishes by causality. This is precisely what happens in our case at $\eta=\eta'$. However, when we take derivatives of this expression to get the correlators involving the conjugate momenta, as we wish to do below, we have to take these derivatives \emph{before} evaluating at $\eta=\eta'$. So we must include the imaginary part also in what follows.

One can do an analogous calculation for the \emph{non-symmetrised} unequal-time correlator
\begin{equation}\label{eqn:ScalarProp}
    \langle \phi(\eta;\bfx)\phi(\eta';\bfx')\rangle=
    \int \frac{d^3k}{(2\pi)^3} w_{\phi}(k,\eta,\eta')e^{i\bfk\cdot(\bfx-\bfx')} \, ,
\end{equation}
where
\begin{align}
    w_{\phi}(k,\eta,\eta')=\frac{H^2}{2} \Biggl[ \Biggl( &\frac{1}{k^3}\cos[k(\eta-\eta')]+\frac{\eta-\eta'}{k^2}\sin[k(\eta-\eta')]+\frac{\eta \eta'}{k}\cos[k(\eta-\eta')] \Biggr)  \nonumber \\ 
    - i \Biggl( &\frac{1}{k^3}\sin[k(\eta-\eta')]-\frac{\eta-\eta'}{k^2}\cos[k(\eta-\eta')]+\frac{\eta \eta'}{k}\sin[k(\eta-\eta')] \Biggr) \Biggr] \, .
\end{align}

From this basic propagator we can obtain the two-point functions involving the canonically conjugate momentum: $\langle \phi\Pi_{\phi}+\Pi_{\phi}\phi\rangle$ and $\langle\Pi_{\phi}\Pi_{\phi}\rangle$. But for that we need to define what we mean by the conjugate momentum.
\begin{equation}
    \Pi_{\phi}=\frac{\delta \mathcal{L}}{\delta (\partial_0\phi)}= a^2(\eta)\phi'\, ,
\end{equation}
and we impose the canonical commutation relation for the fluctuating canonical variables on any given flat slice: $[\phi(\bfx),\Pi_{\phi}(\bfy)]=i\delta(\bfx-\bfy)$.

We see that in order to obtain the correlators involving the conjugate momentum $\Pi_{\phi}$ we we will need to be taking time derivatives of the scalar propagator in \eqref{eqn:ScalarProp}.

The two-point graviton correlator is simply given by
\begin{equation}
    \langle \phi(\bfk)\phi(\bfq)\rangle_{\eta}= w_{\phi}(k,\eta,\eta')|_{\eta'=\eta} \, \delta(\bfk+\bfq) = \frac{H^2}{2}\left(\frac{1+k^2\eta^2}{k^3}\right)\delta(\bfk+\bfq)
\end{equation}

The \emph{symmetrised} two-point mixed scalar-conjugate momentum correlator is given by
\begin{align}
    \langle \phi(\bfk)\Pi_{\phi}(\bfq)+\Pi_{\phi}(\bfk)\phi(\bfq)\rangle_{\eta}&=\Bigg(\frac{1}{(-H\eta)^2}\partial_{\eta}w(k,\eta,\eta')|_{\eta'=\eta} \\
    & \quad +\frac{1}{(-H\eta')^2}\partial_{\eta'}w(k,\eta,\eta')|_{\eta'=\eta} \Bigg) \delta(\bfk+\bfq) \\
    &= \left(\frac{1}{k\eta}\right)\delta(\bfk+\bfq) \, ,
\end{align}

The two-point conjugate momentum correlator is given by
\begin{align}
    \langle \Pi_{\phi}(\bfk)\Pi_{\phi}(\bfq)\rangle_{\eta}&=\left(\frac{1}{2(-H\eta)^2}\partial_{\eta}\left(\frac{1}{2(-H\eta')^2}\partial_{\eta'}w(k,\eta,\eta')\right)\right)|_{\eta'=\eta} \, \delta(\bfk+\bfq)  \\
    &= \frac{1}{2H^2}\left(\frac{k}{\eta^2}\right)\delta(\bfk+\bfq)
\end{align}

Evaluating the correlators on a given flat slice $\Sigma$, by setting $\eta=\eta'$, the results in momentum space are:

\textbf{Without bulk counterterm}
\begin{equation} \label{eqn:ScalarCosmoCorr1}
    \langle \phi(\bfk)\phi(\bfq)\rangle_{\eta}=\frac{1}{2}H^2 \left(\frac{1+k^2\eta^2}{k^3}\right)\delta(\bfk+\bfq) \, ,
\end{equation}
\begin{equation} \label{eqn:ScalarCosmoCorr2}
    \langle \Pi_{\phi}(\bfk)\Pi_{\phi}(\bfq)\rangle_{\eta}=\frac{1}{2}\frac{1}{H^2}\left(\frac{k}{\eta^2}\right)\delta(\bfk+\bfq) \, ,
\end{equation}
\begin{equation}\label{eqn:ScalarCosmoCorr3}
    \langle \phi(\bfk)\Pi_{\phi}(\bfq)+\Pi_{\phi}(\bfk)\phi(\bfq)\rangle_{\eta}= \left(\frac{1}{k\eta}\right)\mathbf\delta(\bfk+\bfq) \, .
\end{equation}

\textbf{With bulk counterterm}
\begin{equation} \label{eqn:ScalarCosmoCorr1CT}
    \langle \phi(\bfk)\phi(\bfq)\rangle_{\eta}=\frac{1}{2}H^2 \left(\frac{1+k^2\eta^2}{k^3}\right)\delta(\bfk+\bfq) \, ,
\end{equation}
\begin{equation} \label{eqn:ScalarCosmoCorr2CT}
    \langle \Pi_{\phi}(\bfk)\Pi_{\phi}(\bfq)\rangle_{\eta}=\frac{1}{2}\frac{1}{H^2}\left(k^3\right)\delta(\bfk+\bfq) \, ,
\end{equation}
\begin{equation}\label{eqn:ScalarCosmoCorr3CT}
    \langle \phi(\bfk)\Pi_{\phi}(\bfq)+\Pi_{\phi}(\bfk)\phi(\bfq)\rangle_{\eta}= \left(-2k\eta\right)\delta(\bfk+\bfq) \, .
\end{equation}
Thus, the \emph{bulk counterterms} ensure that the cosmological correlators are finite at the boundary. More strikingly, after holographic renormalisation the mixed correlator $\langle \phi(\bfk)\Pi_{\phi}(\bfq)+\Pi_{\phi}(\bfk)\phi(\bfq)\rangle_{\eta}$ vanishes as $\eta \to 0$, while the conjugate momentum correlator $\langle \Pi_{\phi}(\bfk)\Pi_{\phi}(\bfq)\rangle_{\eta}$ becomes independent of the bulk time $\eta$, i.e. it is invariant under time evolution.

This behaviour admits a natural interpretation from the perspective of dS/CFT, as well as holographic cosmology more generally speaking. The conjugate momentum $\Pi_\phi$ is the canonical momentum associated with time evolution on a bulk Cauchy slice. After holographic renormalisation, its correlators are reorganised into quantities with definite scaling behaviour under boundary dilatations. As a result, the two-point function of $\Pi_\phi$ is constrained by scale invariance to take the form
\begin{equation}
    \langle \Pi_\phi(\mathbf{k}) \Pi_\phi(-\mathbf{k}) \rangle' \propto k^d \, ,
\end{equation}
which is invariant under renormalisation group flow. In the holographic dictionary, bulk time evolution is identified with RG flow in the dual theory, and RG-invariant quantities therefore become independent of the bulk time coordinate once the appropriate counterterms are included. This explains why the renormalised $\langle \Pi_\phi \Pi_\phi \rangle$ correlator is constant on all Cauchy slices. (See e.g. Appendices B from~\cite{Bzowski:2012ih} for more details).

The vanishing of the mixed correlator $\langle \phi \Pi_\phi + \Pi_\phi \phi \rangle$ at the boundary follows from the same structure. After holographic renormalisation, the renormalised source for $\phi$ is set to zero, while the remaining finite contributions are suppressed by powers of $k\eta$ and hence vanish as $\eta \to 0$. Equivalently, this correlator probes the overlap between operators with different scaling behaviour and is therefore forced to vanish at the RG fixed point.

Together, these features reflect the fact that holographic renormalisation reorganises bulk observables into quantities with definite scaling properties under dilatations, making scale invariance manifest in the late-time limit.\footnote{We thank Kostas Skenderis for helpful discussions regarding this point.}




\subsection{Graviton Correlator}\label{sec:GravitonCorrelator}
\subsubsection{Gravity Wavefunction in Transverse-Traceless Gauge}
To make connection with the duality proposal in Sections \ref{sec:Boundary}-\ref{sec:BoundaryFlow} we need to define cosmological wavefunctions living on the Hilbert space corresponding to this foliation. These would be functionals of the graviton field, $\Psi[h]$, where previously in Section \ref{sec:FreeScalardS} the wavefunctions were functionals of scalar fields. The physical Hilbert space is determined by the constraints inherited from the full quantum gravity description, namely:
\bea
    \mathcal{H}(x)\Psi[g]&:=&\Bigg\{\frac{16\pi G_N}{\sqrt{g}}:\!\Big(\Pi_{ab}\Pi^{ab}-\frac{1}{d-1}\Pi^2\Big)\!:-\frac{\sqrt{g}}{16\pi G_N}(R-2\Lambda)\Bigg\}\Psi[g]=0,\label{eqn:Hconst} \\
    \mathcal{D}^a(x)\Psi[g]&:=&-2\nabla_b \Pi^{ab}\,\Psi[g]=0\label{eqn:Dconst}
\eea
We now expand these constraints to leading-order in $\h_{ab}$, while fixing our gauge to be transverse-traceless. The momentum constraint can be easily seen to impose the preservation of the transverse gauge from slice to slice. If we expand the Hamiltonian operator we get:
\begin{equation}
    \mathcal{H}=-2(16\pi G_N)Hg_{ab}\Pi^{ab}
\end{equation}
which imposes the preservation of the traceless gauge as we move from slice to slice. 

The general solution to this constraint expanded to second-order is
\begin{equation}\label{eqn:Gaussain}
    \Psi[h]=e^{iS[h]}=\exp\left(-\frac{1}{2}\int d^dk\;b(\bfk)\mathbf{P}^{abcd}\h_{ab}(\bfk)\h_{cd}(-\bfk)\right)
\end{equation}

If we Fourier transform the commutation relation we deduce that in momentum space the differential representation for the conjugate momentum is: $\Pi^{ab}(\bfk)=-i\frac{\delta}{\delta \h_{ab}(-\bfk)}$. We further note that the following relation holds if we restrict to the gauge-fixed subspace of metrics:
\begin{equation}
    \h^{ab}(\bfk)\Psi=-\frac{1}{b(\bfk)}\frac{\delta\Psi}{\delta \h_{ab}(-\bfk)}
\end{equation}

In order to compute the relevant expectation values, we first need to define a valid inner product on this Hilbert space. We follow the usual QFT prescription, which should hold in our case since we are in the semiclassical limit, or, equivalently, in the low-energy regime of the gravity EFT. Thus, we define:
\begin{equation}
    \braket{\Psi}{\Phi}=\int \frac{\mathcal{D}h}{\text{Vol(G)}}\Psi^*[h]\Phi[h]
\end{equation}
for any two states on the physical Hilbert space.\footnote{Note that here, prior to gauge fixing, the integration variable $h$ still includes functional degrees of freedom which are pure gauge, not just $\h_{ab}$ from before.} In order to have a well-defined notion of inner product we must divide by the volume of the gauge group. The usual way to proceed is to gauge-fix the integration variable by introducing ghost fields. But because our wavefunctional is Gaussian and because we will be normalising our expectation values, the effect due to these ghost fields will cancel out. Thus, we will ignore this unnecessary complication.

One can compute the following expectation values in the transverse-traceless gauge using \eqref{eqn:Gaussain}
\begin{equation}\label{eqn:hh}
    \langle \h^{ab}(\bfk)\h^{cd}(\bfq)\rangle_{\Psi} = \frac{1}{2\,\text{Re}[b]}\mathbf{P}^{abcd}\delta(\bfk+\bfq)
\end{equation}
\begin{equation}\label{eqn:hpi}
    \langle \Pi^{ab}(\bfk)\Pi^{cd}(\bfq)\rangle_{\Psi}=|b|^2\frac{1}{2\,\text{Re}[b]}\mathbf{P}^{abcd}\delta(\bfk+\bfq)
\end{equation}
\begin{equation}\label{eqn:pipi}
    \langle \h^{ab}(\bfk)\Pi^{cd}(\bfq)+\Pi^{cd}(\bfq)\h^{ab}(\bfk)\rangle_{\Psi}=-\frac{\text{Im}[b]}{\text{Re}[b]}\mathbf{P}^{abcd}\delta(\bfk+\bfq)
\end{equation}

\subsubsection{Graviton Propagator in $d=3$}

We now restrict to $d=3$. In transverse-traceless gauge our dynamical field only has non-vanishing components in the directions along the slice $\Sigma$ (i.e. $\h_{0\mu}=0$). So we only need to focus on the spatial components, which are required to satisfy: $g_{ab}\h^{ab}=0$ (traceless) and $\partial_a\h^{ab}=0$ (transverse). In this case, the only degrees of freedom are the two decoupled scalar mode functions, $\h_\times$ and $\h_+$. Changing basis into circular polarisations we can write a mode expansion for the quantum field on the Cauchy slice $\Sigma$, of constant $\eta$:
\begin{equation}
    \h^{ab}(x)=\int \frac{d^3k}{(2\pi)^{3/2}} [m^a(\bfk)m^b(\bfk)F_R(\bfk,x)a_R + (\text{c.c.})a^\dagger_R] + L \text{ polarisation} \, .
\end{equation}
Here $\bfk$ represents a wavevector on the \emph{flat} slice and we have the usual creation/annihilation algebra. The reason such an expansion is possible is because the theory of linearised gravitons is equivalent to the theory of two minimally coupled, independent, \emph{free} scalars. The mode functions are the same for both polarisations and satisfy $\Box^{(4)} F=0$. The polarisation basis is given by $m^a(\bfk)=\frac{1}{\sqrt{2}}(e^a_1(\bfk)+ie^a_2(\bfk))$ and its complex conjugate, where the set $\{e^a_1(\bfk),e^a_2(\bfk)\}$ forms a RH triad with the wavevector $\bfk$. The orthonormality conditions are with respect to the background metric on the Cauchy slice $\ha_{ab}=a^2\delta_{ab}$. The basis is evidently not unique due to rotational symmetry in the plane orthogonal to $\bfk$. However, there is a particular combination which invariant under rotational symmetry and which is needed to obtain the results that follow:
\begin{equation}
    m^am^bm^{c*}m^{d*} + m^{a*}m^{b*}m^cm^d=\mathbf{P}^{abcd} \, ,
\end{equation}
where we have defined the \emph{projector} onto the transverse-traceless subspace of tensors in the tangent space to $\Sigma$:
\begin{equation}
    \mathbf{P}^{abcd}=\frac{1}{2}(\mathbf{P}^{ac}\mathbf{P}^{bd}+\mathbf{P}^{ad}\mathbf{P}^{bc}-\mathbf{P}^{ab}\mathbf{P}^{cd}) \, ,
\end{equation}
\begin{equation}
    \mathbf{P}^{ab}=\delta^{ab}-\frac{k^ak^b}{k^2} \, ,
\end{equation}
with $\mathbf{P}^{ab}$ being the usual transverse tensor defined in the cosmological correlator literature (see e.g. \cite{WFCtoCorrelators2}).\footnote{In~\cite{Allen:1986dd} a rescaled projection tensor was used  $\mathbf{\tilde{P}}^{ab}=a^2(\eta)\mathbf{P}^{ab}$ corresponding to the metric fluctuations of the rescaled metric $\ha_{ab}=a^2(\eta)\delta_{ab}$.}\footnote{In dS fields generally scale as $\phi \sim \eta^{\Delta +} + \eta^{\Delta -}$ with scaling dimension $\Delta = \frac{d}{2} \pm \sqrt{\frac{d^2}{4}-\frac{m^2}{H^2}}$. Hence for massless fields $\phi \sim \eta^{\Delta -}$ at $\eta_0 = 0$, but in general massive fields die off at $\eta_0 = 0$ and the correlators of such fields have logarithmic divergences of the type $\log(-k \eta_0)$.}

We first want to obtain the graviton propagator in the vacuum state of the Hilbert space corresponding to this flat slicing of the spacetime. As mentioned, in \cite{Allen:1986dd} the symmetrised correlator was computed to be
\begin{equation}\label{eqn:Bruceprop}
    \langle \h^{ab}\h^{a'b'}+\h^{a'b'}\h^{ab}\rangle(\bfx,\eta;\bfx',\eta')=
    \int \frac{d^3k}{(2\pi)^3} \mathbf{P}^{aba'b'}\tilde{w}(k,\eta,\eta')e^{i\bfk\cdot(\bfx-\bfx')} \, ,
\end{equation}
where the primed indices belong to the tangent space at the primed coordinates\footnote{We will temporarily restore the factors of $16\pi G_N$ so as to keep track of dimensional analysis more easily.} and
\begin{equation}
    \tilde{w}(k,\eta,\eta')=(2\pi)^3\frac{2G_NH^2}{\pi^2} \Biggl(\frac{1}{k^3}\cos[k(\eta-\eta')]+\frac{\eta-\eta'}{k^2}\sin[k(\eta-\eta')]+\frac{\eta \eta'}{k}\cos[k(\eta-\eta')] \Biggr) \, .
\end{equation}
For exactly the same reason as explained earlier for the scalar, we must keep track of the anti-symmetric imaginary part. The full propagator is
\begin{equation}\label{eqn:prop}
    \langle \h^{ab}\h^{a'b'}\rangle(\bfx,\eta;\bfx',\eta')=
    \int \frac{d^3k}{(2\pi)^3} \mathbf{P}^{aba'b'}w(k,\eta,\eta')e^{i\bfk\cdot(\bfx-\bfx')} \, ,
\end{equation}
where
\begin{align}
    w(k,\eta,\eta')=(2\pi)^3\frac{2G_NH^2}{\pi^2} \Biggl[ \Biggl( &\frac{1}{k^3}\cos[k(\eta-\eta')]+\frac{\eta-\eta'}{k^2}\sin[k(\eta-\eta')]+\frac{\eta \eta'}{k}\cos[k(\eta-\eta')] \Biggr)  \nonumber \\ 
    - i \Biggl( &\frac{1}{k^3}\sin[k(\eta-\eta')]-\frac{\eta-\eta'}{k^2}\cos[k(\eta-\eta')]+\frac{\eta \eta'}{k}\sin[k(\eta-\eta')] \Biggr) \Biggr] \, .
\end{align}

From this basic propagator we can obtain the two-point functions involving the canonically conjugate momentum: $\langle \h^{ab}\Pi^{a'b'}+\Pi^{a'b'}\h^{ab}\rangle$ and $\langle\Pi^{ab}\Pi^{a'b'}\rangle$. But for that we need to define what we mean by the conjugate momentum. Starting from the momentum corresponding to the full quantum gravity theory and expanding to leading-order in the fluctuating fields we get:\footnote{In this expression we have already applied the gauge condition. This is all we need since these quantities will be inside correlation functions, which we must gauge-fix in order for them to make sense.}
\begin{equation}
    \Pi^{ab}_{\text{(full)}}=-\frac{\sqrt{g}}{16\pi G_N}(K^{ab}_{\text{(back)}}-g^{ab}K_{\text{(back)}})-\frac{\sqrt{g}}{16\pi G_N}(K^{ab}-g^{ab}K)-\frac{\sqrt{g}}{16\pi G_N}\h^{ab}K_{\text{(back)}} + \mathcal{O}(\h^2,hK)
\end{equation}
where $K^{ab}$ is the extrinsic curvature associated to the graviton fluctuations. For our type of foliation the following relation holds:
\begin{equation}
    K^{ab}=-\frac{1}{2a}\frac{\partial \h^{ab}}{\partial \eta}-\frac{2\dot{a}}{a^2}\h^{ab}
\end{equation}

As always there is ambiguity in defining the canonical variables. We note that any term proportional to $\h^{ab}$ can be encapsulated by a canonical transformation so it represents a mere change of coordinates on phase-space. We pick coordinates in which the conjugate momentum takes the form:
\begin{equation}\label{eqn:can_mom}
    \Pi^{ab}=\frac{\sqrt{g}}{16\pi G_N}\frac{1}{2a}\frac{\partial \h^{ab}}{\partial \eta}=\frac{1}{16\pi G_N}\frac{a^2}{2}\frac{\partial \h^{ab}}{\partial \eta}
\end{equation}
and we impose the canonical commutation relation for the fluctuating canonical variables on any given flat slice: $[\h_{ab}(\bfx),\Pi^{cd}(\bfy)]=i\delta_{ab}^{cd}\delta(\bfx-\bfy)$.

We see that in order to obtain the correlators we want we will need to be taking time derivatives of the graviton propagator in \eqref{eqn:prop}. It is useful to note that the tensorial structure $\mathbf{P}^{aba'b'}$ satisfies
\begin{equation}
    \left(\frac{1}{2a}\frac{\partial}{\partial \eta}\right)\mathbf{P}^{aba'b'}=0 \, ,
\end{equation}
\begin{equation}
    \left(\frac{1}{2a'}\frac{\partial}{\partial \eta'}\right)\left(\frac{1}{2a}\frac{\partial}{\partial \eta}\right)\mathbf{P}^{aba'b'}=0 \, ,
\end{equation}
where to obtain this, it is crucial to remember that $\mathbf{P}^{ab}$ has no time-dependence and thus $\partial_{\eta}\mathbf{P}^{ab}=0$ and that the primed indices belong to the primed coordinates. From here, we can obtain recurrence relations for the time derivatives of the combination $\mathbf{P}^{aba'b'}w(k,\eta,\eta')$.

The two-point graviton correlator is simply given by
\begin{equation}
    \langle \h^{ab}\h^{cd}\rangle(\eta;\bfk,\bfq)= w(k,\eta,\eta')|_{\eta'=\eta} \, \mathbf{P}^{abcd}\delta(\bfk+\bfq) = (8\pi G_N)H^2 \left(\frac{1+k^2\eta^2}{k^3}\right)\mathbf{P}^{abcd}\delta(\bfk+\bfq)
\end{equation}

The \emph{symmetrised} two-point mixed graviton-conjugate momentum correlator is given by
\begin{align}
    \langle \h^{ab}\Pi^{cd}+\Pi^{ab}\h^{cd}\rangle(\eta;\bfk,\bfq)&=\frac{1}{16\pi G_N}\Bigg(\frac{1}{2(-H\eta)^2}\partial_{\eta}w(k,\eta,\eta')|_{\eta'=\eta} \\
    & \qquad \qquad \quad +\frac{1}{2(-H\eta')^2}\partial_{\eta'}w(k,\eta,\eta')|_{\eta'=\eta} \Bigg) \mathbf{P}^{abcd}\delta(\bfk+\bfq) \\
    &= \left(\frac{1}{k\eta}\right)\mathbf{P}^{abcd}\delta(\bfk+\bfq) \, ,
\end{align}

The two-point conjugate momentum correlator is given by
\begin{align}
    \langle \Pi^{ab}\Pi^{cd}\rangle(\eta;\bfk,\bfq)&=\left(\frac{1}{16\pi G_N}\frac{1}{2(-H\eta)^2}\partial_{\eta}\left(\frac{1}{16\pi G_N}\frac{1}{2(-H\eta')^2}\partial_{\eta'}w(k,\eta,\eta')\right)\right)|_{\eta'=\eta} \, \mathbf{P}^{abcd}\delta(\bfk+\bfq)  \\
    &= \frac{1}{32\pi G_N}\frac{1}{H^2}\left(\frac{k}{\eta^2}\right)\mathbf{P}^{abcd}\delta(\bfk+\bfq)
\end{align}

Evaluating the correlators on a given flat slice $\Sigma$, by setting $\eta=\eta'$, the results in momentum space are:

\textbf{Without bulk counterterm}
\begin{equation} \label{eqn:cosmocorr1}
    \langle \h^{ab}\h^{cd}\rangle(\eta;\bfk,\bfq)=\frac{1}{2}(16\pi G_N)H^2 \left(\frac{1+k^2\eta^2}{k^3}\right)\mathbf{P}^{abcd}\delta(\bfk+\bfq)
\end{equation}
\begin{equation} \label{eqn:cosmocorr2}
    \langle\Pi^{ab}\Pi^{cd}\rangle(\eta;\bfk,\bfq)=\frac{1}{2}\frac{1}{16\pi G_N}\frac{1}{H^2}\left(\frac{k}{\eta^2}\right)\mathbf{P}^{abcd}\delta(\bfk+\bfq)
\end{equation}
\begin{equation}\label{eqn:cosmocorr3}
    \langle \h^{ab}\Pi^{cd}+\Pi^{ab}\h^{cd}\rangle(\eta;\bfk,\bfq)= \left(\frac{1}{k\eta}\right)\mathbf{P}^{abcd}\delta(\bfk+\bfq)
\end{equation}

\textbf{With bulk counterterm}
\begin{equation} \label{eqn:cosmocorr1CT}
    \langle \h^{ab}\h^{cd}\rangle(\bfk,\bfq;\eta)=\frac{1}{2}(16\pi G_N)H^2 \left(\frac{1+k^2\eta^2}{k^3}\right)\mathbf{P}^{abcd}\delta(\bfk+\bfq)
\end{equation}
\begin{equation} \label{eqn:cosmocorr2CT}
    \langle\Pi^{ab}\Pi^{cd}\rangle(\bfk,\bfq;\eta)=\frac{1}{2}\frac{1}{16\pi G_N}\frac{1}{H^2}\left(k^3\right)\mathbf{P}^{abcd}\delta(\bfk+\bfq)
\end{equation}
\begin{equation}\label{eqn:cosmocorr3CT}
    \langle \h^{ab}\Pi^{cd}+\Pi^{ab}\h^{cd}\rangle(\bfk,\bfq;\eta)= \left(-2k\eta\right)\mathbf{P}^{abcd}\delta(\bfk+\bfq)
\end{equation}
Equations \eqref{eqn:cosmocorr1}--\eqref{eqn:cosmocorr3} give us the Gaussian contribution to the Bunch-Davies vacuum state in the regime of small graviton fluctuations (without counterterm):
\begin{equation}\label{eqn:b_full2}
    b(\bfk)=-\frac{1}{(16\pi G_N)H^2}\left(\frac{ik^2/\eta}{1-ik\eta}\right) \, .
\end{equation}
\begin{equation}\label{eqn:b_real2}
    \text{Re}[b(\bfk)]=\frac{1}{(16\pi G_N)H^2}\frac{1}{\left(\frac{1}{k^3}+\frac{\eta^2}{k}\right)}=\frac{1}{H^2(16\pi G_N)}\left(\frac{k^3}{1+k^2\eta^2}\right) \, .
\end{equation}
\begin{equation}\label{eqn:b_imag2}
    \text{Im}[b(\bfk)]=-\frac{1}{(16\pi G_N)H^2}\frac{1}{\left(\frac{\eta}{k^2}+\eta^3\right)}=-\frac{1}{H^2(16\pi G_N)}\left(\frac{k^2/\eta}{1+k^2\eta^2}\right) \, .
\end{equation}
Our results match what we obtained in Section \ref{sec:FreeGravitondS}. As for the scalar case, the \emph{bulk counterterms} ensure that the cosmological correlators are finite at the boundary and that the $\langle\Pi^{ab}\Pi^{cd}\rangle$ have no dependence on $\eta$. See the end of Section~\ref{sec:ScalarCorrelator} for more details.



\pagebreak[2]


\section{Introduction to Cauchy Slice Holography}\label{sec:Boundary}
In this section we outline the Cauchy Slice Holography (CSH) dictionary, first introduced in \cite{Araujo-Regado:2022gvw}, and then further developed in \cite{Araujo-Regado:2022jpj,Khan:2023ljg,Soni:2024aop,Khan:2025gld}, and explain in detail how to recover the bulk correlators from a boundary dual computation valid at finite time. This is in spirit very aligned with previous computations done in the context of AdS/CFT at finite radial cut-off \cite{Hartman:2018tkw}. Here, we not only adapt it to the language of CSH, but also extend it in several ways. Firstly, working within dS/CFT brings with it several novel features, not seen in the AdS/CFT setup of \cite{Araujo-Regado:2022gvw} and \cite{Hartman:2018tkw}. We will point these out along the way. Secondly, we provide a comprehensive study of the deformation operator for different parameter regimes, finding agreement with the results of \cite{Hartman:2018tkw} when appropriate. In particular, we analyse the deformation across the whole complex-$\Delta$ plane, where $\Delta$ is the scaling dimension of a dual scalar operator, thus checking the proposed duality for both complementary and principal series. Our hope is that the coming sections provide a clear, systematic procedure for doing computations within CSH that could go beyond two-point functions.

In Section \ref{sec:what is it?} we provide a conceptual overview of the main ingredients. This should be a good reference point for the reader working through later sections. After setting notation and conventions in Section \ref{sec:conventions}, from Section \ref{sec:Concrete} onwards we explicitly construct the deformation operators and counterterms relevant for the different regimes of interest in this paper. Section \ref{sec:grav} is dedicated to pure gravity, while Section \ref{sec:matter} addresses the inclusion of scalar matter.

\subsection{What is Cauchy Slice Holography?}\label{sec:what is it?}
CSH provides an explicit isomorphism between the bulk physical Hilbert space and a boundary dual Hilbert space. By the former we will always mean the Hilbert space of solutions to the bulk diffeomorphism constraints. 

\begin{figure}
    \centering
    \includegraphics[width=0.7\textwidth]{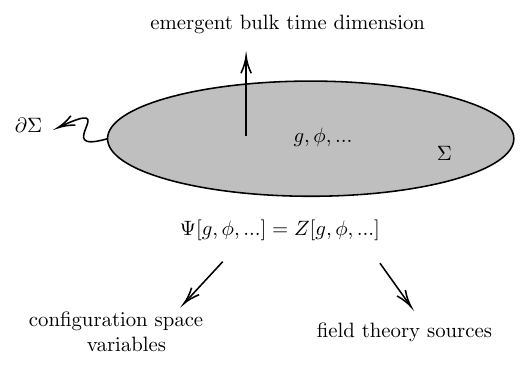}
    \caption{The bulk state $\Psi$ is defined on an abstract spatial slice $\Sigma$ with boundary $\partial\Sigma$. The extra dimension is emergent, via the fact that the state satisfies the Hamiltonian constraint. The fundamental equation of CSH is shown. On the RHS we have a QFT partition function living on the same $\Sigma$. On the boundary $\partial\Sigma$ one has to input a QFT state, which will be mapped to the bulk state $\Psi$ via the evaluation of the partition function $Z$.}
    \label{fig:CSH}
\end{figure}

We have in mind the setup of canonical quantum gravity~\cite{PhysRev.160.1113}, in which a bulk state is given by a coherent superposition of \emph{spatial} metric and scalar configuration eigenstates: $\Psi[g_{ab},\Phi]$, where for simplicity we restrict ourselves to scalar matter in the bulk. The state is defined a priori on an \emph{abstract} slice $\Sigma$, without reference to a bulk spacetime, as shown in Figure \ref{fig:CSH}. It is only for semiclassical states that an emergent notion of spacetime is appropriate~\cite{Khan:2023ljg}. This will, in fact, be the case here, where we will be constructing the bulk Bunch-Davies state and will be effectively working in the $G_N\to 0$ limit of the theory.\footnote{The correct WKB parameter will actually correspond to the size of the spatial slice relative to the cosmological scale set by $\Lambda$. We should really only trust the semiclassical description if we do not go ``too back in time''. On the boundary side, this will be related to the fact that the deformation operator is irrelevant and so we can only really trust the description for small values of the deformation. The deformation parameter will correspond precisely to the bulk time parameter backwards from $\mathcal{I}^{+}$.}  The full quantum state knows about all possible types of slices $\Sigma$ (with or without boundary, with trivial or non-trivial topology), but we will be restricting ourselves to evaluating the state on slices topologically isomorphic to Euclidean space: $\Sigma\cong\mathbb{R}^d$. This is because we want to study the duality with the Poincaré patch of dS. For the case of holographic cosmology on closed spatial slices see~\cite{Araujo-Regado:2022jpj}.
The Hamiltonian constraint in $d+1$-dimensions takes the form:\footnote{We note that the momentum constraints will be implicitly solved throughout by working with functionals $\Psi[g,\Phi]$ that are covariantly constructed from $g_{ab}$ and $\Phi$. This is the case for the partition functions $Z$ we will be using.}
\begin{align}
    \mathcal{H}(x)\Psi[g,\Phi]:=&\Bigg\{\frac{16\pi G_N}{\sqrt{g}}:\!\Big(\Pi_{ab}\Pi^{ab}-\frac{1}{d-1}\Pi^2\Big)\!:-\frac{\sqrt{g}}{16\pi G_N}(R-2\Lambda)\\
    &+ \frac{1}{2}\frac{1}{\sqrt{g}}:\!\Pi_\Phi^2\!:+\frac{1}{2}\sqrt{g}\left((\nabla \Phi)^2+m^2\Phi^2\right)\Bigg\}\Psi[g,\Phi]=0,\label{eqn:ham}
\end{align}
which must hold at every point on $\Sigma$. $\Pi^{ab}=-i\frac{\delta}{\delta g_{ab}}$ and $\Pi_\Phi=-i\frac{\delta}{\delta\Phi}$ are the conjugate momentum variables. 

The $::$ symbol denotes a particular prescription for defining the composite kinetic operator. As usual in quantum theory, there are operator-ordering ambiguities. However, here these are entirely fixed by the proposed duality with a QFT, in which there is a natural definition of composite operators via \emph{point-splitting}. This is described in detail later (see equation \eqref{eq:point-splitting}), but for now all we need to know is that it effectively acts as a normal-ordering prescription, moving all configuration variables to the left of the momentum variables. From the dual side, this is valid only in the strict $N\to\infty$ limit, which on the bulk side corresponds to $G_N\to 0$. In practice, this allows us to manipulate these operators ``classically'', as if they were literally multiplied together.\footnote{See \cite{Araujo-Regado:2022gvw} for a discussion on how to extend this definition for finite, but large $N$.}

\subsubsection{The Large Volume Limit}\label{sec:largevol}
A remarkable feature of this equation, which we briefly sketch below, was explained in \cite{Freidel:2008sh}, where it was shown that if one evaluates a general solution on configurations of large spatial volume (compared to the cosmological scale), it looks like a CFT partition function, up to counterterms. More specifically, for pure gravity in the bulk, it was shown that if one considers the following one-parameter family of geometries
\begin{align}
    \left\{\frac{1}{\rho^2}g_{ab}\Big|\,\rho\in\mathbb{R}_+\right\}\label{eq:freidelfamily}
\end{align}
and expands the wavefunction evaluated on this set, treating $\rho$ as a WKB parameter
\begin{align}
    \lim_{\rho\to 0^+}\Psi\left[\frac{1}{\rho^2}g_{ab}\right] \sim: e^{i\sum_n\rho^{-n}S^{(n)}[g_{ab}]}
\end{align}
then the Hamiltonian constraint equation fixes the functionals $S^{(n)}[g_{ab}]$ associated with the divergent terms, such that the solution looks like
\begin{align}
    \lim_{\rho\to 0^+}\Psi\left[\frac{1}{\rho^2}g\right] \sim e^{\sum_n\rho^{-n}\text{CT}^{(n)}[g]} \, Z_\text{CFT}[g]\,. \label{eqn:IR}
\end{align}
Here, $\text{CT}^{(n)}[g]$ are some explicit, universal local functionals, which we will write down later. They are the same for all states. The functional $Z_\text{CFT}[g]$ satisfies the Weyl anomaly equation
\begin{align}
    g_{ab}\frac{\delta}{\delta g_{ab}}(x) \log Z_\text{CFT}[g] \sim \mathcal{A}(x)\,.\label{eqn:anomaly}
\end{align}
where $\mathcal{A}(x)$ is a local functional of the metric called the \emph{anomaly}, entirely determined by the Hamiltonian constraint itself. For this reason, the proposal is that one should equate $Z_\text{CFT}[g]$ with the partition function of an actual CFT. Note that because the metric $g_{ab}$ ``sourcing'' the family \eqref{eq:freidelfamily} was arbitrary, \eqref{eqn:anomaly} holds as an operator equation under this equivalence. Finally, as discussed in \cite{Freidel:2008sh}, and as we will see explicitly later, $n\in\{d,d-2,...\}$, with the important caveat that for even $d$ there is also a term that goes like $\sim \log \rho$. This corresponds precisely to the case that the anomaly $\mathcal{A}(x)\neq 0$ and so $\rho\to 0^+$ acts as a UV-cutoff for the CFT, meaning that we should really write $Z_\text{CFT}^{(\rho)}[g]$. For odd $d$, $Z_\text{CFT}[g]$ is independent of $\rho$ and $\mathcal{A}(x)=0$. What we have said so far works for both $\Lambda<0$ and $\Lambda>0$.\footnote{Crucially, this conclusion fails for $\Lambda=0$.}

This connection is known to be actually exact in AdS/CFT, if one considers the wavefunction that lives on radial cut-off slices as done in \cite{Freidel:2008sh}, since we have explicit CFT dual theories living on the asymptotic boundary. Because we lack explicit duals to dS, a precise equivalence of this type is only a conjecture in the context of dS/CFT. That being said, there has been substantial progress in studying the properties of such hypothetical duals and it was shown in \cite{Chakraborty:2023yed, Chakraborty:2023los, Chakraborty:2025izq, Dey:2024zjx} that the cosmological correlators satisfy the appropriate CFT Ward identities in this limit (after removing the counterterms). This gives strong evidence to trust the conjecture.

It is interesting to note that for $\Lambda>0$ we have that $\text{CT}^{(n)}[g,\Phi]$ are all \emph{purely imaginary}, unlike AdS/CFT where they are all purely real. This means that they correspond to rapid oscillations of the phase of the wavefunction, a feature of being in the WKB limit. This also means that there are two equally good WKB branches, related by CPT conjugation. One of them describes an emergent expanding spacetime region around $\mathcal{I}^+$, while the other the time-reversed contracting region around $\mathcal{I}^-$, see \cite{Goodhew:2024eup,Araujo-Regado:2022jpj}. In this paper, we will consider only the expanding branch since we want to compute correlators in one single Poincaré patch. But taking superpositions of both branches becomes important to, for example, reproduce the Hartle-Hawking and Vilenkin wavefunctions, as done in~\cite{Araujo-Regado:2022jpj}. But from now on, we work with a single branch.

\subsubsection{The $T^2$ Deformation}
CSH provides the mechanism through which one can recover the full solution $\Psi[g,\Phi]$ from this asymptotic limit. 

Starting from the \emph{seed} CFT associated with $Z_\text{CFT}[g,\Phi]$, one applies an \emph{irrelevant deformation}, generated by the operator $\hat{O}$ (determined by the Hamiltonian constraint itself) to get the full solution
\begin{align}
\Psi^{(\rho)}[g,\Phi]=Z^{(\rho)}[g,\Phi]:=e^{\text{CT}^{(\rho)}[g,\Phi]}\left( \text{P} \exp \int_{\epsilon}^\rho\frac{d\lambda}{\lambda}\hat{O}(\lambda)\right)Z_{\text{CFT}}^{(\epsilon)}[g,\Phi]\,.\label{deformation}
\end{align}
The RHS is now to be interpreted as the partition function of some QFT, which we will describe in detail later. As illustrated in Figure \ref{fig:CSH}, to fully specify the partition function one has to input a QFT state at the spatial boundary $\partial\Sigma$. In our case, because the spatial boundary in the Poincaré slicing lies at $\mathcal{I}^+$ (see Figure \ref{fig:dS}), the state to be inputted is a CFT state. Thus, equation \eqref{deformation} provides an explicit \emph{boundary-to-bulk} map, from which for each boundary CFT state, one gets a bulk state satisfying the Hamiltonian constraint.

$\text{CT}^{(\rho)}[g,\Phi]$ denotes the combined $\sum_n\rho^{-n}\text{CT}^{(n)}[g]$ from before. We will, from now on, call these \emph{counterterms}, because, from the perspective of the partition function $Z^{(\rho)}[g,\Phi]$, they remove power-law divergences. Further, because the deformation operator is irrelevant in an RG sense, the theory is only defined with a UV-cutoff scale. This is provided by the dimensionless deformation parameter $\rho$ via the length scale $L_\text{dS}\cdot\rho$. We will work with a choice of RG-scheme such that $L_\text{dS}\cdot\rho$ is the \emph{only} dimensionful coupling in the theory. In fact, for Poincaré slicing we will identify this parameter with the conformal time
\begin{align}
    \rho = -\eta\,. \label{eqn:cutoff}
\end{align}
This is analogous to what is done in AdS/CFT at finite radial cutoff, as in \cite{Hartman:2018tkw}. $\epsilon$ corresponds to the UV-cutoff in the CFT limit, if the theory is anomalous. Otherwise, we can take it to $0$.

The deformation operator $\hat{O}(\lambda)$ will turn out to always be quadratic in the stress-tensor $T^{ab}$ and in the operator $\mathcal{O}$ dual to the bulk scalar field. For this reason, it is generically called a $T^2$ operator\footnote{Note that it is a double-trace operator, if the seed CFT is a large-$N$ gauge theory, as we assume.}. It generalises the ``$T\Bar{T}$''-operator \cite{Mazenc:2019cfg,Taylor:2018xcy,Hartman:2018tkw} to the dS/CFT setup, higher dimensions and general scaling dimension $\Delta$, as we will discuss below. 

\begin{figure}[ht]
    \centering
    \includegraphics[width=0.75\textwidth]{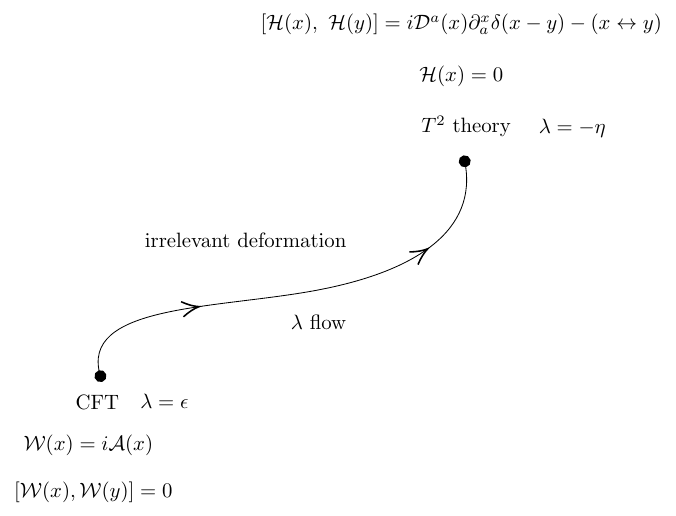}
    \caption{The seed CFT is deformed by an irrelevant operator quadratic in the stress-tensor. The flow effectively deforms the algebra of local (anomalous) scale invariance to the ``algebra'' of local diffeomorphism invariance. The Weyl anomaly constraint gets deformed to the Hamiltonian constraint. The value of the deformation parameter is in one-to-one correspondence with the conformal time labelling bulk slices in Figure \ref{fig:dS}.}
    \label{fig:RGflow}
\end{figure}

The picture one should have in mind is the following, shown in Figure \ref{fig:RGflow}. Starting from the seed CFT, we follow a particular flow in the space of theories, by turning on the operator $\hat{O}(\lambda)$. This new theory describes low-energy bulk quantum gravity, because it satisfies the diffeomorphism constraints.\footnote{Here, by low-energy we mean that the bulk degrees of freedom are given by low curvature fluctuating geometry and low energy-density fluctuating matter.} We should think of this dual theory as lying on an RG flow line connecting some unknown UV fixed-point to the seed CFT, which here lives in the IR limit. This is in line with the bulk side of the story. Whatever the UV-complete theory of quantum gravity is, there is a limit in which one recovers the quantum theory of bulk fluctuating geometry, whose states satisfy the Hamiltonian constraint (possibly with higher-curvature corrections, suppressed in powers of $G_N$). If we go further into the IR, the constraint becomes the one in \eqref{eqn:ham} and eventually we reach the true IR (large-volume) limit, given by states of the form \eqref{eqn:IR}. By recasting everything in holographic language, accessing UV features of quantum gravity now becomes identified with the problem of UV-completing the $T^2$-deformed theory.\footnote{We reiterate that we are \emph{already} accessing quantum gravity effects, albeit at low-energy, because our prescription generates general quantum bulk states, and is not restricted to semiclassical ones. In particular, we can generate arbitrary superpositions of semiclassical states in the bulk, as well as states that have no semiclassical interpretation at all. Our map \eqref{deformation} provides a bulk state for any boundary CFT state, not just states known to have a good semiclassical bulk dual.}

By comparing with Figure \ref{fig:dS}, we see that, in this setup, there is a one-to-one correspondence between the deformation flow of the dual QFT and the time flow in the bulk. It should be pointed out that this correspondence is only exact in the limit $N\to\infty$ limit of the boundary theory, or, equivalently, the $G_N\to 0$ limit of the bulk theory. More generally, we take the point of view that we should \emph{define} what we mean by emergent bulk time evolution by this dual flow, even when working at finite $N$, or when the bulk state is not semiclassical. 

However, for the purposes of this paper, we will be working in the bulk semiclassical limit. On the boundary of the spatial slice, we will input the CFT vacuum state. Together, this will generate the Bunch-Davies state $\Psi_\text{BD}[g,\Phi]$ in the bulk, as obtained by quantising scalar and graviton fluctuations on a fixed dS background. One of the main points of this paper is to show precisely this by computing the corresponding wavefunction coefficients at finite conformal time from dual computations.

\subsubsection{Preliminaries and Conventions} \label{sec:conventions}
In mapping quantities from the bulk to the boundary, it is conveninet to rescale by powers of the cutoff so that things transform appropriately under boundary Weyl rescalings. This is analogous to what is done in AdS/CFT. We will from now on set $L_\text{dS}=1$.

For the Poincaré slicing in conformal time $\eta$, the induced metric is $\ha_{ab}=a^2(\eta)\delta_{ab}$, for $a^2(\eta)=1/(-\eta)^2$. The boundary metric is defined to be
\begin{equation}
    \h_{ab}:=(-\eta)^2\ha_{ab}=\delta_{ab} \, .
\end{equation}
The boundary stress-tensor operator is then defined to be
\begin{equation}
    T^{ab}:=\frac{2}{\sqrt{\h}}\frac{\delta}{\delta \h_{ab}}=(-\eta)^{-(d+2)}\frac{2}{\sqrt{\ha}}\frac{\delta}{\delta \ha_{ab}} \, .
\end{equation}
With these definitions we have that, under Weyl rescalings on the spatial slice
\begin{eqnarray}
    \h_{ab}&\mapsto& \Omega^2\h_{ab}\\
    T^{ab}&\mapsto& \Omega^{-(d+2)} T^{ab} \, .
\end{eqnarray}
This means that $T:=\h_{ab}T^{ab}$ has conformal dimension $d$.

Similarly, for scalar fields we define their boundary source and operator counterparts in terms of their bulk induced values $\Phi$ via
\begin{eqnarray}
    \phi&:=& (-\eta)^{\Delta-d}\Phi\\
    {\cal O}&:=& \frac{1}{\sqrt{\h}}\frac{\delta}{\delta \phi} = (-\eta)^{-\Delta} \frac{1}{\sqrt{\mathfrak{h}}}\frac{\delta}{\delta \Phi} \,.
\end{eqnarray}
So we are defining $\Delta$ to be the scaling dimension of the dual operator $\mathcal{O}$.
Under Weyl rescalings we have, by definition
\begin{align}
    \phi &\mapsto \Omega^{\Delta-d}\phi \, , \\
    \mathcal{O} &\mapsto \Omega^{-\Delta} \mathcal{O} \, .
\end{align}

The Weyl generator of space-dependent rescalings on $\Sigma$ is 
\begin{align}
    \mathcal{W}:=2 \pi -(d-\Delta)\phi\, \pi_\phi
\end{align}
where $\pi^{ab}:=-i\frac{\delta}{\delta h_{ab}}$ and $\pi_\phi:=-i\frac{\delta}{\delta\phi}$. A uniform Weyl rescaling is equivalent to a change of the cutoff, which we argued around \eqref{eqn:cutoff} is given by the conformal time 
\begin{align}
    \int_\Sigma\mathcal{W} = i (-\eta)\frac{\partial}{\partial(-\eta)}\label{eqn:uniformWeyl}
\end{align}

In the new variables, \eqref{deformation} becomes
\begin{align}
    Z^{(\eta)}[h,\phi]:=e^{\text{CT}^{(\eta)}[h,\phi]}\left( \text{P} \exp \int_{\epsilon}^{-\eta}\frac{d\lambda}{\lambda}\hat{O}(\lambda)\right)Z_{\text{CFT}}^{(\epsilon)}[h,\phi]\,. \label{eqn:deformationnew}
\end{align}
We will often leave the $\eta$ dependence implicit. Defining
\begin{align}
    \Tilde{Z}[h,\phi]:=e^{-\text{CT}[h,\phi]}Z[h,\phi]\,,
\end{align}
from \eqref{eqn:uniformWeyl} and \eqref{eqn:deformationnew} we have that
\begin{align}
    \left(-\mathcal{W} + i \hat{X}(-\eta)\right) \Tilde{Z}[h,\phi]=0\label{eqn:RG}
\end{align}
where
\begin{align}
    \hat{O}(\lambda)=:\int_\Sigma d^dx\,\sqrt{h}\,\hat{X}(\lambda)\,.
\end{align}
Equation \eqref{eqn:RG} is the Hamiltonian constraint equation after having removed the counterterms. It looks like an RG-flow equation along the operator $\hat{X}$.

In terms of the re-scaled variables, the Hamiltonian constraint for pure Einstein gravity with a free minimally-coupled scalar field in $d+1$ dimensions reads
\begin{align}
    \mathcal{H}&= (-\eta)^d \frac{16\pi G_N}{\sqrt{h}}\left(\pi^{ab}\pi_{ab}-\frac{1}{d-1}\pi^2\right)-(-\eta)^{-d}\frac{\sqrt{h}}{16\pi G_N}\left((-\eta)^2\mathcal{R}-\frac{d(d-1)}{2}\right) \nonumber \\
    &+ \frac{1}{2}(-\eta)^{2\Delta-d}\frac{1}{\sqrt{h}}\pi_\phi^2+\frac{1}{2}(-\eta)^{d-2\Delta}\sqrt{h}\left((-\eta)^2(\nabla_h\phi)^2+m^2\phi^2\right)\,, \label{eqn:rescaledH}
\end{align}
where $m^2=\Delta(d-\Delta)$ and $\mathcal{R}$ is the Ricci scalar for the metric $h_{ab}$.

Finally, we define the following classification of terms appearing in \eqref{eqn:rescaledH}, based on the powers of the $(-\eta)$ cutoff they come with:
\begin{itemize}
    \item a term is called \emph{relevant} if it has a $(-\eta)^p$ coefficient with $\text{Re}(p)<0$;
    \item a term is called \emph{irrelevant} if it has a $(-\eta)^p$ coefficient with $\text{Re}(p)>0$;
    \item a term is called \emph{marginal} if it has no $(-\eta)$ dependence;
    \item a term is called \emph{lateral} if it has a $(-\eta)^p$ coefficient with $\text{Re}(p)=0$, but with $\text{Im}(p)\neq0$.
\end{itemize}

\subsubsection{Complex $\Delta$ Strips} \label{sec:Deltastrips}

As can be seen from equation \eqref{eqn:rescaledH}, the character of the different matter terms depends heavily on the dimension $\Delta$, or, equivalently, on the mass of the dual bulk field, via
\begin{align}
    m^2 = \Delta (d-\Delta)\,.
\end{align}
Thus, the prescription for constructing the deformation operator will vary as we change $\Delta$. In dS, unitary bulk scalar fields fall into two classes
\begin{itemize}
    \item \emph{complementary series}: $\Delta\in [0,d]$, and
    \item \emph{principal series}: $\Delta=\frac{d}{2}+i\nu$, $\nu\in\mathbb{R}$.
\end{itemize}
One may think that only the principal series requires a special treatment. But as it turns out, even the complementary series breaks up into different deformation regimes, that need to be treated separately. This seems in tension with the fact that from the bulk point of view one does not encounter such problems and can just compute the wavefunction coefficient for general (unitary) value of $\Delta$, as done in Section \ref{sec:FreeScalardS}. As we will show, these seemingly different regimes of the dual theory are related by analytic continuation. This will be explained in Section \ref{sec:FlowAnalyticContinuationPrincipalSeries} after we gain some intuition about the structure of the deformation operator in the sections in-between.

As an aside, we note that under the holographic paradigm, given a holographic CFT the bulk theory is uniquely determined. This means, in particular, its low-energy dynamical field content. If, for instance, a minimally-coupled scalar field is present in the bulk, the CFT knows about its mass. Then, the appropriate $T^2$-deformation is uniquely determined. So, we should bear in mind that when we discuss varying $\Delta$ of the dual scalar operator we are actually talking about changing both the bulk and (hypothetical) dual theories together. The point is that there is a (complex) one-parameter family of possible bulk/dual theories, parametrised by $\Delta\in\mathbb{C}$.\footnote{Since we are working bottom-up, there is no guarantee that all of these CFTs actually exist from a top-down perspective---this might only be the case for specific values of $\Delta$, and with specific other bulk matter fields also included.  But this does not prevent us from calculating the hypothetical formulae as if they all did exist.}  What we will argue is that the deformed partition function $Z(\Delta)$ is analytic in $\Delta$, except for singularities at isolated points along the real axis. These correspond precisely to values of $\Delta$ for which the seed CFT picks up extra anomaly terms. 

This means that, in practice, we can compute the flow of the two-point function in a given, convenient range of $\Delta$ and analytically continue the result to the whole complex-$\Delta$ plane. In Sections \ref{sec:flowsimple} and \ref{sec:flownextstrip} we show that we obtain exactly the same function at the end.

The simplest range of $\Delta$ is the one for which the required counterterm $\text{CT}$ is minimal, in a sense to be made clear in Section \ref{sec:Concrete}. This is given by
\begin{align}
    \frac{d}{2}<\text{Re}(\Delta)<\frac{d+2}{2}\,.
\end{align}
As it will turn out, it is simpler to go right than left, see Figure \ref{fig:strips}, because we get to keep the standard holographic form of the dictionary, in which the bulk field acts as a source for the dual theory, like in standard AdS/CFT. It is a well-known problem in dS/CFT how to map dual sources and operators to bulk fields for the whole range of unitary $\Delta$ (see \cite{Dey:2024zjx} for a recent proposal for the principal series). Here, we circumvent this issue by computing our observables in a regime in which the dictionary has a standard interpretation and then analytically continuing the result.

\section{Building up the Deformation Operator} \label{sec:Concrete}

Here we construct explicit examples of deformation operators and the necessary counterterms for theories of Einstein gravity coupled to scalar fields in various dimensions, before extracting the general features that will be relevant for the two-point functions of interest.

Following the algorithm of \cite{Araujo-Regado:2022gvw}, here is a summary of the steps required to obtain the operator:
\begin{enumerate}
    \item We first pick a counterterm implementing a canonical transformation such that the Hamiltonian constraint has the (dual) Weyl generator $\mathcal{W}$ as its marginal piece. We label these counterterms jointly by $\text{CT}_\mathcal{W}$ and they are local functionals of the induced bulk fields, or equivalently, of the dual sources. We will see their explicit form below. Under this canonical transformation the constraint gets modified
    \begin{align}
        \mathcal{H}\mapsto e^{-\text{CT}_\mathcal{W}}\,\mathcal{H}\,e^{\text{CT}_\mathcal{W}}=:\mathcal{H}_{\text{CT}_\mathcal{W}}\,.
    \end{align}
    $\text{CT}_\mathcal{W}$ is chosen such that the combination\footnote{The particular prefactor here is a convenient choice that makes the prescription that follows look simpler. We will explain this in a bit more detail later, but had we picked a different prefactor at this point we would need to add additional counterterms at the end of the three steps. Nevertheless, the dual partition function obtained at the end would be the same as the one obtained here. So we simply choose the cleanest route to the answer.}
    \begin{align}
        -\mathcal{W}+i\mathcal{A}
    \end{align}
    corresponds to the marginal term of the modified constraint. Here, $\mathcal{A}$ stands for any local function of the sources that is marginal. It will correspond to the trace anomaly in the CFT limit.
    \item We then pick further counterterms chosen judiciously to eliminate any terms in the Hamiltonian constraint that are relevant. This is an iterative process and how many steps are required (or whether it terminates at all) depends on the bulk theory in question. We will give some examples below and we will discuss this step extensively in the following subsection.
    \item The deforming operator is then the total irrelevant $+$ marginal operator in the (final) modified Hamiltonian constraint (up to a factor of $i$). 
\end{enumerate}

The first choice of counterterms is a universal one that must be carried out in any dimension and for any field content.  For pure Einstein gravity in $d+1$ dimensions we have
\begin{equation}
    \text{CT}_\mathcal{W}\supset i\,2\frac{d-1}{(16\pi G_N)}(-\eta)^{-d}\int_\Sigma\sqrt{h}  =i\, 2(d-1)\,\alpha\, (-\eta)^{-d}\,\int_\Sigma\sqrt{h} \, , \label{eqn:ct_grav}
\end{equation}
where we define the dimensionless parameter
\begin{equation}
    \alpha:=\frac{1}{16\pi G_N}\,.
\end{equation}

If we further have a massive minimally-coupled real scalar field, then a new counterterm is required to obtain the scalar part of the Weyl generator. The correct counterterm depends on the mass. For example, for $\text{Re}(\Delta)>\frac{d}{2}$ we additionally have
\begin{equation}
    \text{CT}_\mathcal{W}\supset i\,\frac{d-\Delta}{2}\,(-\eta)^{d-2\Delta}\int_\Sigma\sqrt{h} \, \phi^2 \, .
    \label{eqn:ct_scalar}
\end{equation}

Importantly, the counterterms in \eqref{eqn:ct_grav} and \eqref{eqn:ct_scalar} are purely imaginary, a feature of dS/CFT, unlike the case of AdS/CFT with real counterterms, see \cite{Araujo-Regado:2022gvw}.\footnote{Also note that as $\eta \to 0^-$ they diverge, which will correspond to rapid oscillations of the phase of the bulk wavefunction, signalling the expected classical regime at future asymptotic infinity.}

Under the joint counterterm $\text{CT}_\mathcal{W}$ obtained by combining \eqref{eqn:ct_grav} and \eqref{eqn:ct_scalar}, the Hamiltonian constraint in \eqref{eqn:rescaledH} gets transformed to
\begin{align}
    \mathcal{H}_{\text{CT}_\mathcal{W}}&= \mathcal{H} -\mathcal{W} - \alpha(-\eta)^{-d}\sqrt{h}\frac{d(d-1)}{2} - \frac{1}{2}(-\eta)^{d-2\Delta}\sqrt{h} \Delta(d-\Delta)\phi^2 \nonumber \\
    &-\frac{1}{\alpha}(-\eta)^{2d-2\Delta}\frac{d-\Delta}{2(d-1)} \phi^2\pi - \frac{1}{\alpha}(-\eta)^{3d-4\Delta}\frac{d(d-\Delta)^2}{16(d-1)}\sqrt{h}\phi^4 \, .
\end{align}
Both the cosmological constant and the mass terms cancel to yield
\begin{align}
\mathcal{H}_{\text{CT}_\mathcal{W}}&=-\mathcal{W}+\frac{1}{\alpha}(-\eta)^d \frac{1}{\sqrt{h}}\left(\pi^{ab}\pi_{ab}-\frac{1}{d-1}\pi^2\right)+ \frac{1}{2}(-\eta)^{2\Delta-d}\frac{1}{\sqrt{h}}\pi_\phi^2  \nonumber \\
    &-\frac{1}{\alpha}(-\eta)^{2d-2\Delta}\frac{d-\Delta}{2(d-1)} \phi^2\pi  \nonumber \\
    &+\frac{1}{2}(-\eta)^{d-2\Delta+2}\sqrt{h}(\nabla_h\phi)^2-\alpha(-\eta)^{-d+2}\sqrt{h}\mathcal{R} - \frac{1}{\alpha}(-\eta)^{3d-4\Delta}\frac{d(d-\Delta)^2}{16(d-1)}\sqrt{h}\phi^4 \, .
\end{align}

To proceed, we need to separate relevant/marginal/irrelevant terms. For that we simply look at the power of $-\eta$ in front of each term. Relevant terms have a negative power, marginal terms have a zero power, while irrelevant terms have a positive power. For any relevant term we add a new counterterm that eliminates it. This will build up the combined counterterm required in the second step mentioned above. 

\subsection{Operator Scalings}

We will be conducting our analysis for the complex strip 
\begin{equation}
    \frac{d}{2}<\text{Re}(\Delta)<d \, .
\end{equation}
The lower bound is chosen so that we have the cleanest structure of counterterms. The upper bound is there for unitarity of the bulk theory.

The $(-\eta)$ scaling in front of each operator and counterterm is precisely given by its Weyl scaling dimension. For example, the following pair may appear
\begin{align}
    (-\eta)^{(n-1)d-n\Delta}\, \sqrt{h}\phi^n
\end{align}
Such a term is irrelevant for
\begin{equation}
    \text{Re}(\Delta)<\frac{n-1}{n}d \, .
\end{equation}
Terms with higher powers of $\phi$ are irrelevant until ``later'' as we increase $\Delta$ in the real direction. We may also have terms of the form
\begin{equation}
    (-\eta)^{n(d-\Delta)}\,\phi^n\pi \quad \text{and}\quad (-\eta)^{(n-1)(d-\Delta)}\,\phi^n\pi_\phi \, ,
\end{equation}
both of which are irrelevant for
\begin{equation}
    \text{Re}(\Delta)<d \, .
\end{equation}

\subsection{Terms Generated by New Counterterms} \label{sec:counterterms}

We can re-write the Hamiltonian constraint in a convenient form
\begin{align} \label{eqn:HCTW}
    \mathcal{H}_{\text{CT}_\mathcal{W}}&=-\mathcal{W} \nonumber  \\
    &+ \frac{1}{\alpha}(-\eta)^d\frac{1}{\sqrt{h}}\mathcal{G}_{abcd}\Tilde{\pi}^{ab}\Tilde{\pi}^{cd}
    + \frac{1}{2}(-\eta)^{2\Delta-d}\frac{1}{\sqrt{h}}\Tilde{\pi}_\phi^2 \nonumber \\
    &+\frac{1}{2}(-\eta)^{d-2\Delta+2}\sqrt{h}(\nabla_h\phi)^2-\alpha(-\eta)^{-d+2}\sqrt{h}\mathcal{R} \, ,
\end{align}
where 
\begin{align}
    \mathcal{G}_{abcd}:=\frac{h_{ac}h_{bd}+h_{ad}h_{bc}}{2}-\frac{h_{ab}h_{cd}}{d-1}
\end{align}
is the de-Witt metric and
\begin{align}
    \Tilde{\pi}^{ab}:=\pi^{ab}+ (-\eta)^{d-2\Delta}\frac{d-\Delta}{4}\sqrt{h}\,\phi^2h^{ab}\,,
\end{align}
while $\Tilde{\pi}_\phi=\pi_\phi$ at this stage. The tilde denotes that conjugate momenta have been shifted from their original definition.

We now need to study the presence of any relevant terms (those with coefficient $(-\eta)^q,\,\text{Re}(q)<0$). If such a term exists, we need to add a new counterterm to eliminate it. This will be a recursive process, which for the simplest range we are considering will terminate. Let us imagine we are at the $n$-th step of this recursion, meaning we are about to add the $n$-th counterterm on top of $\text{CT}_\mathcal{W}$.

Consider a general (local) function of both metric and matter: $\mathcal{F}_n(h,\phi)$, which could well involve derivatives. Using the property that
\begin{align}
    \left[\mathcal{W}(x),\int d^dy\,\sqrt{h}\,\mathcal{F}_n(h,\phi)\right] =i(\Delta_{\mathcal{F}_n}-d)\sqrt{h}\,\mathcal{F}_n(h,\phi)(x)\,,
\end{align}
where $\Delta_{\mathcal{F}_n}$ is the Weyl-scaling dimension of $\mathcal{F}_n$, we can eliminate the would-be relevant term $(-\eta)^{\Delta_{\mathcal{F}_n}-d}\sqrt{h}\,\mathcal{F}_n(h,\phi)(x)$ by adding the counterterm
\begin{align}
    \text{CT}_{\mathcal{F}_n}=-i \frac{1}{\Delta_{\mathcal{F}_n}-d}(-\eta)^{\Delta_{\mathcal{F}_n}-d}\int d^dy\,\sqrt{h}\,\mathcal{F}_n(h,\phi)\,.
\end{align}

This will generate a shift in the dual operators as follows\footnote{If the function $\mathcal{F}(h,\phi)$ includes derivatives, then boundary terms will be generated by the functional derivatives $\frac{\delta}{\delta h_{ab}(x)}$ and $\frac{\delta}{\delta \phi(x)}$. However, these do not contribute as in the Hamiltonian constraint we are always evaluating our fields and conjugate momenta at a point $x$ in the bulk. So the result of the shift is again a local function of the sources.}
\begin{align}
    \Tilde{\pi}^{ab}(x)&\mapsto \Tilde{\pi}^{ab}(x)-i\frac{\delta\text{CT}_{\mathcal{F}_n}}{\delta h_{ab}(x)}=:\Tilde{\pi}^{ab}(x) + (-\eta)^{\Delta_{\mathcal{F}_n}-d}\sqrt{h}\,\mathbf{H}^{ab}_{\mathcal{F}_n}(x) \\
    \Tilde{\pi}_\phi(x)&\mapsto\Tilde{\pi}_\phi(x) -i\frac{\delta \text{CT}_{\mathcal{F}_n}}{\delta\phi(x)}=:\Tilde{\pi}_\phi(x) + (-\eta)^{\Delta_{\mathcal{F}_n}-d}\sqrt{h}\,\mathbf{\Phi}_{\mathcal{F}_n}(x)\,.
\end{align}
Here, the explicit functional form of $\mathbf{H}^{ab}_{\mathcal{F}_n}(h,\phi)$ and $\mathbf{\Phi}_{\mathcal{F}_n}(h,\phi)$ depends on the details of the function $\mathcal{F}_n$. So we can write that, after all counterterms up to the $n$-th one, the conjugate momenta take the form
\begin{align}
    \Tilde{\pi}^{ab}_{(n)} &= \pi^{ab} + \underbrace{(-\eta)^{2(d-\Delta)-d}\frac{d-\Delta}{4}\sqrt{h}\,\phi^2h^{ab}+ \sum_{m=1}^n(-\eta)^{\Delta_{\mathcal{F}_m}-d}\sqrt{h}\,\mathbf{H}^{ab}_{\mathcal{F}_m}}_{=:\sqrt{h}\mathbf{H}^{ab}_n} \label{eqn:tildepi} \\
    \Tilde{\pi}_{\phi(n)}&=\pi_{\phi} + \underbrace{\sum_{m=1}^n(-\eta)^{\Delta_{\mathcal{F}_m}-d}\sqrt{h}\,\mathbf{\Phi}_{\mathcal{F}_m}}_{=:\sqrt{h}\mathbf{\Phi}_n}\,,\label{eqn:tildepiphi}
\end{align}
where we have included the shifts from all the previous steps. Notice that the middle term in \eqref{eqn:tildepi} is the shift induced by the counterterm $\mathcal{F}=\phi^2$ in $\text{CT}_\mathcal{W}$, which has $\Delta_\mathcal{F}=2(d-\Delta)$. 

Expanding the terms $\frac{1}{\alpha}(-\eta)^d\frac{1}{\sqrt{h}}\mathcal{G}_{abcd}\Tilde{\pi}^{ab}_{(n)}\Tilde{\pi}^{cd}_{(n)}$ and $\frac{1}{2}(-\eta)^{2\Delta-d}\frac{1}{\sqrt{h}}\Tilde{\pi}_{\phi(n)}^2$ at the $n$-th step we get two categories of terms:
\begin{itemize}
    \item Terms linear in the conjugate momenta:
    \begin{align}
        &\frac{2}{\alpha}(-\eta)^d\mathcal{G}_{abcd}\,\mathbf{H}^{ab}_n\,\pi^{cd}
        \label{eqn:linearpi} \\
        \text{and } \quad &(-\eta)^{2\Delta-d}\,\mathbf{\Phi}_n\,\pi_\phi \, .
        \label{eqn:linearpiphi}
    \end{align}
    \item Terms zeroth-order in the conjugate momenta, i.e. purely source terms:
    \begin{align}
        &\frac{1}{\alpha}(-\eta)^d\sqrt{h}\,\mathcal{G}_{abcd}\,\mathbf{H}^{ab}_n\,\mathbf{H}^{cd}_n\label{eqn:sourcepi} \\
        \text{and } \quad &\frac{1}{2}(-\eta)^{2\Delta-d}\sqrt{h}\,\mathbf{\Phi}_n^2 \, .
        \label{eqn:sourcepiphi}
    \end{align}
\end{itemize}

At this point, we highlight two important lessons. 

Firstly, by inspecting \eqref{eqn:HCTW} we see that, in the strip $\frac{d}{2}<\text{Re}(\Delta)<d$, the terms that can potentially become relevant at the first step are
\begin{equation}
    (-\eta)^{-d + (2(d-\Delta)+2)}\sqrt{h}(\nabla_h \phi)^2\,,\quad (-\eta)^{-d+2}\sqrt{h}\mathcal{R}\,,\quad (-\eta)^{-d+(4(d-\Delta))}\sqrt{h}\phi^4 \, .
\end{equation}
So we have that $\text{Re}(\Delta_{\mathcal{F}_1})>0$ and, for the counterterms for which $\mathbf{\Phi}_{\mathcal{F}_1}$ is non-trivial, we have that $\text{Re}(\Delta_{\mathcal{F}_1})>2(d-\text{Re}(\Delta))$. This means that the terms linear in the operators that would be generated by trying to eliminate any such $\mathcal{F}_1$ (see eqns \eqref{eqn:linearpi} and \eqref{eqn:linearpiphi}) would be guaranteed to be irrelevant, as can be seen from the $(-\eta)$ scalings in equations \eqref{eqn:tildepi} and \eqref{eqn:tildepiphi}. Thus, we can focus entirely on new potentially-relevant terms generated by equations \eqref{eqn:sourcepi} and \eqref{eqn:sourcepiphi}.

Secondly, because of the ``initial'' conditions for the range of $\Delta_{\mathcal{F}_1}$, we have that the source terms generated by $\mathcal{F}_1$ can only be \emph{more} irrelevant. This property holds at every step of the recursion. This means that given a value for $\Delta$, the process of eliminating all the relevant terms terminates. This would not be the case had we started outside the strip $\frac{d}{2}<\text{Re}(\Delta)<d$.

\subsection{Pure Gravity} \label{sec:grav}
To illustrate the procedure we will consider some concrete examples, starting from pure gravity and adding matter later. The starting point is always equation \eqref{eqn:HCTW}.

Beginning with pure gravity in $2+1$-dimensions, we see that there are no more relevant terms in $\mathcal{H}_{\text{CT}_\mathcal{W}}$ to be eliminated and so we can directly read off the deforming operator\footnote{Note that our definition for the stress-tensor is $T^{ab}=\frac{2}{\sqrt{h}}\frac{\delta}{\delta h_{ab}}$, which differs from the standard convention used in $d=2$ CFT. We make this choice here in order to have a single definition for all dimensions.}
\begin{align}
    \hat{O}_{\text{pure grav}}^{d=2}(-\eta) &= \int_\Sigma d^2x \left(-i \frac{1}{\alpha}(-\eta)^2 \frac{1}{\sqrt{h}}\mathcal{G}_{abcd}\pi^{ab}\pi^{cd} + i \alpha \sqrt{h}\mathcal{R}\right) \\
    &= i \int_\Sigma d^2x \sqrt{h} \left(\frac{(-\eta)^2}{\alpha} :\frac{1}{4}\mathcal{G}_{abcd}T^{ab}T^{cd}:(x) + \alpha\mathcal{R}(x)\right)\,. \label{eq:Opuregravd=2}
\end{align}
This is the dS/CFT equivalent of the $T\Bar{T}$ flow, also written in \cite{Araujo-Regado:2022jpj}. We notice that the operator is \emph{purely imaginary}, unlike the case of deforming the CFT dual to $\text{AdS}_3$. Further note that the curvature term has no $\eta$-scaling, signalling the fact that it is marginal. This means that the integral $\text{Pexp}\left(\int_0^{-\eta}\frac{d\lambda}{\lambda}\hat{O}(\lambda)\right)$ has a log-divergence and so we must introduce a lower cut-off. This is, however, consistent with the fact that, in $d=2$, the starting point of the flow \emph{already} necessitates a UV cut-off due to the presence of a conformal anomaly. Thus, the deformed theory reads
\begin{align}
    \Tilde{Z}(-\eta) =\text{Pexp}\left(\int_{\epsilon}^{-\eta}\frac{d\lambda}{\lambda}\hat{O}(\lambda)\right) Z_{\text{CFT}_2}(\epsilon)\,,
\end{align}
where $\epsilon>0$ is a UV cut-off length scale for the CFT. The final result is independent of $\epsilon$.

Moving on to pure gravity in $3+1$-dimensions, we need to add a new counterterm to eliminate the now relevant term $\sim (-\eta)^{2-d}\sqrt{h}\mathcal{R}$. Following the recipe above, the required counterterm (which we write for general $d>2$) is given by
\begin{align}
    \text{CT}_{\mathcal{R}}=-i\, \alpha \frac{1}{d-2} (-\eta)^{2-d} \int_\Sigma d^2x\,\sqrt{h}\mathcal{R}\,. \label{eqn:CTR}
\end{align}
This counterterm will be needed for all $d>2$. This shifts the conjugate momentum to
\begin{align}
    \Tilde{\pi}^{ab} = \pi^{ab} + \alpha \frac{1}{d-2} (-\eta)^{2-d}\sqrt{h}\, G^{ab}\,,
\end{align}
where $G^{ab}=\mathcal{R}^{ab}-\frac{1}{2}\mathcal{R}h^{ab}$ is the Einstein tensor for the metric $h_{ab}$. The new Hamiltonian operator, for pure gravity, is then
\begin{align}
    \mathcal{H}_{\text{CT}_\mathcal{W}+\text{CT}_\mathcal{R}} &= -\mathcal{W} + \frac{1}{\alpha} (-\eta)^d \frac{1}{\sqrt{h}}\mathcal{G}_{abcd}\Tilde{\pi}^{ab}\Tilde{\pi}^{cd}\\
    &= -\mathcal{W} + \frac{1}{\alpha} (-\eta)^d \frac{1}{\sqrt{h}}\mathcal{G}_{abcd}\pi^{ab}\pi^{cd}\\ 
    &+\frac{2}{d-2} (-\eta)^{2} \mathcal{G}_{abcd}G^{ab}\pi^{cd} + \alpha \frac{1}{(d-2)^2} (-\eta)^{4-d} \sqrt{h}\, \mathcal{G}_{abcd}G^{ab}G^{cd}\,. \label{eq:HCTWR}
\end{align}
We see that only the last term has a chance of becoming relevant as we change $d$. In particular, it will become marginal at $d=4$, in which case it matches with the conformal anomaly for $d=4$, similarly to how in $d=2$ the marginal term above was given by the $d=2$ conformal anomaly. But for $d=3$ all the terms are irrelevant and no new counterterm is needed. The deformation operator then reads
\begin{align}
    \hat{O}^{d=3}_\text{pure grav}(-\eta) &= -i\int_\Sigma d^3x \frac{1}{\alpha} (-\eta)^3 \frac{1}{\sqrt{h}}\mathcal{G}_{abcd}\Tilde{\pi}^{ab}\Tilde{\pi}^{cd}\\
    &=i \int_\Sigma d^3x \sqrt{h} \Bigg(\frac{(-\eta)^3}{\alpha} :\frac{1}{4}\mathcal{G}_{abcd}T^{ab}T^{cd}:(x)\\
    &+i (-\eta)^2\mathcal{G}_{abcd}G^{ab}T^{cd}(x)-\alpha  (-\eta)\, \mathcal{G}_{abcd}G^{ab}G^{cd}(x)\Bigg)\,. \label{eq:Opuregravd=3}
\end{align}
As expected, there are no log-divergences because there are no marginal terms. The deformed partition function is defined as
\begin{align}
    \Tilde{Z}(-\eta) =\text{Pexp}\left(\int_{0}^{-\eta}\frac{d\lambda}{\lambda}\hat{O}(\lambda)\right) Z_{\text{CFT}_3}\,,
\end{align}
without the need to introduce an extra cut-off. 

From here we can proceed in increasing $d$. The case of $d=4$ is straightforward since no new counterterms are required. All that happens is that the source term in \eqref{eq:HCTWR} becomes marginal, signalling the appearance of an anomaly in the CFT limit, which means an $\epsilon$ cut-off is again needed. But the form of $\hat{O}^{d=4}_\text{pure grav}$ is identical to $\hat{O}^{d=3}_\text{pure grav}$, only with the appropriate $\eta$ scalings. Going to $d>4$ requires a new counterterm. We leave it as an exercise for the reader to follow the recipe.

It is interesting to note the \emph{complex sign} of the conformal anomalies obtained via our prescription. As explained in Section \ref{sec:largevol}, the $\eta \to \epsilon$ limit of the deformation recovers the CFT anomaly equation
\begin{align}
    (-\mathcal{W} + i\mathcal{A})Z_\text{CFT}=0\,,
\end{align}
where $\mathcal{A}$ simply corresponds to the marginal terms in the deformation operator. In our conventions, this equation is equivalent to
\begin{align}
    \expval{T}_\text{CFT} = - \frac{1}{\sqrt{h}}\mathcal{A}\,.
\end{align}
We thus have that
\begin{align}
    \expval{T}_{\text{CFT}_2} &= -i\,\alpha\, \mathcal{R}\\ \label{eqn:anomalyd=2}
    \expval{T}_{\text{CFT}_4} &= i\, \alpha\,\frac{1}{4}\mathcal{G}_{abcd}G^{ab}G^{cd}\,.
\end{align}

\begin{figure}[H]
    \centering
    \includegraphics[width=0.7\textwidth]{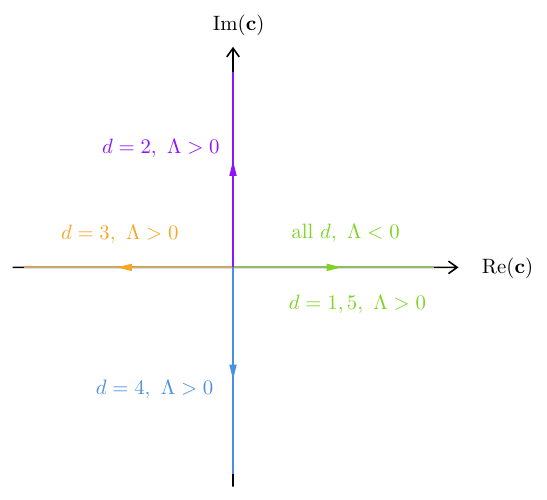}
    \caption{Phase of the tree-level contribution (i.e. leading in $1/N$) to the central charge $\mathbf{c}$ in the complex plane for different dimensions and signs of the cosmological constant. For $\Lambda>0$ the central charge $\mathbf{c}_{\text{dS}}$ rotates counterclockwise by $\pi/2$ with each increment in dimension.}
    \label{fig:central_charge}
\end{figure}

This is part of a general pattern of a putative dS/CFT duality, in which the ``number of degrees of freedom'', loosely defined to be given by the central charge $\mathbf{c}$ (i.e. the coefficient of the stress-tensor two-point function), is complex and rotates by $\pi/2$ (at leading order in $1/N$) with every increment in dimension. This is studied in detail in \cite{Goodhew:2024eup} and  shown in Figure \ref{fig:central_charge}.

\subsection{Adding Matter}\label{sec:matter}

We now add matter to our bulk theory and see how the deformation gets modified. So we go back to the starting equation \eqref{eqn:HCTW} with all the matter terms included. We begin by considering the strip $\frac{d}{2}<\text{Re}(\Delta)<d $ and how this breaks up into different deformation regimes. Then, we will study the analytic continuation to and past the principal series axis at $\text{Re}(\Delta)=\frac{d}{2}$.

Imagine we have added all the necessary pure gravity counterterms for dimension $d$ (so, for example, for $d=3$ we are at the stage of $\mathcal{H}_{\text{CT}_\mathcal{W}+\text{CT}_\mathcal{R}}$ of \eqref{eq:HCTWR}). We need to keep track of two things: 1. any matter--gravity interaction terms generated by the pure gravity counterterms; 2. given $\Delta$ in the strip $\frac{d}{2}<\text{Re}(\Delta)<d$, if there are any matter terms that are relevant, in which case matter counterterms are necessary.

\begin{enumerate}
    \item For the first issue, such terms are indeed generated by equation \eqref{eqn:sourcepi} because of the term $\sim \phi^2$ in the definition of $\mathbf{H}^{ab}_n$ in \eqref{eqn:linearpi}. As an illustration, consider the case of $d>2$ in which case we have (at least) the pure gravity counterterm $\text{CT}_\mathcal{R}$ in \eqref{eqn:CTR}. This induces a coupling term of the form
    \begin{equation}
        \frac{d-\Delta}{4(d-1)}(-\eta)^{2(d-\Delta)+2-d}\,\sqrt{h}\,\phi^2\mathcal{R} \, .
    \end{equation}
    We note that this is precisely a $d$-dimensional conformal coupling of $\phi$ to the metric $h_{ab}$. If the dimension is $d>4$, then the next pure gravity counterterm will ``kick in'' and a new coupling (still quadratic in the matter) will be induced
    \begin{equation}
        \frac{1}{(-d+2)^2}\frac{d-\Delta}{4(d-1)} (-\eta)^{2(d-\Delta)+4-d}\,\sqrt{h}\,\phi^2\,\mathcal{G}_{abcd}G^{ab}G^{cd} \, . 
    \end{equation}

    In general, at dimension $d$, if we restrict attention to terms \emph{quadratic} in $\phi$ (or if we restrict to a range of $\Delta$ for which no matter counterterms are necessary, see below), the matter--gravity coupling terms take the schematic form
    \begin{equation}
        (-\eta)^{2(d-\Delta)-d}\,\phi^2\,\sum_{d_e<d} (-\eta)^{d_e}\mathcal{A}_{d_e} \, ,
    \end{equation}
    where $d_e$ stands for even dimensions and $\mathcal{A}_{d_e}$ is the dual holographic anomaly at dimension $d_e$ for a pure gravity bulk theory. For instance, $\mathcal{A}_2\sim \sqrt{h}\,\mathcal{R}$, $\mathcal{A}_4\sim \sqrt{h}\,\mathcal{G}_{abcd}\,G^{ab}G^{cd}$, \emph{without} the pre-factor scaling like the number of degrees of freedom (aka the central charges).
    \item Having done the previous step, we now look at the matter terms that can become relevant as we vary $\Delta$. These are
    \begin{equation}
        (-\eta)^{2(d-\Delta)+2-d}\sqrt{h}(\nabla_h\phi)^2\,,\quad (-\eta)^{2(d-\Delta)+d_e-d}\phi^2\mathcal{A}_{d_e}\,,\quad (-\eta)^{4(d-\Delta)-d}\sqrt{h}\,\phi^4 \, ,
    \end{equation}
    where the latter comes from \eqref{eqn:sourcepi} due to the presence of the $\sim\phi^2$ term in \eqref{eqn:tildepi}, while the middle term runs over all even $d_e<d$. We first note that the middle term can only be \emph{more} irrelevant than the first. Their $(-\eta)$ scalings only match for $d_e=2$. The lowest upper bound on $\Delta$ that keeps the first two terms irrelevant is then
    \begin{equation}
        \text{Re}(\Delta)<\frac{d+2}{2}\,.
    \end{equation}
    The last term is irrelevant if $\text{Re}(\Delta)<\frac{3d}{4}$.

Following the argument in Section \ref{sec:counterterms}, when any of the above become relevant, the new required counterterm will generate new matter--gravity and matter--matter couplings from equations \eqref{eqn:sourcepi} and \eqref{eqn:sourcepiphi}. These will always be more irrelevant than the term eliminated, and so for any given $\Delta$ in our strip, the recursion terminates. 
    
\end{enumerate}

\subsubsection{The Simplest Strip: $\frac{d}{2}<\text{Re}(\Delta)<\frac{d+2}{2}$} \label{sec:simplestrip}

As motivated in Section \ref{sec:Deltastrips}, we will first take the cleanest and simplest route and just consider the complex strip of $\Delta$ for which no new matter counterterms are needed. As will become evident in the next section, for the purposes of computing the two-point functions of the theory, it suffices to compute the deforming operator up to $\mathcal{O}(\phi^4)$ and $\mathcal{O}(\phi^3\pi_\phi)$ terms. This means that the result we will compute will be valid in the strip
    \begin{equation}
        \frac{d}{2}<\text{Re}(\Delta)<\frac{d+2}{2}\,,
    \end{equation}
    which corresponds to the strip in which all the terms \emph{quadratic} in $\phi$ are irrelevant.
    
    The matter part of the final Hamiltonian constraint written up to $\mathcal{O}(\phi^4, \phi^3\pi_\phi)$ reads
    \begin{align}
        \mathcal{H}_\text{matter}&=\frac{1}{2}(-\eta)^{2\Delta-d}\frac{1}{\sqrt{h}}\pi_\phi^2
    +\frac{1}{2}(-\eta)^{d-2\Delta+2}\sqrt{h}(\nabla_h\phi)^2 \nonumber \\ 
    &+ (-\eta)^{2(d-\Delta)-d}\phi^2\sum_{d_e<d} (-\eta)^{d_e}\mathcal{A}_{d_e} - \frac{1}{\alpha}\frac{d-\Delta}{2(d-1)}(-\eta)^{2(d-\Delta)}\,\phi^2\,\pi\,.
    \end{align}
The corresponding matter part of the deforming operator is thus
\begin{align}
    \hat{O}_\text{matter}^{\left(\frac{d}{2},\frac{d+2}{2}\right)} (-\eta) &= i\int_\Sigma d^dx \sqrt{h} \Bigg( \frac{1}{2}(-\eta)^{2\Delta-d}:\mathcal{O}^2:-\frac{1}{2}(-\eta)^{d-2\Delta+2}(\nabla_h\phi)^2\\
    &-(-\eta)^{2(d-\Delta)-d}\phi^2\sum_{d_e<d} (-\eta)^{d_e}\mathcal{A}_{d_e}-i\frac{1}{\alpha}\frac{d-\Delta}{4(d-1)}(-\eta)^{2(d-\Delta)}\,\phi^2\,T\Bigg)\\
    &+ O(\phi^4,\phi^3\mathcal{O})\,. \label{eq:Omattersimple}
\end{align}
This is valid for all dimensions. Notice that the coupling to the stress-tensor is suppressed by $1/\alpha$.

\subsubsection{Moving Past $\text{Re}(\Delta)=\frac{d+2}{2}$}

At the line $\text{Re}(\Delta)=\frac{d+2}{2}$ the terms
\begin{equation}
    (-\eta)^{d-2\Delta+2}\sqrt{h}\left(\frac{1}{2}(\nabla_h\phi)^2+\frac{d-\Delta}{4(d-1)}\,\phi^2\mathcal{R}\right)
\end{equation}
become \emph{lateral}, as defined in Section \ref{sec:new aspects}. If we move to the right of this line, they become relevant and we will need to add the new counterterm
\begin{equation}
    \text{CT}_{\frac{d+2}{2}} = -i\, \frac{1}{d-2\Delta+2}(-\eta)^{d-2\Delta+2}\int_\Sigma d^dx\,\sqrt{h}\left(\frac{1}{2}(\nabla_h\phi)^2+ \frac{d-\Delta}{4(d-1)}\,\phi^2\mathcal{R}\right) \, . \label{eq:CT1stbarrier}
\end{equation}

\begin{figure}[ht]
    \centering
    \includegraphics[width=\textwidth]{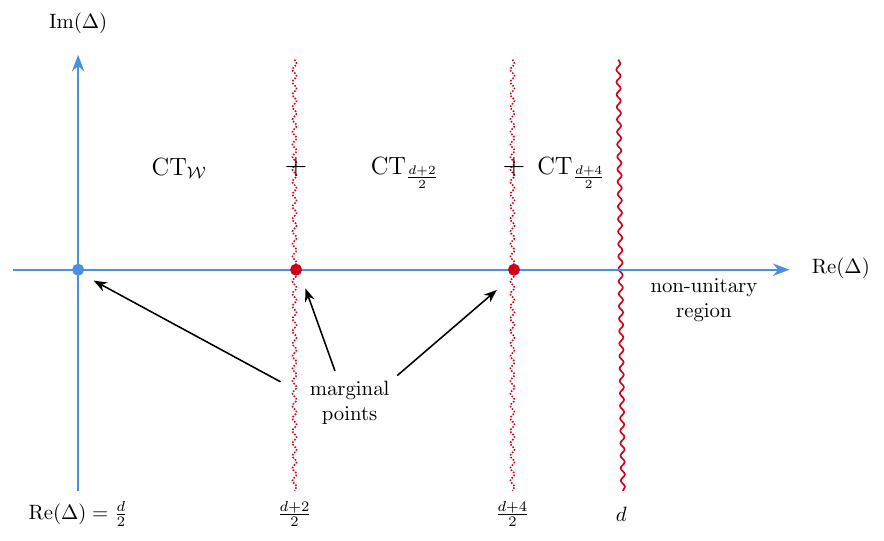}
    \caption{Strips in the complex-$\Delta$ plane. The blue lines indicate the principal and complementary series. The thick curvy red line at $\text{Re}(\Delta)=d$ indicates the unitarity bound. The thick dots indicate points at which terms in the deformation become marginal. At these points there will be anomaly contributions in the CFT limit. The dashed curvy red lines indicate the presence of lateral terms in the deformation. As we move across lateral lines from left to right, new counterterms are needed to remove terms that transition into being relevant. For even $d$ the unitarity bound coincides with a lateral line. For odd $d$ it sits halfway between two lateral lines. Crucially, the picture shows the situation when we work to order $O(\phi^4,\phi^3\mathcal{O})$, as appropriate for computing two-point functions. If we care about higher-order terms, the structure gets increasingly refined.}
    \label{fig:strips}
\end{figure}

It is important to note that whether or not we should cross this ``barrier'' depends on the dimension $d$, since we are always limiting ourselves to be inside the strip $\frac{d}{2}<\text{Re}(\Delta)<d$ for bulk unitarity. For example, for $d=2$, this line coincides precisely with the bulk unitarity bound. For $d>2$ there will always be a unitary region to the right of the line. This pattern is shown in Figure \ref{fig:strips}.\footnote{In any case, we can always imagine being interested in what is beyond, even if the bulk is not unitary, and classify the various strips in the complex-$\Delta$ plane. This turns out to \emph{immediately} introduce new complications, as seen right below.}

The counterterm in \eqref{eq:CT1stbarrier} will induce a not-so-simple shift of the metric conjugate momentum, of the schematic form
\begin{equation}
    \Tilde{\pi}^{ab}\mapsto \Tilde{\pi}^{ab} + (-\eta)^{d-2\Delta+2}\sqrt{h}\left(\phi^2 G^{ab} + \nabla_h^a\phi\nabla_h^b\phi + \phi\nabla_h^a\nabla_h^b\phi + (\nabla_h\phi)^2g^{ab} + \phi\Box_h \phi\, g^{ab}\right)\,. \label{eq:pishift2ndstrip}
\end{equation}
However, the shift of the momentum conjugate to the scalar is simpler and, as it will turn out, it will be all we need to compute the flow of the scalar two-point function. The shift is given by
\begin{equation}
    \Tilde{\pi}_\phi \mapsto \Tilde{\pi}_\phi + \frac{1}{d-2\Delta+2}(-\eta)^{d-2\Delta+2}\sqrt{h}\left(\Box_h-\frac{d-\Delta}{2(d-1)}\mathcal{R}\right)\phi\,.
\end{equation}

Thus, the matter part of the deformation reads
\begin{align}
    \hat{O}_\text{matter}^{\left(\frac{d+2}{2},\frac{d+4}{2}\right)}(-\eta)&=i\int_\Sigma d^dx \sqrt{h} \Bigg( \frac{1}{2}(-\eta)^{2\Delta-d}:\mathcal{O}^2: \\ 
    &+i \frac{1}{d-2\Delta+2}(-\eta)^2\left[\left(\Box_h-\frac{d-\Delta}{2(d-1)}\mathcal{R}\right)\phi\right]\,\mathcal{O}\\
    &-\frac{1}{2(d-2\Delta+2)^2}(-\eta)^{d-2\Delta+4}\left[\left(\Box_h-\frac{d-\Delta}{2(d-1)}\mathcal{R}\right)\phi\right]^2\\
    &+  (-\eta)^{2(d-\Delta)-d}\phi^2\sum_{2<d_e<d} (-\eta)^{d_e}\mathcal{A}_{d_e}\\
    &+ \frac{1}{\alpha}(-\eta)^{2(d-\Delta)}F\left[\phi^2 \,T, (-\eta)^2 \phi^2 \mathcal{R}\, T, (-\eta)^2(\nabla_h^2\phi^2)\, T  \right]\Bigg)\\
    &+O(\phi^4, \phi^3\mathcal{O})\,.\label{eq:O2ndstrip}
\end{align}
The operator is again written up to quadratic order in $\phi$. The schematic terms in the penultimate line are those generated by the shift in \eqref{eq:pishift2ndstrip}. They come at sub-leading order in $1/\alpha$. They are all irrelevant in the whole range $\frac{d}{2}<\text{Re}(\Delta)<d$, so they do not ``induce" any more counterterms.\footnote{We see that one \emph{major} consequence of moving beyond the bulk unitarity bound is that these mixed terms involving the stress-tensor would themselves become relevant. This would necessitate the introduction of counterterms that are functionals of not only the sources but also the dual operators. This would make the interpretation of such ``counterterms'' unclear at best. Perhaps this is a sign for us to stay away from such non-unitary quests.} This will always be the case for the terms involving couplings of the matter source to the stress-tensor. We see from the third line that the next ``barrier'' happens at $\text{Re}(\Delta) = \frac{d+4}{2}$. Hence, the deformation operator written above is valid in the strip
\begin{equation}
    \frac{d+2}{2}<\text{Re}(\Delta)<\frac{d+4}{2} \, ,
\end{equation}
as indicated by its superscript, provided the dimension $d$ is high enough such that $\frac{d+4}{2}<d$. Otherwise, it is simply valid \emph{up to} the line $\text{Re}(\Delta)=d$. See Figure \ref{fig:strips} for clarification.

\subsubsection{Analytic Continuation to Generic $\Delta$ and the Principal Series}\label{sec:FlowAnalyticContinuationPrincipalSeries}
Although formal convergence requires picking a specific strip, as we shall see it is actually possible to analytically continue our results to generic complex values of $\Delta$.  This can be seen from the fact that, when we e.g.~derive our explicit formula for the scalar two-point function in \eqref{eqn:dSCSH2ptCorrelatorFlowEquationSolutionCT} (for the strip $\text{Re}(\Delta)\in \left(\frac{d}{2},\frac{d+2}{2}\right)$) and \eqref{eqn:solnnextstrip} (for the strip $\text{Re}(\Delta)\in \left(\frac{d+2}{2},\frac{d+4}{2}\right)$) in Section \ref{sec:BoundaryFlow}, this agrees with our previous calculations from the pure cosmology side in \eqref{eqn:dSscalar2ptEtaFactors} (valid for all $\Delta$) in Section \ref{sec:FreeScalardS}.  
 
The interpretation of this analytic continuation can be seen by comparison to Fig.~\ref{fig:strips}.  For example, let us consider the case of continuing from the $\text{Re}(\Delta)\in\left(\frac{d}{2},\frac{d+2}{2}\right)$ window to the one on the right: $\text{Re}(\Delta)\in\left(\frac{d+2}{2},\frac{d+4}{2} \right)$.  In this case, convergence requires the introduction of a new counterterm $\text{CT}_{\frac{d+2}{2}}$, associated with $(\partial \phi)^2$.  Without this counterterm, there is a relevant term in the deformation, which means that inside of \eqref{eqn:deformationnew}, the integral
\begin{equation}\label{eqn:deform_integral}
\int_{\epsilon}^{(-\eta)}\frac{d\lambda}{\lambda}\hat{O}(\lambda)
\end{equation}
will have a power law divergence as $\epsilon \to 0$.  We can just throw out this power-law divergence by hand, 
since typically in renormalisation theory, power-law divergences are pure scheme.  This is equivalent to analytically continuing from a window where the corresponding term is irrelevant.  The anomaly matching conditions are just the analytic continuation of those in the convergent window.\footnote{We note however that in our experience it is very easy to get confused when deriving the anomaly matching conditions when throwing out the power law divergence, without reference to a convergent region.  So it is best to do the calculation properly in at least one region, and analytically continue from there.}

Suppose now that we are right on the boundary between the two regions $\text{Re}(\Delta)=\frac{d+2}{2}$, which is shown as a curvy red line in Fig.~\ref{fig:strips}, with a nonzero imaginary part of $\Delta$. In this case, the integral \eqref{eqn:deform_integral} will have an oscillatory character:
\begin{equation}
\int_{\epsilon}^{(-\eta)}\frac{d\lambda}{\lambda} e^{ia \log \lambda},\qquad a \in \mathbb{R}_{\ne 0},
\end{equation}
where the value of $a$ depends on the location along the line.  This corresponds to a lateral operator in our classification from the Introduction.  Now this integral is right on the edge of being convergent in the limit as $\epsilon \to 0$, and with an arbitrarily tiny damping factor $\varepsilon$,
\begin{equation}
\lim_{\varepsilon \to 0} \int_{\epsilon}^\mu\frac{d\lambda}{\lambda} e^{(ia + \varepsilon) \log \lambda},\qquad a \in \mathbb{R}_{\ne 0},
\end{equation}
it becomes convergent.  In other words, this case may be defined simply by continuity with the convergent region, without the need for a true analytic continuation.

More interestingly, because it is a unitary case, we can instead continue to the region $\text{Re}(\Delta)=\frac{d}{2}$, corresponding to the principal series.  This case, shown as the vertical blue line in Fig.~\ref{fig:strips}, is also a lateral case.  Since our cosmological results in Section \ref{sec:FreeScalardS} are valid for the principal series as well, this means that it is also valid to analytically continue to the principal series case.\footnote{In the principal series case, it is necessary to choose between a conformal vs.~ a self-adjoint boundary condition, in the initial dS/CFT boundary condition.  In selecting the source to have a fixed value of $\Delta$, we are implicitly selecting the conformal case here.}

The only cases that require special care are the cases where a counterterm becomes \emph{marginal}, corresponding to the blue and red dots in Fig.~\ref{fig:strips}.  In this case there is a true log divergence, and it is necessary to introduce a scale in order to eliminate it.  The seed CFT has a new type of trace anomaly, and there is a non-trivial anomaly matching condition between this trace anomaly and the bulk Hamiltonian constraint.

\subsubsection{Matter at $\Delta=d$}

It is interesting to note that the graviton is massless. This means that it behaves like a scalar field with $\Delta=d$ for each polarisation. Thus, the fact that the right deformation valid for $\Delta=d$ depends on dimension is not entirely surprising and is consistent with the fact that the pure gravity deformation operators written in \eqref{eq:Opuregravd=2} and \eqref{eq:Opuregravd=3} were also dimension dependent. In fact, we see that \emph{exactly} at the point $\Delta=d$ the matter deformation operators greatly simplify. 

For instance, for $d=2$
\begin{equation}
    \hat{O}_\text{matter}^{\Delta=d=2}  = i\int_\Sigma d^dx \sqrt{h} \Bigg( \frac{1}{2}(-\eta)^{2}:\mathcal{O}^2:-\frac{1}{2}(\nabla_h\phi)^2\Bigg) + O(\phi^4,\phi^3\mathcal{O}) \, .
\end{equation}
The only matter counterterm needed is given by \eqref{eqn:ct_scalar}, which identically vanishes at $\Delta=d$. So, $\text{CT}_\text{matter}^{\Delta=d=2}=0$. Furthermore, this case happens to lie at a marginal point. The corresponding conformal anomaly includes a matter contribution
\begin{align}
    \expval{T}_{\text{CFT}_2}^{\Delta=2}= i\left(-\,\alpha\, \mathcal{R} +\frac{1}{2}(\nabla_h\phi)^2\right)\,.
\end{align}

For $d=3$
\begin{align}
    \hat{O}_\text{matter}^{\Delta=d=3} &=i\int_\Sigma d^dx \sqrt{h} \Bigg( \frac{1}{2}(-\eta)^{3}:\mathcal{O}^2: - i (-\eta)^2 (\Box_h\phi) \, \mathcal{O} -\frac{1}{2}(-\eta)\left(\Box_h\phi\right)^2 \\
    &-i\frac{1}{\alpha}\frac{1}{2}(-\eta)^2\left(\partial_a\phi\partial_b\phi\, T^{ab}-\frac{1}{4}(\nabla_h\phi)^2\,T\right)\Bigg) + O(\phi^4,\phi^3\mathcal{O}) \, .
\end{align}
If we are to take the scalar at $\Delta=d$ to correspond to the graviton, then the term coupling $\phi$ quadratically to the stress-tensor is to be interpreted as a graviton self-interaction, which is suppressed by $1/\alpha$. The matter counterterm for $\Delta=d=3$ is now not trivial since we have the additional one given by \eqref{eq:CT1stbarrier}. This becomes
\begin{equation} \label{eqn:CTDelta=3}
    \text{CT}_\text{matter}^{\Delta=d=3} = i \frac{1}{2}(-\eta)^{-1}\int_\Sigma d^3x\, \sqrt{h}\, (\nabla_h\phi)^2 \, .
\end{equation}

\section{Flow of Correlation Functions}\label{sec:BoundaryFlow}
In this section we compute the flow of various two-point functions of the dual theory along the $T^2$-deformation, taking the CFT result as an initial condition. Under the CSH duality, this will give us the wavefunction coefficients discussed in the Introduction, which we will show precisely match with the bulk computations in section. We start by providing a general algorithm for the computation, before going to the specific examples of scalar and stress-tensor operators. 

\subsection{General Structure}
We have the starting definitions
\bea
    Z&=&\,\exp{\text{CT}}\,\tilde{Z} \, , \\
    \tilde{Z}&=&\text{P exp}\left(\int_0^{-\eta} \frac{d\lambda}{\lambda}\hat{O}(\lambda)\right)\,Z_\text{CFT} \, ,
\eea
where $\hat{O}=\int_\Sigma\, \hat{X}(x)$. $Z$ is a functional of the set of all field theory sources, collectively denoted by $\mathbb{J}$: $Z=Z[\mathbb{J}]$. Note that $\hat{X}(x)=\hat{X}(\mathbb{J}(x),\delta/\delta\mathbb{J}(x))$ and so $\hat{O}=\hat{O}[\mathbb{J}]$ is also a functional of the sources.\footnote{More appropriately, it is a vector field on the theory space, whose points label source profiles $\mathbb{J}(x)$.} This is nothing but the statement that the deformation operator is defined ``locally'' in theory space.

We want to compute the flow of the connected two-point function $\expval{\mathbf{O}(x)\mathbf{O}(y)}_c$, where $\mathbf{O}$ is the operator dual to the (abstract) source $J\in \mathbb{J}$. The task splits into two steps: 1. we compute the flow under the $T^2$-deformation operator, i.e. we compute $\widetilde{\expval{\mathbf{O}(x)\mathbf{O}(y)}_c}(-\eta)$, where the tilde indicates that the correlation function is evaluated in the $\Tilde{Z}$ theory; 2.  we shift the result by the appropriate counterterms to get $\expval{\mathbf{O}(x)\mathbf{O}(y)}_c(-\eta)$ in the $Z$ theory. In equations,
\begin{align}
    \expval{\mathbf{O}(x)\mathbf{O}(y)}_c(-\eta):&=\left(\frac{1}{\sqrt{h(x)}}\frac{\delta}{\delta J(x)}\right)\left(\frac{1}{\sqrt{h(y)}}\frac{\delta}{\delta J(y)}\right)\log Z \Big|_{-\eta} \\
    &=\left(\frac{1}{\sqrt{h(x)h(y)}}\frac{\delta^2}{\delta J(x)\delta J(y)}\text{CT}\right)(-\eta) + \widetilde{\expval{\mathbf{O}(x)\mathbf{O}(y)}_c}(-\eta) \, , \label{eqn:2-point}
\end{align}
where 
\begin{align}
    \widetilde{\expval{\mathbf{O}(x)\mathbf{O}(y)}_c}(-\eta) :=\left(\frac{1}{\sqrt{h(x)}}\frac{\delta}{\delta J(x)}\right)\left(\frac{1}{\sqrt{h(y)}}\frac{\delta}{\delta J(y)}\right)\log \Tilde{Z} \Big|_{-\eta}\,.
\end{align}

To compute the non-trivial flow induced by the deformation, we first compute the $(-\eta)\frac{\partial}{\partial(-\eta)}$ derivative of the $\Tilde{Z}$ correlator
\begin{align}
    (-\eta)\frac{\partial}{\partial(-\eta)}\widetilde{\expval{\mathbf{O}(x)\mathbf{O}(y)}_c}
    &=\frac{1}{\sqrt{h(x)h(y)}}\frac{\delta^2}{\delta J(x)\delta J(y)}\left((-\eta)\frac{\partial}{\partial(-\eta)}\log \tilde{Z}\right)\\
    &=\frac{1}{\sqrt{h(x)h(y)}}\frac{\delta^2}{\delta J(x)\delta J(y)}\left(-i\int_\Sigma\mathcal{W}\log \tilde{Z}\right)\\
    &=\frac{1}{\sqrt{h(x)h(y)}}\frac{\delta^2}{\delta J(x)\delta J(y)}\left(\frac{1}{\tilde{Z}}\hat{O}(-\eta)\tilde{Z}\right) \label{eq:tildeflow}
\end{align}
where we used the defining (RG-type) equation of $\tilde{Z}$ to get to the last line \eqref{eqn:RG}. 

To proceed we need to remember that the general structure of the deforming operator $\hat{O}$ is
\bea
    \hat{O}[\mathbb{J}] = \sum_{i\in I_2}\hat{O}_i^{(2)}[\mathbb{J}] + \sum_{i\in I_1}\hat{O}_i^{(1)}[\mathbb{J}] + \sum_{i\in I_0}\hat{O}_i^{(0)}[\mathbb{J}],
\eea    
where the superscript indicates the degree of field theory operators in each term and we have the corresponding index sets $I_2,\,I_1,\,I_0$. In particular, every $\hat{O}_i^{(2)}$ is a quadratic (double-trace) operator defined via \emph{normal-ordering}
\bea
\hat{O}_i^{(2)}[\mathbb{J}]:=\int_\Sigma d^dz \sqrt{h}\,\lim_{w\to z} \left( \hat{A}_i(z)\hat{B}_i(w)-\expval{\hat{A}_i(z)\hat{B}_i(w)}_c\big|_{\mathbb{J}}\right) \,, \label{eq:point-splitting}
\eea
for some local operators $\hat{A}_i(x)$ and $\hat{B}_i(x)$, which can (and will), in general, also depend explicitly on $\mathbb{J}(x)$. This definition is valid only in the strict $N\to\infty
$ limit, in which the theory becomes generalised free. Notice that the normal-ordering prescription is such that $\widetilde{\expval{\hat{O}_i^{(2)}}}\big|_{\mathbb{J}}:=\frac{1}{\Tilde{Z}}\hat{O}_i^{(2)}\Tilde{Z}\big|_{\mathbb{J}}=0\,,\forall \mathbb{J}$. Furthermore, the point-splitting definition of $\hat{O}_i^{(2)}$ effectively moves any explicit  dependence on sources to the left of all functional derivatives.\footnote{This is because we take the derivatives \emph{before} taking the limit.} The terms $\hat{O}_i^{(1)}$ are linear in the local operators, while $\hat{O}_i^{(0)}$ are proportional to the identity. 

Then we can write
\begin{align}
    \frac{1}{\tilde{Z}}\hat{O}\tilde{Z}=\sum_{i\in I_2}\int_\Sigma d^dx \,\sqrt{h}\, \left( \hat{A}_i(x)\log \Tilde{Z}\right)\left(\hat{B}_i(x)\log \Tilde{Z}\right)
    + \sum_{i\in I_1}\hat{O}_i^{(1)}\log\Tilde{Z} + \sum_{i\in I_0}\hat{O}_i^{(0)}\,. \label{eq:Ohataction}
\end{align}

We now act with $\delta^2/\delta J(x)\delta J(y)$ and evaluate the final result for $\mathbb{J}=0$.\footnote{The set of sources includes scalars, as well as tensors, like the metric. So when we say $\mathbb{J}=0$ we mean some choice of ``zero-point'', which could entail $J=0$ for a scalar source, $h_{ab}=h^{0}_{ab}$ for some reference metric, together with a corresponding assignment for all the sources. Because we are working in the framework of conformal perturbation theory, the ``zero-point'' is taken to preserve the CFT symmetries.} To do this, we note that the local single-trace operators appearing in the deformation operator generically involve some multiplicative polynomial function of the sources (and their spatial derivatives)
\bea
    \hat{A}_i(x) = \mathbf{A}_i(\mathbb{J}(x))\frac{1}{\sqrt{h}}\frac{\delta}{\delta J_{A_i}(x)}=:\mathbf{A}_i(\mathbb{J}(x)) \hat{\mathcal{A}}_i(x)\,.
\eea
This is because our deformations are really only defined in conformal perturbation theory, around the CFT fixed-point corresponding to $\mathbb{J}=0$. Thus, we are expanding our operators in a neighbourhood of their CFT limit. In practice, the operator arising from the Hamiltonian constraint will correspond to a truncated expansion, which is a consequence of the fact that our Hamiltonian included only the leading irrelevant terms.\footnote{In an EFT expansion, we could include higher-curvature corrections to the Hamiltonian constraint, but these would be Planck-suppressed, or in QFT language, they would not contribute at leading-order in the $O(1/N)$ expansion.} In the end, for the computation of interest here, none of these higher-order terms will contribute, which justifies the truncation. For convenience, we also define $\hat{B}_i(x)=:\mathbf{B}_i(\mathbb{J}(x))\hat{\mathcal{B}}_i(x)$, as well as $\hat{O}^{(1)}_i=:\int_\Sigma d^dx\,\sqrt{h}\, \mathbf{C}_i(\mathbb{J}(x))\hat{\mathcal{C}}_i(x)$ and $\hat{O}^{(0)}_i=:\int_\Sigma d^dx\,\sqrt{h}\,\mathbf{D}_i(\mathbb{J}(x))$, with the same meaning as described above. 
The operators $\hat{\mathcal{A}}_i(x)$, $\hat{\mathcal{B}}_i(x)$ and $\hat{\mathcal{C}}_i(x)$ are defined to have vanishing one-point function when $\mathbb{J}=0$, for all $i$.

We get the following flow induced by the deformation
\begin{align}
    &(-\eta)\frac{\partial}{\partial(-\eta)}\widetilde{\expval{\mathbf{O}(x)\mathbf{O}(y)}_c}=\frac{1}{\sqrt{h(x)h(y)}}\frac{\delta^2}{\delta J(x)\delta J(y)}\left(\frac{1}{\tilde{Z}}\hat{O}\tilde{Z}\right)\big|_{\mathbb{J}=0} \\
    &=\int_\Sigma d^dz\,\sqrt{h(z)}\,\Bigg\{\sum_{i\in I_2} \mathbf{A}_i\mathbf{B}_i(\mathbb{J}(z))\left(\widetilde{\expval{\mathbf{O}(x)\hat{\mathcal{A}}_i(z)}_c}\big|_{\mathbb{J}=0}\widetilde{\expval{\mathbf{O}(y)\hat{\mathcal{B}}_i(z)}}_c\big|_{\mathbb{J}=0} + (x\leftrightarrow y)\right)  \nonumber \\
    &+\sum_{i\in I_1}\left(\frac{1}{\sqrt{h(x)}}\frac{\delta \mathbf{C}_i(\mathbb{J}(z))}{\delta J(x)}\widetilde{\expval{\mathbf{O}(y)\hat{\mathcal{C}}_i(z)}}_c\big|_{\mathbb{J}=0}+(x\leftrightarrow y) + \mathbf{C}_i(\mathbb{J}(z))\widetilde{\expval{\mathbf{O}(x)\mathbf{O}(y)\hat{\mathcal{C}}_i(z)}}_c\big|_{\mathbb{J}=0}\right)  \nonumber \\
    &+\sum_{i\in I_0}\frac{1}{\sqrt{h(x)h(y)}}\frac{\delta^2 \mathbf{D}_i(\mathbb{J}(z))}{\delta J(x)\delta J(y)}\Bigg\} \,. \label{eqn:flow}
\end{align}

Importantly, we are only allowed to trust this equation at leading order in $1/N$. This is because the point-splitting definition in \eqref{eq:point-splitting} is only valid for $N\to\infty$. To work to sub-leading order we would have to take the finiteness of $N$ into account in the definition of the double-trace operators as well. This would, however, be heavily dependent on the precise details of the holographic dual in question, via the OPE. To understand how each term scales with $N$, we do a small digression into the behaviour of large-$N$ gauge theories in Section \ref{sec:largeN}. But first we argue that, at this order, mixed two-point correlators exactly vanish for all $\eta$, when evaluated on $\phi=0$ and $h_{ab}=\delta_{ab}$ (this won't be the case on generic curved backgrounds). This simplifies the analysis considerably.

\subsubsection{CFT Limit and Off-Diagonal Correlators}

It will be important to briefly discuss the CFT limit of certain correlation functions. 

The conformal symmetry determines the two-point functions of primary operators, up to normalisation. So we have that
\begin{align}
    \expval{\mathcal{O}(x)\mathcal{O}(y)}_c^\text{CFT}\sim\frac{1}{|x-y|^{2\Delta}}\\
    \expval{T^{ab}(x)T^{cd}(y)}_c^\text{CFT}\sim \frac{\mathbb{I}^{abcd}(|x-y|)}{|x-y|^{2d}}\,,
\end{align}
where $\mathbb{I}^{abcd}(|x-y|)$ is an appropriate tensor structure. Furthermore, it dictates that ``off-diagonal'' two-point functions vanish, in particular 
\begin{equation}
    \expval{\mathcal{O}(x)T^{ab}(y)}_c^\text{CFT}=0\,,
\end{equation}
This last equation is particularly important for what follows, because it implies that, at leading order in the $1/N$ expansion, we have that 
\begin{equation}
    \widetilde{\expval{\mathcal{O}(x)T^{ab}(y)}}_c
    (-\eta)=0 \, , \quad \forall \quad \eta \, .
    \label{eqn:mixed}
\end{equation}
Mixed two-point functions vanish, without flowing. We can see this by inspecting \eqref{eqn:flow} and the form of the deformation operators in \eqref{eq:Opuregravd=2}, \eqref{eq:Opuregravd=3}, \eqref{eq:Omattersimple}, \eqref{eq:O2ndstrip}, as well as the discussion in Section \ref{sec:counterterms}. If we take $J(x)\sim \phi(x)$ and $J(y)\sim h_{ab}(y)$ (and hence, $\mathbf{O}(x)\sim \mathcal{O}(x)$ and $\mathbf{O}(y)\sim T^{ab}(y)$) we conclude that at $\eta=0$\footnote{It is useful to keep in mind that there is a $\phi\to -\phi$ symmetry in the game. This means that in the deformation operator we always have combinations of the form $\sim\phi^n\mathcal{O}^m$ with $n+m$ even.}
\begin{itemize}
    \item the first line contributes $0$ because each term is  proportional to $\widetilde{\expval{\mathcal{O}(x)T^{ab}(y)}}_c$;
    \item in the second line, the terms that go like two-point functions contribute $0$ because each term is either proportional to $\widetilde{\expval{\mathcal{O}(x)T^{ab}(y)}}_c$ or to a first-order derivative with respect to $\phi$. The latter is, however, always $0$ because the various terms go either like $\sim \phi^0$ or $\sim \phi^2$ and we set $\phi=0$ at the end. The 3-point function term also does not contribute because we set $\phi=0$ and $h_{ab}=\delta_{ab}$;
    \item the third line contributes $0$ because it involves a first-order derivative with respect to $\phi$;
\end{itemize}
So, if the initial condition is that $\widetilde{\expval{\mathcal{O}(x)T^{ab}(y)}}_c(-\eta=0)=0$, then the flow equation dictates that $(-\eta)\frac{\partial}{\partial(-\eta)}\widetilde{\expval{\mathcal{O}(x)T^{ab}(y)}}_c\Big|_{-\eta=0}=0$ and the conclusion follows. Lastly, there is no counterterm contribution for the same reason as the last bullet point. 

We note that, if we work around a curved background, then all the contributions to the flow vanish except the 3-point function term $\sim \expval{\mathcal{O}(x)T^{ab}(y)T^{cd}(z)}$. This is \emph{not} 0 in the CFT limit and so it induces a non-trivial flow for the mixed two-point correlator, even when we set $\phi=0$. This is an example of non-trivial effects coming from coupling to the curvature. As we will see below, this term contributes at the correct leading-order in the $1/N$ expansion.

\subsubsection{Large-$N$ Scalings} \label{sec:largeN}

In a large-$N$ gauge theory, like our $T^2$-deformed holographic CFT, connected n-point functions scale with $N$. The scaling, of course, depends on a choice of normalisation for our dual operators. This choice on the boundary side is related to an analogous choice of normalisation for the bulk fields. In the bulk, we are taking our fields to be normalised in a ``canonical'' fashion, meaning that we have a $\frac{1}{16\pi G_N}$ factor in front of the graviton kinetic term and a $\frac{1}{2}$ factor in front of the kinetic term of the real scalar field. This means that their dual operators are $N$-normalised such that
\begin{align}
    \widetilde{\expval{\mathcal{O}(x_1)\mathcal{O}(x_2)}}_c &\sim O(1)\\
    \widetilde{\expval{T^{ab}(x_1)T^{cd}(x_2)}}_c &\sim \mathbf{c}\sim O(N^p)\,,
\end{align}
where $p$ is a power that, from experience in the AdS/CFT case, depends on the particular duality in question.  Here we assume only that $p > 0$, so that large $N$ is the same as large number of degrees of freedom (or, equivalently, large central charge $\mathbf{c}$).\footnote{From a boundary perspective, $N$ is the number of colours in the gauge theory, while in the bulk perspective it is usually the number of flux units in the bulk.  In the most famous example, ${\cal N} = 4$ Super Yang Mills in $d=4$, we have $c \sim N^2$, but in the ABJM model $d=3$, we have $c \sim N^{3/2}$, and in the $(2,0)$ model $d=6$ we have $c \sim N^3$.  It is interesting that all of these cases have $c \sim N^{d/2}$, but from the bulk perspective based on flux dimensional analysis, this relies in a subtle numerical way on the fact that the total number of spacetime dimensions is $D = 10$ (in the string theory examples) or $D = 11$ (in the $M$-theory examples), and seems to be related to special properties of supergravity in $D = 10,11$.  Hence, the relation probably does not generalise to dS/CFT, with no large extra dimensions.  So we keep the power $p$ arbitrary here.  We thank Juan Maldacena for discussions related to this point.}

This implies a rule for the $N$-scaling of higher-point functions
\begin{align}
    \widetilde{\expval{\mathcal{O}(x_1)...\mathcal{O}(x_n)T^{a_1b_1}(y_1)...T^{a_mb_m}(y_m)}}_c \sim O(N^{p-np/2})\,.
\end{align}
There is a relative suppression by $N^{p/2}$ for every scalar insertion.

In general holographic settings, the $N$ parameter in the gauge theory is related to the ratio of bulk scales
\bea
    \alpha\sim \left(\frac{L_{\text{dS}}}{L_\text{Pl}}\right)^{d-1} \sim N^p\sim \mathbf{c}.
\eea
For $N\gg1$, we are in a regime of the bulk theory in which the characteristic curvature length scale of the cosmology is much larger than the Planck scale. This corresponds to the semiclassical regime of the theory. Furthermore, as shown by the large-$N$ suppression of higher n-point functions outlined above, the interactions are also suppressed, meaning that the matter is in a free/classical regime around some fixed background geometry. 

We should also remember that the source polynomials  $(\mathbf{A}_i, \mathbf{B}_i, \mathbf{C}_i, \mathbf{D}_i)$ themselves come with certain powers of $N$ (or, equivalently, $\alpha$), as can be seen, for example, in \eqref{eq:Opuregravd=2}, \eqref{eq:Opuregravd=3}, \eqref{eq:Omattersimple} and \eqref{eq:O2ndstrip}. The possible $\alpha$-scalings belong to the set $\{\frac{1}{\alpha},1,\alpha\}$. 

There is now a distinction between scalars and gravitons with regards to the 3-point function contribution on the third line of \eqref{eqn:flow}. 
\begin{itemize}
    \item For the scalar two-point function, this is always suppressed by $1/\alpha$ relative to the other terms. This is because $\widetilde{\expval{\mathcal{O}(x)\mathcal{O}(y)\mathcal{O}(z)}}\sim O(1/\alpha)$, while $\widetilde{\expval{\mathcal{O}(x)\mathcal{O}(y)T^{ab}(z)}}\sim O(1)$ but is always accompanied by a coefficient going as $O(1/\alpha)$. Therefore, the 3-point functions \emph{never} contribute to the flow at leading order.
    \item  This is not the case for the stress-tensor correlator. We have that $\widetilde{\expval{T^{ab}(x)T^{cd}(y)T^{ef}(z)}}\sim O(\alpha)$ with a coefficient going as either $O(1)$ (from pure-gravity terms) or $O(1/\alpha)$ (from gravity--matter coupling terms). The former case contributes at \emph{the same} order $O(\alpha)$ as $\widetilde{\expval{T^{ab}(x)T^{cd}(y)}}$. The coefficient happens to vanish when evaluated on certain backgrounds, like the case of interest $h_{ab}=\delta_{ab}$. So it does not contribute in our case. However, for generic curved backgrounds the stress-tensor 3-point function \emph{does} contribute to the flow of the two-point function, even at leading order. This is a non-linear gravitational effect. There are also other sub-leading terms going like $\widetilde{\expval{T^{ab}(x)T^{cd}(y)\mathcal{O}(z)}}\sim O(\sqrt{\alpha})$ with $O(1)$ coefficients. As expected, terms coming from coupling to matter are the first ones to kick in at the next order.
\end{itemize}
To summarise, in the case we will be interested in of $\phi=0$ and $h_{ab}=\delta_{ab}$ (corresponding to Poincaré slicing), there is no contribution from any 3-point function to either the scalar or the stress-tensor two-point function. So the flow equations dramatically simplify and can actually be solved without knowing any detailed information about the holographic CFT.

\hfill \break

All in all, this gives a first-order flow equation for $\widetilde{\expval{\mathbf{O}(x)\mathbf{O}(y)}}_c\big|_{\mathbb{J}=0}(-\eta)$, as a function of $\eta$, at leading order in $1/N$.  We will then integrate this equation with the initial condition provided by the CFT two-point function $\widetilde{\expval{\mathbf{O}(x)\mathbf{O}(y)}}_c\big|_{\mathbb{J}=0}(0)=\expval{\mathbf{O}(x)\mathbf{O}(y)}_c^\text{CFT}\big|_{\mathbb{J}=0}$, to get the function $\widetilde{\expval{\mathbf{O}(x)\mathbf{O}(y)}}_c\big|_{\mathbb{J}=0}(-\eta)$ to leading-order in $N$. Adding the counterterm shift in \eqref{eqn:2-point}, we end up with the desired function $\expval{\mathbf{O}(x)\mathbf{O}(y)}_c\big|_{\mathbb{J}=0}(-\eta)$.


\subsection{The Scalar Field Case}
We now analyse the flow of the two-point function for concrete bulk dynamical content coupled to gravity, following the field theory computation just described. 

We begin by considering the two-point function of the scalar operator dual to a real, minimally-coupled, massive scalar field in the bulk in $d+1$ dimensions. This example already highlights most of the relevant features and it makes transparent the similarities and differences with an analogous computation in AdS spacetimes (see \cite{Hartman:2018tkw}).

In what follows we have the identification
\begin{align}
    J &\leftrightarrow \phi\\
    \mathcal{O} &\leftrightarrow \frac{1}{\sqrt{h}}\frac{\delta}{\delta\phi}\,.
\end{align}

\subsubsection{Computation in the $\text{Re}(\Delta)\in\left(\frac{d}{2},\frac{d+2}{2}\right)$ Strip} \label{sec:flowsimple}

As shown before in Section \ref{sec:simplestrip}, at leading-order in the $1/N$ expansion, the contributing matter part of the deforming operator in this region of parameter space is given by \eqref{eq:Omattersimple}.

The quadratic operator reads
\bea
    \hat{O}^{(2)} = i\frac{1}{2}(-\eta)^{2\Delta-d}\int_\Sigma d^dz\,\sqrt{h}:\mathcal{O}^2:(z)\,,
\eea
meaning it comes at $O(N^0)$. There is no operator in $\hat{O}^{(1)}$ that survives at this order. From the terms proportional to the identity the following one contributes at $O(N^0)$
\bea
    \hat{O}^{(0)}=-i (-\eta)^{2(d-\Delta)-d} \int_\Sigma d^dz\,\left((-\eta)^2\frac{1}{2}\sqrt{h}\,(\nabla_h\phi)^2+\phi^2\sum_{d_e<d} (-\eta)^{d_e}\mathcal{A}_{d_e}\right)(z)\,.
\eea

Using the formula in \eqref{eqn:flow},  we get that 
\bea
    (-\eta)\frac{\partial}{\partial(-\eta)}\widetilde{\expval{\mathcal{O}(x)\mathcal{O}(y)}}_c\big|_{\mathbb{J}=0} &=& i(-\eta)^{2\Delta-d}\int_\Sigma d^dz\, \sqrt{h}\,\widetilde{\expval{\mathcal{O}(x)\mathcal{O}(z)}}_c\big|_{\mathbb{J}=0}\widetilde{\expval{\mathcal{O}(y)\mathcal{O}(z)}}_c\big|_{\mathbb{J}=0} \nonumber \\
    &+&i(-\eta)^{2(d-\Delta)+2-d}  \frac{1}{\sqrt{h}}\,\Box_h\delta^d(x-y) \nonumber \\
    &-&2i(-\eta)^{2(d-\Delta)-d}\frac{1}{\sqrt{h}}\, \delta^d(x-y)\sum_{d_e<d} (-\eta)^{d_e}\frac{\mathcal{A}_{d_e}}{\sqrt{h}} \, ,
\eea
where to get the second line we performed an integration by parts. The last line encodes local contact terms from coupling to the curvature of the background geometry. The first line introduces a \emph{non-local} effect coming from the quadratic (double-trace) nature of the deformation operator.

We can simplify the expression further if we evaluate the correlation function in Fourier space on the flat background $h_{ab}=\delta_{ab}$. This corresponds to the case of Poincaré slicing. 

On such a background geometry we have that $\mathcal{A}_{d_e}=0$, for all $d_e$. Because the slice has a boundary, we have freedom to input a QFT state there. Assuming we take the vacuum, then we have that $\widetilde{\expval{\mathcal{O}(x)\mathcal{O}(y)}}_c\big|_{\mathbb{J}=0}=F(|x-y|)$. Defining $\Tilde{F}(k)\delta^d(\bfk+\bfq):=\int_{\mathbb{R}^{2d}}d^dxd^dy\, F(|x-y|)e^{i\bfx\cdot\bfk}e^{i\bfy\cdot\bfq}$,  where $k=|\bfk|$, we then get
\begin{equation}\label{eqn:dSCSH2ptCorrelatorFlowEquation1}
        i(-\eta)\frac{\partial}{\partial(-\eta)}\Tilde{F}(k;\eta)= (-\eta)^{2(d-\Delta)+2-d}\, k^2-  (-\eta)^{2\Delta-d}\,\Tilde{F}^2(k;\eta) \, .
\end{equation}
We note that this is the same flow equation as previously obtained in~\cite{Hartman:2018tkw}, after the appropriate analytic continuation from Euclidean AdS to the dS Poincaré patch. The general solution for \eqref{eqn:dSCSH2ptCorrelatorFlowEquation1} can be written as
\begin{equation}
    \Tilde{F}(k;\eta) = \frac{i k (-\eta)^{d-2 \Delta +1} \left(Y_{\frac{1}{2} (d-2 \Delta +2)}(-k\eta)+c_1 J_{\frac{1}{2} (d-2 \Delta +2)}(-k\eta)\right)}{Y_{\frac{1}{2} (d-2 \Delta )}(-k\eta)+c_1 J_{\frac{1}{2} (d-2 \Delta )}(-k\eta)} \, ,
\end{equation}
where $J_{\alpha}$ is the Bessel function of the first kind with order $\alpha$, $Y_{\beta}$ is the Bessel function of the second kind with order $\beta$ and $c_1$ is an arbitrary complex number. 

As an initial condition for the flow equation we input the CFT result
\begin{equation}\label{eqn:dSCSHFlowEquationInitialCondition}
    \Tilde{F}(k;0) = \frac{C(d,\Delta)}{k^{d-2\Delta}} \, ,
\end{equation}
where $C(d,\Delta)$ includes a non-trivial and important phase (see~\cite{Goodhew:2024eup,Thavanesan:2025kyc} for the derivation of the closed form expression of this phase)
\begin{equation}\label{eqn:CPTphase}
    C(d,\Delta)= e^{i\frac{\pi}{2}\!\left((d-1)-2d+2\Delta\right)} \, .
\end{equation}
Once this has been accounted for, we can write the explicit solution for \eqref{eqn:dSCSH2ptCorrelatorFlowEquation1} satisfying the initial condition \eqref{eqn:dSCSHFlowEquationInitialCondition} as
\begin{equation}\label{eqn:dSCSH2ptCorrelatorFlowEquationSolution1}
    \tilde{F}(k;\eta) = \frac{i k \, (-\eta)^{d-2 \Delta +1} H_{\frac{1}{2} (d-2 \Delta +2)}^{(1)}(k \eta)}{H_{\frac{1}{2} (d-2 \Delta)}^{(1)}(k\eta)} \, .
\end{equation}
After the convenient change of variables $\Delta=\frac{d}{2}+\nu$, we can re-write this as\footnote{This requires using the following Hankel function identities (see e.g. 10.4 and 10.11 of~\cite{NIST:DLMF})
\begin{align}
    H_{\alpha}^{(1)}(e^{-i\pi}x) \equiv -e^{i\pi\alpha}H_{\alpha}^{(2)}(x) \quad \text{and} \quad H_{-\alpha}^{(2)}(x) \equiv e^{-i\pi\alpha}H_{\alpha}^{(2)}(x) \qquad &\text{for } x>0 \label{eqn:HankelIdentity1} \\ 
    \implies H_{-\alpha}^{(1)}(e^{-i\pi}x)=-e^{-i\pi\alpha}H_{-\alpha}^{(2)}(x)=-e^{-i\pi\alpha}e^{-i\pi\alpha}H_{\alpha}^{(2)}(x) = -e^{-i\pi(2\alpha)}H_{\alpha}^{(2)}(x) \qquad &\text{for } x>0 \label{eqn:HankelIdentity2} \\
    \implies H_{-\alpha}^{(1)}(e^{-i\pi}x) \equiv -e^{-i\pi(2\alpha)}H_{\alpha}^{(2)}(x) \qquad &\text{for } x>0 \, . \label{eqn:HankelIdentity3}
\end{align}
where we have been careful to rotate in the lower-half of the complex plane as dictated by the Bunch-Davies vacuum condition in \eqref{eqn:BDFreeScalarField} (see Section 7 of~\cite{Goodhew:2024eup} for more details).}
\begin{equation}\label{eqn:dSCSH2ptCorrelatorFlowEquationSolution3}
    \tilde{F}(k;\eta)= \frac{ik (-\eta )^{1-2\nu} H_{\nu-1}^{(2)}(-k \eta )}{H_{\nu}^{(2)}(-k \eta )} \, .
\end{equation} 
Finally, we must shift by the correction coming from the counterterm in \eqref{eqn:ct_scalar},
giving us the final answer (restoring the factors of $L_{\text{dS}}$ for comparison) \begin{equation}\label{eqn:dSCSH2ptCorrelatorFlowEquationSolutionCT}
    F(k;\eta)= iL_{\text{dS}}^{d-1}(d - \Delta) \left(-\eta\right)^{-2\nu} + iL_{\text{dS}}^{d-1} \frac{k (-\eta )^{1-2\nu} H_{\nu-1}^{(2)}(-k \eta )}{H_{\nu}^{(2)}(-k \eta )} \, .
\end{equation}
This matches precisely the bulk result in \eqref{eqn:dSscalar2ptEtaFactors}. Here we obtained it purely from boundary structures. 

Since we already know that \eqref{eqn:dSscalar2ptEtaFactors} in Section \ref{sec:FreeScalardS} is true for all values of $\Delta$ and $d$, i.e. for the full range of the complementary $\Delta \in (\frac{d}{2},d)$ and principal series $\Delta \in \frac{d}{2}+i\epsilon$, this suggests that we can in fact analytically continue the result we have obtained here for the strip $\text{Re}(\Delta) \in (\frac{d}{2},\frac{d+2}{2})$ to arbitrary $\Delta$. We argued for this in Section \ref{sec:FlowAnalyticContinuationPrincipalSeries}. But we will now explicitly check that the boundary prescription is consistent with this analytic continuation. We will perform the same steps in a different strip, in which both the counterterms and the deformation operator will be different. Nevertheless, the two-point function we obtain at the end will be the same as the analytic continuation of \eqref{eqn:dSCSH2ptCorrelatorFlowEquationSolutionCT}.

\subsubsection{Computation in the $\text{Re}(\Delta)\in\left(\frac{d+2}{2},\text{min}\left\{\frac{d+4}{2},d\right\}\right)$ Strip} \label{sec:flownextstrip}

Proceeding analogously, we can write down the flow equation for the next strip, evaluated on $h_{ab}=\delta_{ab}$, starting from the deformation operator in \eqref{eq:O2ndstrip} 
\begin{align}
    (-\eta)\frac{\partial}{\partial(-\eta)}\widetilde{\expval{\mathcal{O}(x)\mathcal{O}(y)}}_c\big|_{\mathbb{J}=0} &= i(-\eta)^{2\Delta-d}\int_\Sigma d^dz\, \sqrt{h}\,\widetilde{\expval{\mathcal{O}(x)\mathcal{O}(z)}}_c\big|_{\mathbb{J}=0}\widetilde{\expval{\mathcal{O}(y)\mathcal{O}(z)}}_c\big|_{\mathbb{J}=0} \nonumber \\
    &- \frac{1}{d-2\Delta+2}(-\eta)^2 (\Box_h^{(x)}+\Box_h^{(y)})\widetilde{\expval{\mathcal{O}(x)\mathcal{O}(y)}}_c\big|_{\mathbb{J}=0}\nonumber \\
    &-i\frac{1}{(d-2\Delta+2)^2}(-\eta)^{d-2\Delta+4}\frac{1}{\sqrt{h}}(\Box_h)^2\delta^d(x-y) \, ,
\end{align}
plus $O(1/N)$ corrections. There is now a linear term, compared to the previous case. 

Again, going to Fourier space, it simplifies to
\begin{align}
    i(-\eta)\frac{\partial}{\partial(-\eta)}\Tilde{F}(k;\eta)&= \frac{(-\eta)^{d-2\Delta+4}}{(d-2\Delta+2)^2}\,k^4+i\frac{2(-\eta)^2}{d-2\Delta+2}\,k^2\Tilde{F}(k;\eta) -  (-\eta)^{2\Delta-d}\,\Tilde{F}^2(k;\eta)\\
    &=-(-\eta)^{(d+4)-2\Delta}\left((-\eta)^{2\Delta-(d+2)}\Tilde{F}(k;\eta)+i\frac{k^2}{2\Delta-(d+2)}\right)^2\,.\label{eqn:FlowEquationNextStrip}
\end{align}
The general solution is
\begin{equation}
    \Tilde{F}(k;\eta) = \frac{-i k^2 (-\eta)^{2+d-2 \Delta } \left(Y_{\frac{1}{2} (d-2 \Delta +2)}(k\eta)+c_2 J_{\frac{1}{2} (d-2 \Delta +2)}(k\eta)\right)}{(2+d-2\Delta)\left(Y_{\frac{1}{2} (d-2 \Delta )}(ik\mu^{1/d})+c_2 J_{\frac{1}{2} (d-2 \Delta )}(k\eta)\right)} \, .
\end{equation}
Inputting the initial conditions \eqref{eqn:dSCSHFlowEquationInitialCondition} and \eqref{eqn:CPTphase} determines $c_2=-i$ allowing us to write the solution, again after using the identities \eqref{eqn:HankelIdentity1}-\eqref{eqn:HankelIdentity3}, as
\begin{equation}\label{eqn:FlowEquationNextStripSolution2}
    \Tilde{F}(k;\eta) = \frac{-i k^2 (-\eta )^{d-2 \Delta +2} H_{\frac{1}{2} (2 \Delta -d -4)}^{(2)}(-k \eta )}{H_{\frac{1}{2} (2 \Delta -d)}^{(2)}(-k \eta )} \, .
\end{equation}

Adding the counterterm contributions from \eqref{eqn:ct_scalar} and \eqref{eq:CT1stbarrier} we get
\begin{equation}
    F(k;\eta)= i\left((d-\Delta)(-\eta)^{d-2\Delta}-\frac{(-\eta)^{d-2\Delta+2}}{d-2\Delta+2}\,k^2\right) + \Tilde{F}(k;\eta) \, , \label{eqn:solnnextstrip}
\end{equation}
which is simply the analytic continuation of \eqref{eqn:dSCSH2ptCorrelatorFlowEquationSolutionCT} to this complex strip.\footnote{Setting again $\Delta=\frac{d}{2}+\nu$ makes this evident.}

It will be of interest for later reference to write down the flow equation for the particular case of $\Delta=d=3$, which falls in this strip. It is 
\begin{align}
    i(-\eta)\frac{\partial}{\partial(-\eta)}\Tilde{F}(k;\eta)&= (-\eta)\,k^4-2i(-\eta)^2\,k^2\Tilde{F}(k;\eta) - (-\eta)^3\,\Tilde{F}^2(k;\eta)\\
    &= (-\eta)  \left(k^2-i (-\eta) \Tilde{F}(\eta )\right)^2 \, .\label{eqn:flowDelta=d=3}
\end{align}
In Section \ref{sec:gravd=3}, we will show that this matches precisely the corresponding flow of the stress-tensor two-point function.

The solution to \eqref{eqn:flowDelta=d=3} satisfying the initial conditions \eqref{eqn:dSCSHFlowEquationInitialCondition} and \eqref{eqn:CPTphase} can be written as
\begin{equation}\label{eqn:solutionDelta=d=3}
    \Tilde{F}(k;\eta)=\frac{k^3}{(1-ik\eta)}
\end{equation}
Adding in the counterterm in \eqref{eqn:CTDelta=3} we thus find 
\begin{equation}
    F(k;\eta)= \frac{i k^2}{(-\eta)} + \Tilde{F}(k;\eta) = \frac{i k^2}{(-\eta)\left(1+ik(-\eta)\right)} \, .
\end{equation}
which agrees with the previously obtained solution \eqref{massless-no-CT} from the bulk computation.  

\subsection{The Graviton Case}
Next, we compute the flow for the stress-tensor two-point function, in the presence of this massive scalar field in the bulk. We will relate this to the graviton two-point function later.

Here, we make the identification
\begin{align}
    J &\leftrightarrow h_{ab}\\
    \mathcal{O} &\leftrightarrow \frac{2}{\sqrt{h}}\frac{\delta}{\delta h_{ab}}=T^{ab}
\end{align}

Due to equation \eqref{eqn:mixed} and the discussion in Section \ref{sec:largeN}, the contributing terms from the deformation operator include only pure-gravity terms. At leading order in $1/N$ there is, therefore, no gravity--matter coupling contribution and we can compute the flow of the stress-tensor correlator as if we were dealing with a pure gravity bulk and its corresponding holographic deformation
\begin{align}
    \hat{O}=-i\frac{1}{\alpha}(-\eta)^d\int_\Sigma d^dz \frac{1}{\sqrt{h}}\mathcal{G}_{abcd}\,\Tilde{\pi}^{ab}_{(n)}\Tilde{\pi}^{cd}_{(n)}
\end{align}
where we remind the reader that  $\Tilde{\pi}^{ab}_{(n)}=\pi^{ab}+\sqrt{h}\,\mathbf{H}^{ab}_n=\pi^{ab}+\sqrt{h}\sum_{m=1}^n(-\eta)^{\Delta_{\mathcal{F}_m}-d}\mathbf{H}^{ab}_{\mathcal{F}_m}$ where $\{\mathbf{H}^{ab}_{\mathcal{F}_m}: m=1,...,n\}$ are the shifts generated by all the necessary pure-gravity counterterms for dimension $d$. Here we do not include the $\sim \phi^2$ shift (present in \eqref{eqn:tildepi}) because this does not contribute, for the reason in the preceding paragraph.

We note that pure gravity counterterms scale like
\begin{align}
    \mathcal{F}_m \sim O(\alpha)\sim O(N^p) \implies \mathbf{H}^{ab}_{\mathcal{F}_m}\sim O(N^p)
\end{align}
This guarantees that all terms contribute at the same order $O(N^p)$  to the flow. 

The quadratic operator reads
\begin{align}
    \hat{O}^{(2)} = i\frac{1}{\alpha}(-\eta)^d\int_\Sigma\,\sqrt{h}\, :\frac{1}{4}\mathcal{G}_{abcd}\,T^{ab}T^{cd}:(z) =:i\frac{1}{\alpha}(-\eta)^d\int_\Sigma\,\sqrt{h}\, :T^2:(z)\,, 
\end{align}
which we call the ``$T^2$ operator'' by virtue of it being quadratic in the stress-tensor.  Looking at the first line of \eqref{eqn:flow} we see that $\hat{O}^{(2)}$ will contribute the following term to the flow
\begin{align}
    i\frac{2}{\alpha} (-\eta)^d \int_\Sigma d^dz \;\sqrt{h(z)}\frac{1}{4}\mathcal{G}_{pqmn}\,\widetilde{\expval{T^{ab}(x)T^{pq}(z)}}_c\widetilde{\expval{T^{cd}(y)T^{mn}(z)}}_c\,.
\end{align}
Because $\widetilde{\expval{T^{ab}(x)T^{pq}(z)}}_c\sim O(N^p)$, then this whole term goes like $\sim O(N^p)\sim O(\alpha)$.

The linear operator reads
\begin{align}
    \hat{O}^{(1)}=-\frac{2}{\alpha}(-\eta)^d \int_\Sigma d^dz \,\sqrt{h}\,\frac{1}{2}\mathcal{G}_{abcd}\,\mathbf{H}^{ab}_n\,T^{cd}(z)\,.
\end{align}
Looking at the second line of \eqref{eqn:flow} we see that it contributes with
\begin{align}
    -\frac{2}{\alpha}(-\eta)^d\int_\Sigma d^dz\, \frac{2}{\sqrt{h(x)}}\frac{\delta}{\delta h_{ab}(x)}\left(\sqrt{h(z)}\frac{1}{2}\mathcal{G}_{pqmn}\mathbf{H}^{pq}_n(z)\right)\widetilde{\expval{T^{cd}(y)T^{mn}(z)}}_c + (x\leftrightarrow y)
\end{align}
Again this goes like $\sim O(N^p)\sim O(\alpha)$.

Finally, the operator proportional to the identity reads
\begin{align}
    \hat{O}^{(0)}=-i\frac{1}{\alpha}(-\eta)^d\int_\Sigma d^dz \sqrt{h}\,\mathcal{G}_{abcd}\mathbf{H}^{ab}_n\mathbf{H}^{cd}_n\,,
\end{align}
which contributes the term (from the third line of \eqref{eqn:flow})
\begin{align}
    -i\frac{1}{\alpha}(-\eta)^d \int_\Sigma d^dz \frac{4}{\sqrt{h(x)h(y)}}\frac{\delta^2}{\delta h_{ab}(x)\delta h_{cd}(y)}\left(\sqrt{h}\,\mathcal{G}_{pqmn}\mathbf{H}^{pq}_n\mathbf{H}^{mn}_n(z)\right)\,,
\end{align}
which goes like $\sim O(N^p)\sim O(\alpha)$. In the end, all the three types of terms contribute to the $(-\eta)\frac{\partial}{\partial(-\eta)}$ derivative of the stress-tensor two-point function at the same order of the correlator itself. This means they generate a non-trivial flow at leading-order in the $1/N$ expansion.

The counterterm contribution is
\begin{align}
    i\sum_{m=1}^n \frac{\#_m}{\Delta_{\mathcal{F}_m}-d}(-\eta)^{\Delta_{\mathcal{F}_m}-d} \frac{4}{\sqrt{h(x)h(y)}}\frac{\delta^2}{\delta h_{ab}(x)h_{cd}(y)}\left(\int_\Sigma d^dz \sqrt{h}\,\mathcal{F}_m\right)\sim O(N^2)\sim O(\alpha)\,.
\end{align}

To proceed, we will again evaluate the result on $h_{ab}=\delta_{ab}$ and go to Fourier space. Under the $SO(d)$ isometry subgroup of $\mathbb{R}^d$, the stress-tensor transforms in the symmetric representation, which decomposes into the trivial (trace part) and the symmetric traceless irreps: $T^{ab}=T\delta^{ab} + \left(T^{ab}-\frac{1}{d}T\delta^{ab}\right)$. In momentum space, the conservation equation $\partial_a T^{ab}=0$ becomes $k_a \Tilde{T}^{ab}=0$. Imposing this and translation invariance in the vacuum state, we have that, in momentum space, the two-point function takes the form
\begin{align}
    \widetilde{\expval{\Tilde{T}^{ab}(\mathbf{k})\Tilde{T}^{cd}(\mathbf{q})}}_c(-\eta)= \left(\Xi(k;\eta)\, \mathbf{P}^{abcd}+\Upsilon(k;\eta)\,\mathbf{P}^{ab}\mathbf{P}^{cd}\right)\delta^{d}(\mathbf{k}+\mathbf{q})\,,
\end{align}
where
\begin{align}
    \mathbf{P}^{ab}&:=\delta^{ab}-\frac{k^ak^b}{k^2}\,,\\
    \mathbf{P}^{abcd}&:= \frac{1}{2}(\mathbf{P}^{ac}\mathbf{P}^{bd}+\mathbf{P}^{ad}\mathbf{P}^{bc})-\frac{1}{d-1}\mathbf{P}^{ab}\mathbf{P}^{cd}\,,
\end{align}
and $\Xi(k;\eta)$ and $\Upsilon(k;\eta)$ are two undetermined scalar functions. 

Using the relations $\mathcal{G}_{pqmn}\mathbf{P}^{abpq}\mathbf{P}^{cdmn}=\mathbf{P}^{abcd}$, $\mathcal{G}_{pqmn}\mathbf{P}^{pq}\mathbf{P}^{mn}=0$ and $\mathcal{G}_{pqmn}\mathbf{P}^{abpq}\mathbf{P}^{mn}=0$ we get that the quadratic term contributes as 
\begin{align}
    i\frac{1}{2\alpha}(-\eta)^d\,\Xi(k;\eta)^2\,\mathbf{P}^{abcd}\,\delta^d(\mathbf{k}+\mathbf{q})\,.
\end{align}
To get the other terms we need to plug in the explicit expressions for $\mathbf{H}^{ab}_n$ in order to compute the functional derivatives. We now write down the answer in low dimensions to illustrate the result. The procedure becomes increasingly complicated as $d$ increases. 

\subsubsection{$d=2$}

We have no counterterm beyond $\text{CT}_\mathcal{W}$. Thus, $\hat{O}^{(1)}=0$. The source term is proportional to the Einstein-Hilbert action in $d=2$, see \eqref{eq:Opuregravd=2}. Thus, its second derivative is given by the graviton propagator. Including all the factors, in momentum space it contributes as
\begin{align}
    i \frac{\alpha}{2}\,k^2\, \mathbf{P}^{abcd}\delta^2(\mathbf{k}+\mathbf{q})\,.
\end{align}
So we see that the flow equation only appears to be non-trivial for $\Xi(k;\eta)$. However, in $d=2$ the transverse-traceless projector is trivial: $\mathbf{P}^{abcd}_{(d=2)}=0$. This means that
\begin{align}
    \widetilde{\expval{\Tilde{T}^{ab}(\mathbf{k})\Tilde{T}^{cd}(\mathbf{q})}}_c^{(d=2)}(-\eta)&= \Upsilon(k;\eta)\,\mathbf{P}^{ab}\mathbf{P}^{cd}\delta^{2}(\mathbf{k}+\mathbf{q})\,,\\
    (-\eta)\frac{\partial}{\partial(-\eta)}\Upsilon(k;\eta)&=0\,.
\end{align}

The fact that the two-point function does not flow is in correspondence with the fact that pure gravity in $2+1$-dimensions has no graviton degrees of freedom. We fix the answer by the initial condition
\begin{align}
    \widetilde{\expval{\Tilde{T}^{ab}(\mathbf{k})\Tilde{T}^{cd}(\mathbf{q})}}_c^{\text{CFT}_2}=i\,4\pi^2\alpha\,k^2\,\mathbf{P}^{ab}\mathbf{P}^{cd}\delta^2(\mathbf{k}+\mathbf{q})=\widetilde{\expval{\Tilde{T}^{ab}(\mathbf{k})\Tilde{T}^{cd}(\mathbf{q})}}_c(-\eta)\,,\quad\forall\eta\,,
\end{align}
which follows from the trace anomaly in the $2d$ CFT, see \eqref{eqn:anomalyd=2}. Note the important imaginary sign required by dS/CFT. The counterterm contribution vanishes.

\subsubsection{$d=3$} \label{sec:gravd=3}

Looking at the deformation operator in \eqref{eq:Opuregravd=3}, we can compute the contribution from the $\hat{O}^{(1)}$ term in momentum space
\begin{align}
    -2(-\eta)^2 \, k^2\,\Xi(k;\eta)\,\mathbf{P}^{abcd}\delta^3(\mathbf{k}+\mathbf{q})\,.
\end{align}
Similarly, for the $\hat{O}^{(0)}$ term we get the contribution
\begin{align}
    -2\,i\,(-\eta)\alpha\,k^4\,\mathbf{P}^{abcd}\,\delta^{3}(\mathbf{k}+\mathbf{q})\,.
\end{align}
Combining everything together we see that, again, only the coefficient in front of the transverse-traceless projector flows at leading order
\begin{align}
    i(-\eta)\frac{\partial}{\partial(-\eta)}\Xi(k;\eta)&=2\alpha(-\eta)\,k^4-2i(-\eta)^2\,k^2\,\Xi(k;\eta) -\frac{1}{2\alpha}(-\eta)^3\,\Xi(k;\eta)^2\\
    &=-\frac{(-\eta)}{2\alpha}\left((-\eta)\Xi(k;\eta)+2i\alpha\,k^2\right)^2\\ \label{eqn:flowgravitond=3}
    (-\eta)\frac{\partial}{\partial(-\eta)}\Upsilon(k;\eta)&=0\,.
\end{align}

We now note that the flow equation above matches exactly the flow of the scalar two-point function for $\Delta=d=3$ once we correct for the different normalisations of the kinetic terms in the action. Namely, with the identification
\begin{align}
    \Xi(k;\eta)\leftrightarrow2\alpha \Tilde{F}(k;\eta)\label{eqn:identify}
\end{align}
equation \eqref{eqn:flowgravitond=3} reproduces exactly \eqref{eqn:flowDelta=d=3}. This is telling us that one can, equivalently, compute the flow for a massless scalar (i.e. with $\Delta=d$), multiply the result by $2\alpha$ and then simply append the tensor structure $\mathbf{P}^{abcd}$ to get the transverse-traceless component of the stress-tensor two-point function. Because it is much easier to compute the flow of the scalar, this is a good trick to follow in higher dimensions. The tensor $\mathbf{P}^{abcd}$ automatically takes into account the right number of graviton polarisations for each $d$. The component proportional to $\mathbf{P}^{ab}\mathbf{P}^{cd}$ we get from the CFT initial condition. We note that this equivalence between graviton and massless scalar only holds on certain backgrounds, like $h_{ab}=\delta_{ab}$, because of the non-linear effects coming from the gravity deformation discussed towards the end of Section \ref{sec:largeN}, which are not captured by the matter part of the deformation.

In the CFT limit, we have that there is no anomaly in $d=3$, so we get that 
\begin{align}
    \delta_{ab}\widetilde{\expval{\Tilde{T}^{ab}(\mathbf{k})\Tilde{T}^{cd}(\mathbf{q})}}_c^{\text{CFT}_3} = 2\Upsilon(k;0)\, \mathbf{P}^{cd}\,\delta^3(\mathbf{k}+\mathbf{q})\stackrel{!}{=}0 \implies \Upsilon(k;\eta)=0\,.
\end{align}
The transverse-traceless mode $\Xi(k;\eta)$ is obtained from the scalar two-point function in \eqref{eqn:solutionDelta=d=3} via the identification \eqref{eqn:identify}. So we have the final answer
\begin{equation}\label{eqn:StressTensord=3}
    \widetilde{\expval{\Tilde{T}^{ab}(\mathbf{k})\Tilde{T}^{cd}(\mathbf{q})}}_c(-\eta)= 2\alpha\, \frac{k^3}{(1-ik\eta)}\, \mathbf{P}^{abcd}\,\delta^3(\mathbf{k}+\mathbf{q})\,.
\end{equation}
This matches precisely with equation \eqref{eqn:b_full_CT} obtained from the bulk.\footnote{We must remember that $T^{ab}=\frac{2}{\sqrt{h}}\frac{\delta}{\delta h_{ab}}$ to get the right factor of $2$, i.e.~it differs from $b(\bfk)$ in Section \ref{sec:FreeGravitondS} by a factor of $4$.} Including the now non-trivial counterterm contribution we get
\begin{equation}
    \expval{\Tilde{T}^{ab}(\mathbf{k})\Tilde{T}^{cd}(\mathbf{q})}_c(-\eta) = 2\alpha \,\frac{i k^2}{(-\eta)\left(1+ik(-\eta)\right)}\, \mathbf{P}^{abcd}\,\delta^3(\mathbf{k}+\mathbf{q})\,.
\end{equation}

\section{Discussion}\label{sec:Discussion}
Assuming the existence of dS/CFT, we have showed how to deform it so as to obtain a dual that lives on a finite $\eta$ Cauchy slice.  The resulting theory satisfies the Hamiltonian constraint equation, and therefore has an emergent holographic time dimension. By considering small perturbations of sources around the flat case, we have showed in Sections  \ref{sec:Boundary} to \ref{sec:BoundaryFlow} how to recover the correct cosmological two-point function for a scalar field $\phi$ and the metric $h_{ab}$, which we previously calculated in Sections \ref{sec:Wavefunction} and \ref{sec:BulkPropagator} from the bulk.  Presumably, higher point functions also agree. This theory could also in principle be studied on spatial slices that are not very close to flat, thus exhibiting the ``many fingered time'' aspect of spacetime diffeomorphisms.

\paragraph{Future Directions.} To further strengthen the correspondence we advocate here, it would be important to recover higher-point wavefunction coefficients from similar flow computations at large $N$. This could be achieved for example for the three-point functions (which have previously been computed at the late-time boundary~\cite{Maldacena:2011nz,Pajer:2020wxk,Cabass:2021fnw,Cabass:2022jda,Jazayeri:2025vlv}), following the general prescription provided in Section \ref{sec:BoundaryFlow}. It would also be interesting to conduct a similar analysis, even for the two-point function, on closed spatial slices. Furthermore, going beyond exactly-dS dual spacetimes to more general FLRW spacetimes would be another non-trivial check of the duality (see e.g~\cite{Arkani-Hamed:2023bsv,Arkani-Hamed:2023kig} where the late-time wavefunction coefficients were computed for general flat FLRW spacetimes). A less straightforward, but important task, is to incorporate $1/N$ corrections to the flow equations. We also find it worthwhile to show consistency between the Cauchy Slice Holography dS/CFT dictionary and the $T\Bar{T}+\Lambda$ proposal of \cite{Gorbenko:2018oov,Shyam:2021ciy,Coleman:2021nor,Batra:2024kjl}, by computing the deformation of AdS/CFT correlators along this combined flow. More conceptually, the Cauchy Slice Holography paradigm has time emerging from the RG flow of the dual theory. Here, we see this very explicitly on a state with a semiclassical dual geometry. How should we make sense of this when the bulk state is truly quantum-gravitational?

\paragraph{Are we living in a generic RG flow?} Unlike the spacetimes considered when renormalising AdS/CFT, a flat FLRW universe is a good approximation to our own universe.  This raises the question of whether we can, or should, think of our own universe as sitting in some type of RG flow.  Perhaps, gravity and time are a \emph{generic} emergent consequence of renormalising some class of Euclidean field theories?

Suppose we assume that the $T^2$-deformed theory can somehow be UV completed into a consistent field theory.  As we have seen, this field theory cannot be unitary in the usual sense (reflection positivity of Euclidean correlators), but it should have some other properties which imply that the bulk is unitary, which have been discussed in~\cite{Goodhew:2020hob,Goodhew:2024eup,Thavanesan:2025kyc,Thavanesan:2025ibm}. Let us assume that this UV completed theory is approximately local at distances much smaller than the dS scale.

So long as the bulk theory is Lorentz invariant (which is a natural consequence of flowing to a boundary theory which has conformal symmetry at late times), it seems that a Hamiltonian formulation of such a bulk theory would necessarily result in a Hamiltonian constraint which satisfies the standard closure relation
\begin{equation}
\Big[ \mathcal{H}(x), \mathcal{H}(y)\Big]=i\left(\mathcal{D}^a(x)\partial_a^{(x)}-\mathcal{D}^a(y)\partial_a^{(y)}\right)\delta(x-y),
\end{equation}
as this equation is the implementation of Lorentz-invariance in the Hamiltonian formalism.  Here, $\mathcal{D}^a$ is the spatial diffeomorphism constraint, defined in \eqref{eqn:Dconst} for pure gravity.  This Hamiltonian constraint $\cal H$ could be very complicated, but as we flow away from the UV fixed point, terms with a sufficiently high number of bulk derivatives should become less important, and so it seems plausible that in a large $N$ limit,\footnote{This assumption is needed to ensure that one can take products of single-trace operators, without large quantum corrections.} one universally ends up with the simplest Hamiltonian constraint \eqref{eqn:Hconst} with up to two bulk derivatives satisfying the closure relation, which in the gravitational sector is:
\begin{equation}
{\cal H} = \kappa{\sqrt{g}}:\!\Big(\Pi_{ab}\Pi^{ab}-\frac{1}{d-1}\Pi^2\Big)\!:-\sqrt{g}(\kappa^{-1}R-2\Lambda) = 0.
\end{equation}
for some coefficients $\kappa,\Lambda$.\footnote{The overall multiplicative factor of ${\cal H}$ doesn't matter since we set it equal to zero.}  If the field theory is chosen to satisfy the reality conditions in~\cite{Goodhew:2020hob,Goodhew:2024eup,Thavanesan:2025ibm}, this will ensure that these parameters do not have random complex phases, and then there is perhaps a 50\% chance of getting each real sign correct.  We might be left with the usual fine-tuning problem for why $\Lambda$ is smaller than the Planck scale,\footnote{In AdS/CFT this is usually a consequence of the number of fields $N$ being large, but this might be misleading since AdS/CFT is typically supersymmetric, and thus $\Lambda$ is protected from quantum loop corrections.  Since dS/CFT does not allow supersymmetry with a unitary bulk, it might turn out to be much harder to find examples of dS/CFT with $\Lambda \ll 1$ in Planck or string units.} but as we cannot solve all the problems in one go, perhaps it would be better to modify our thesis to say that we live in a generic RG flow \emph{apart from this well-known naturalness problem}, i.e.~that only a single tuning is necessary to see gravity energy.

\paragraph{Naturalness in Holographic Cosmology.}  Actually it is a nontrivial question whether dS/CFT respects our physical intuitions about naturalness in the first place.  In principle, a sufficiently weird duality between field theories might map a proposition that looks ``natural'' to physicists studying one side, to something that looks ``unnatural'' to physicists studying the other side.  Of course, discovering such a duality would then problematise these notions of naturalness, since the duality would not respect it.

In the usual perspective on holographic renormalisation, points on the boundary of the bulk (${\cal I}^+$) are associated with the UV of the field theory, while points deep in the bulk are associated with IR dynamics.  Hence in the case of dS/CFT, we have:
\begin{eqnarray}
\text{early times} &=& \text{IR};\label{earlyIR}\\
\text{late times} &=& \text{UV}.\label{lateUV}
\end{eqnarray}
Now suppose we turn on some single-trace sources in the CFT, so that we have a nontrivial FLRW cosmology.  Let us suppose now we wish to find a dual field theory which accounts for the cosmology we live in, including the Second Law of Thermodynamics.\footnote{In a closed universe, it is plausible that dS/CFT will always describe a Hartle-Hawking state which maximises entropy in every static patch, and thus cannot have a thermodynamic arrow.  But that would make it impossible to use dS/CFT to describe our world.  Hence, let us assume that the universe is either i) open, or ii) described by a dual field theory where one obtains a more general state.}  Now, in our own cosmology, the Second Law of Thermodynamics tells us that the state at late times arises as a result of a \emph{low entropy} initial condition at very early times; thus we have the following arrow of scientific explanation:
\begin{equation}\label{arrow1}
\text{early times} 
\longrightarrow 
\text{late times},
\end{equation}
but not vice versa.  In our world, this arrow is irreversible because the final boundary condition is disordered and does not satisfy any additional restrictions, besides coming from the initial condition.  On the other hand, in a QFT, the Wilsonian picture of renormalisation tells us to think of the long distance field theory as an \emph{effective field theory} whose properties are determined by the theory at short scales.  This gives us another arrow:
\begin{equation}\label{arrow2}
\text{UV} 
\longrightarrow 
\text{IR},
\end{equation}
but not vice versa.  Reversing this order would involve a failure of Wilsonian ``naturalness'' in the boundary field theory.  But it can be seen that the four relations \eqref{earlyIR}--\eqref{arrow2} together add up to a philosophical conflict: the usual order for doing scientific explanations does not agree between the bulk cosmology and the boundary field theory.  Thus, any dS/CFT account of our cosmos would have an apparent failure of ``naturalness'', either with respect to bulk thermodynamics or with respect to boundary RG flow.\footnote{The arrows could be made to coincide by placing the dual field theory at ${\cal I}^-$ instead of ${\cal I}^+$, but this would describe a cosmology very different from our own where thermodynamic arrow goes the opposite direction from the cosmological arrow of expansion.}

But the $T^2$ formalism described in this paper suggests a completely different picture from the above.  In our work, the holographic CFT is an \emph{IR limit} of the deformed theory, as follows from the fact that it is an irrelevant deformation.\footnote{For those who find this switch counter-intuitive, it may help to try thinking of the $T^2$ deformation first as a \emph{UV cutoff scale} of a usual AdS/CFT.  If we place a cutoff very near the boundary, at $z = \epsilon$, it is evident that this cutoff will be most important for bulk physics at small $z$, in the UV of the field theory. But that means that if we alternatively think of the wall at $z = \epsilon$ as an \emph{actual physical effect}, that effect will be more important in the UV than in the IR; hence it is described by an irrelevant deformation, and hence we recover the original holographic CFT in the IR.  The same argument applies to $T^2$ deformed dS/CFT, after doing the replacement $z \to -\eta$.}  Hence, the late time limit corresponds to the IR, not the UV:
\begin{equation}
\text{late times} = \text{IR},\label{lateIR}
\end{equation}
contrary to the associations \eqref{earlyIR}--\eqref{lateUV} in the single-trace RG picture.  However, the UV does not correspond to early times (unless the Cauchy slice $\Sigma$ is itself at early times); instead it corresponds to going to sub-dS scales, on the boundary time slice, which is the UV of the bulk theory.  

Thus, by thinking of the RG flow in terms of a $T^2$ deformed theory, we thus avoid the unnatural dictionary mapping that occurs in the usual dS/CFT dictionary, when the CFT is taken to be at ${\cal I}^+$.  While this might not be as attractive for those hoping to solve fine-tuning problems, it makes the overall picture of holographic cosmology more physically plausible.

\section*{Acknowledgements}
AT is grateful to Victor Gorbenko for helpful discussions. GAR thanks Rifath Khan for initial collaboration. The authors are also grateful to Ahmed Almheiri, Matthew Blacker, Laurent Freidel, Monica Guica, Tom Hartman, Simon Lin, Juan Maldacena.  Paul McFadden, Vasudev Shyam, Eva Silverstein, Kostas Skenderis, and Ronak Soni for helpful discussions. 

GAR was supported by a Harding Distinguished Postgraduate Scholarship at the University of Cambridge, as well as funding from Okinawa Institute of Science and Technology Graduate University and support from ID\# 62312 grant from the John Templeton Foundation, as part of \href{https://www.templeton.org/grant/the-quantum-information-structure-of-spacetime-qiss-second-phase}{‘The Quantum Information Structure of Spacetime’ Project (QISS)}. The opinions expressed in this project are those of the authors and do not necessarily reflect the views of the John Templeton Foundation.  GAR, AT and AW were supported by the AFOSR grant FA9550-19-1-0260 “Tensor Networks and Holographic Spacetime”.  GAR and AW are grateful for the hospitality of the KITP in early 2020, during the ``Gravitational Holography'' program, during which this work was supported in part by the National Science Foundation under Grant No. NSF PHY-1748958.  AW was also supported by the STFC grant ST/P000681/1 “Particles, Fields and Extended Objects”, and by the subsequent STFC grant ST/X000664/1 “Quantum Fields, Quantum Gravity and Quantum Particles.  AT was supported in part by the Heising-Simons Foundation, the Simons Foundation, the Bell Burnell Graduate Scholarship Fund, the Cavendish (University of Cambridge) and a KITP Graduate fellowship. AT also acknowledges support from the SNF starting grant “The Fundamental Description of the Expanding Universe.  AT also gratefully acknowledges hospitality from the Perimeter Institute while working on this article. AW was also partly supported by NSF grant PHY-2207584 while working on this paper during his sabbatical at the IAS.

\newpage

\bibliographystyle{JHEP}
\bibliography{refsdSCSH}

\end{document}